\documentclass[preprint,12pt,authoryear]{elsarticle}
\usepackage[a4paper, top=2cm, bottom=2cm, left=2cm, right=2cm]{geometry}
\usepackage{framed,multirow}\setcitestyle{square,numbers,sort&compress}
\biboptions{numbers,sort&compress}
\usepackage{amssymb}
\usepackage{amsmath}
\usepackage{latexsym}
\usepackage{bm}
\usepackage{amssymb}
\usepackage{subfigure}
\usepackage{booktabs}
\usepackage{url}
\usepackage[colorlinks=true,linkcolor=black,citecolor=blue,urlcolor=blue,]{hyperref}
\usepackage[doipre={doi:~}]{uri}
\usepackage{xcolor}
\usepackage{listings}
\usepackage{threeparttable}
\numberwithin{equation}{section}

\begin{document}

\begin{frontmatter}
\title{A Unified Framework for Bell-Shaped Gaussian Quadratures in Deterministic Boltzmann Solvers} 

\author[inst1]{Shanshan Dong}
\author[inst1,inst2]{Lu Wang\corref{cor}}
\ead{wanglu@hdu.edu.cn}
\author[inst1]{Guanqing Wang}

\cortext[cor]{Corresponding author.}

\affiliation[inst1]{
	organization={School of Physics, Hangzhou Dianzi University},
	city={Hangzhou},
	postcode={310018},
	country={China}
}

\affiliation[inst2]{
	organization={Zhejiang Key Laboratory of Industrial Carbon Metrology Technology},
	city={Hangzhou},
	postcode={310018},
	country={China}
}
\begin{abstract}
	
In this work, we develop a unified theoretical framework that constructs Bell-shaped Gaussian quadratures for deterministic Boltzmann solvers.
The proposed framework establishes a general class of bell-shaped weight functions together with the corresponding integral transformations, revealing the common mathematical structure underlying existing generalized Gaussian quadratures.
Based on this framework, a unified procedure for constructing discrete velocity sets is developed by combining radial quadratures with appropriate directional quadratures in different velocity-space dimensions.
Several existing generalized Gaussian quadratures are naturally recovered as special cases, while the proposed framework provides a systematic route for constructing new quadrature rules based on different bell-shaped weight functions.
Extensive benchmark simulations over a broad range of flow regimes demonstrate that the proposed framework achieves high integration accuracy and computational efficiency with significantly fewer discrete velocities than conventional quadrature strategies.
Overall, the proposed framework not only unifies existing generalized Gaussian quadratures under a common mathematical foundation but also establishes a general design principle for the systematic development of efficient velocity-space quadratures in deterministic Boltzmann solvers.

\end{abstract}
\end{frontmatter}

\section{Introduction}

Accurate and efficient simulation of multiscale gas flows remains a fundamental challenge in computational fluid dynamics, as the local Knudsen number may vary by several orders of magnitude within a single flow field. This multiscale behavior arises in a broad range of applications, including hypersonic aerodynamics \cite{hypersonic1,hypersonic2}, spacecraft re-entry \cite{spacecraft1,spacecraft2}, microelectromechanical systems (MEMS) \cite{MEMS1,MEMS2}, and vacuum technologies \cite{vacuum1,vacuum2}. The Boltzmann equation provides a unified kinetic description of gas dynamics from the continuum to the rarefied regime \cite{Cercignani1988,Bird1994,Sone2007}. However, developing an efficient Boltzmann solver that remains accurate over a wide range of Knudsen numbers is a long-standing challenge. Traditionally, numerical methods for solving the Boltzmann equation can be broadly classified into stochastic and deterministic approaches. Stochastic methods, typified by the Direct Simulation Monte Carlo (DSMC) method \cite{DSMC1,Bird1994,DSMC2,DSMC3}, simulate molecular transport and collisions using computational particles and have become the standard approach for rarefied gas dynamics. However, the statistical noise inherent in particle sampling limits their efficiency for low-speed and steady-state flows, where a large number of particles and extensive averaging are required. Conversely, deterministic approaches, represented by the Discrete Velocity Method (DVM) \cite{DVM1,DVM2,DVM3}, directly solve the Boltzmann equation in a discretized velocity space, thereby eliminating statistical noise. Nevertheless, discretizing the global velocity space incurs substantial computational and memory costs owing to the curse of dimensionality, particularly for hypersonic and highly nonequilibrium flows, where both a large velocity domain and fine velocity resolution are required. Moreover, both conventional DSMC and DVM rely on operator splitting to treat molecular transport and collisions separately, leading to stringent restrictions on the mesh size and time step.

To overcome the limitations of conventional Boltzmann solvers, a variety of asymptotic-preserving (AP) \cite{AP1,AP2,AP3} multiscale methods have been developed by coupling molecular transport and collisions within a unified evolution framework. The unified gas-kinetic scheme (UGKS) \cite{UGKS1,UGKS2} constructs numerical fluxes from the integral solution of the kinetic equation, thereby enabling multiscale simulations without explicitly resolving the microscopic spatial and temporal scales. The discrete unified gas-kinetic scheme (DUGKS) \cite{DUGKS1,DUGKS2,DUGKS3} further simplifies the flux evaluation through the characteristic solution of the kinetic equation while retaining the multiscale capability of UGKS. Alternatively, the unified gas-kinetic wave-particle (UGKWP) \cite{UGKWP1,UGKWP2} method combines wave and particle representations through an adaptive decomposition to improve computational efficiency, whereas the general synthetic iterative scheme (GSIS) \cite{GSIS1,GSIS2,GSIS3} achieves both asymptotic-preserving and fast-converging properties by tightly coupling kinetic equations with macroscopic synthetic equations. These developments have greatly expanded the efficiency and applicability of multiscale Boltzmann solvers across different flow regimes.

Among these AP methods, deterministic approaches based on the discrete velocity framework remain particularly attractive because they provide noise-free solutions with high numerical accuracy. While AP methods have largely overcome the stringent constraints imposed by spatial and temporal discretizations, the discretization of the velocity space has emerged as the primary computational bottleneck in deterministic Boltzmann solvers. Since macroscopic moments are evaluated through numerical integration in velocity space, the computational efficiency and accuracy of these methods critically depend on the choice of discrete velocity sets and the associated quadrature rules. Accordingly, a variety of quadrature strategies have been developed. Uniform tensor-product discretizations based on Newton–Cotes quadrature \cite{NCq1,NCq2} remain widely used because of their simplicity and robustness for general nonequilibrium flows. Gaussian quadratures, including Gauss–Hermite \cite{GH1,GH2,GH3} and half-range Gauss–Hermite formulations \cite{HRGH1,HRGH2,HRGH3}, provide higher-order integration accuracy for weighted moments and have been extensively adopted in kinetic methods. Alternatively, spherical-coordinate formulations \cite{sphericalcoordinate2012} decompose the velocity-space integration into radial and angular components, allowing different quadrature rules to be combined for each subspace. In addition, sparse-grid techniques and adaptive velocity discretizations \cite{Chen2012,Brull2014} have been proposed to alleviate the computational cost associated with high-dimensional velocity spaces. Despite these advances, existing quadrature strategies still involve fundamental trade-offs among integration accuracy, computational efficiency, and general applicability, motivating the development of more flexible Gaussian quadratures based on problem-specific weight functions.

Recently, an alternative strategy has emerged by designing problem-specific weight functions for Gaussian quadratures. Instead of using fixed quadrature rules, these methods determine the quadrature nodes and weights according to the characteristics of the kinetic distribution function, thereby reducing the number of discrete velocities required for accurate moment integration. Representative examples include the parametric Gaussian quadrature (PGQ) \cite{PGQ}, the arctangent-based Gauss–Jacobi (ATGJ) \cite{ATGJ} quadrature, and the generalized Gauss–Jacobi quadrature (GGJQ) \cite{GGJQ}, which have demonstrated promising performance in multiscale Boltzmann solvers. Although these quadratures are derived from different weight functions, their successful applications suggest the existence of a common mathematical principle governing the construction of generalized Gaussian quadratures. However, existing methods have been developed independently, each tailored to a particular class of weight functions, and their mathematical relationships have yet to be established within a unified theoretical framework. As a result, the construction of new generalized Gaussian quadratures still relies on deriving each formulation separately, limiting both theoretical understanding and the systematic extension to broader classes of weight functions and multidimensional velocity spaces. Therefore, a unified theoretical framework is needed to reveal the common mathematical foundation underlying existing generalized Gaussian quadratures and to provide a systematic design principle for developing new quadrature rules.

To address this gap, we develop a unified framework for generalized Gaussian quadratures applicable to deterministic Boltzmann solvers. The proposed framework establishes a general class of bell-shaped weight functions and derives the corresponding integral transformations in velocity spaces of different dimensions, thereby revealing the unified mathematical structure underlying existing generalized Gaussian quadratures. Based on this framework, a unified procedure is developed for systematically constructing discrete velocity sets by combining radial and angular quadratures in different velocity-space dimensions. Furthermore, the proposed framework unifies several existing generalized Gaussian quadratures within a single theoretical formulation, clarifying their mathematical relationships while providing a systematic basis for developing new quadrature rules. Extensive numerical simulations are performed to validate the accuracy, efficiency, and generality of the proposed framework across a wide range of benchmark problems.

\section{Mathematical formulations}
\label{sec2}

This section presents the kinetic formulation adopted in the present study. The BGK-Shakhov kinetic model is first introduced for gas flow simulations. To reduce the dimensionality of the velocity space and improve computational efficiency, a reduced distribution function formulation is subsequently employed. The governing equations are then nondimensionalized, and the associated dimensionless parameters are introduced. Finally, the boundary conditions prescribed at solid walls and far-field boundaries are described.

\subsection{BGK-Shakhov kinetic model}

In kinetic theory, gas transport phenomena are described by the molecular velocity distribution function. The evolution of the distribution function \(f\) in phase space is governed by the Boltzmann equation. Because the full Boltzmann collision operator is computationally expensive, the BGK-Shakhov model is employed in the present work to approximate intermolecular collisions \cite{Shakhov1968}:
\begin{equation}
	\frac{\partial f}{\partial t}
	+
	\boldsymbol{\xi}\cdot\nabla_{\boldsymbol{x}} f
	=
	\Omega(f)
	=
	-\frac{1}{\tau}
	\left(
	f-f^{S}
	\right),
	\label{eq:Shakhov}
\end{equation}
where \(f=f(\boldsymbol{x},\boldsymbol{\xi},\boldsymbol{\eta},\boldsymbol{\zeta},t)\) is the molecular velocity distribution function, \(\boldsymbol{x}\) denotes the spatial coordinate, \(\boldsymbol{\xi}\) is the translational molecular velocity, \(\boldsymbol{\eta}=(\xi_{D+1},\ldots,\xi_3)\) denotes the remaining components of the three-dimensional molecular velocity, while \(\boldsymbol{\zeta}\) represents the internal degrees of freedom of the molecules, \(\Omega(f)\) is the collision operator, \(\tau\) is the relaxation time, and \(f^{S}\) denotes the Shakhov equilibrium distribution.

The relaxation time \(\tau\) is related to the dynamic viscosity \(\mu\) and pressure \(p\) through \(\tau=\mu/p\). For an ideal gas, the equation of state reads
\(p=\rho RT\), with \(\rho\), \(T\), and \(R\) denoting the gas density,
temperature, and specific gas constant, respectively. The viscosity follows a power-law dependence on temperature.
\begin{equation}
	\mu
	=
	\mu_{\mathrm{ref}}
	\left(
	\frac{T}{T_{\mathrm{ref}}}
	\right)^{\omega},
\end{equation}
with \(\mu_{\mathrm{ref}}\) and \(T_{\mathrm{ref}}\) representing the
reference viscosity and temperature, respectively, and \(\omega\) denoting
the viscosity index.

To recover the correct Prandtl number, the Shakhov model supplements the Maxwellian equilibrium distribution \(f^M\) with a heat-flux correction:
\begin{equation}
	f^{S}
	=
	f^{M}
	\left[
	1
	+
	(1-\mathrm{Pr})
	\frac{
		\boldsymbol{c}\cdot\boldsymbol{q}
	}
	{5pRT}
	\left(
	\frac{
		|\boldsymbol{c}|^{2}
		+
		|\boldsymbol{\eta}|^{2}
	}
	{RT}
	-5
	\right)
	\right],
\end{equation}

\begin{equation}
f^{M}
=
\frac{\rho}
{(2\pi RT)^{(K+3)/2}}
\exp
\left(
-\frac{
	|\boldsymbol{c}|^{2}
	+
	|\boldsymbol{\eta}|^{2}
	+
	|\boldsymbol{\zeta}|^{2}
}
{2RT}
\right),
\end{equation}
where \(\mathrm{Pr}\) denotes the Prandtl number, \(K\) is the number of internal degrees of freedom, and \(\boldsymbol{c}=\boldsymbol{\xi}-\boldsymbol{u}\) is the peculiar velocity relative to the macroscopic flow velocity \(\boldsymbol{u}\).

The macroscopic flow variables are obtained from velocity moments of the distribution function.
In particular, the conservative variables are defined as
\begin{equation}
	\boldsymbol{W}
	=
	\begin{pmatrix}
		\rho \\
		\rho\boldsymbol{u} \\
		\rho E
	\end{pmatrix}
	=
	\int
	\begin{pmatrix}
		1 \\
		\boldsymbol{\xi} \\
		\dfrac{
			|\boldsymbol{\xi}|^{2}
			+
			|\boldsymbol{\eta}|^{2}
			+
			|\boldsymbol{\zeta}|^{2}
		}{2}
	\end{pmatrix}
	f
	\,
	d\boldsymbol{\xi}
	d\boldsymbol{\eta}
	d\boldsymbol{\zeta},
\end{equation}
where \(E\) denotes the total energy per unit mass.
The heat-flux vector is obtained from the third-order moment of the distribution function:
\begin{equation}
	\boldsymbol{q}
	=
	\frac{1}{2}
	\int
	\boldsymbol{c}
	\left(
	|\boldsymbol{c}|^{2}
	+
	|\boldsymbol{\eta}|^{2}
	+
	|\boldsymbol{\zeta}|^{2}
	\right)
	f
	\,
	d\boldsymbol{\xi}
	d\boldsymbol{\eta}
	d\boldsymbol{\zeta}.
\end{equation}
The temperature is recovered from the total energy and macroscopic flow velocity according to
\begin{equation}
	T
	=
	\frac{\gamma-1}{R}
	\left(
	E
	-
	\frac{1}{2}
	|\boldsymbol{u}|^{2}
	\right),
\end{equation}
where \(\gamma=(K+5)/(K+3)\) is the specific heat ratio.

\subsection{Reduced kinetic formulation}

The BGK--Shakhov model is formulated in an extended phase space that includes the translational molecular velocity and internal degrees of freedom. Although this formulation provides a complete kinetic description, resolving the resulting high-dimensional velocity space is computationally expensive. To reduce the dimensionality while preserving the correct macroscopic moments, the internal degrees of freedom are analytically integrated out, leading to two reduced distribution functions \cite{DVM2, DUGKS2},
\begin{equation}
	g(\boldsymbol{x},\boldsymbol{\xi},t)
	=
	\int
	f(\boldsymbol{x},\boldsymbol{\xi},\boldsymbol{\eta},\boldsymbol{\zeta},t)
	\,
	d\boldsymbol{\eta}\,
	d\boldsymbol{\zeta},
\end{equation}
and
\begin{equation}
	h(\boldsymbol{x},\boldsymbol{\xi},t)
	=
	\int
	\left(
	|\boldsymbol{\eta}|^{2}
	+
	|\boldsymbol{\zeta}|^{2}
	\right)
	f(\boldsymbol{x},\boldsymbol{\xi},\boldsymbol{\eta},\boldsymbol{\zeta},t)
	\,
	d\boldsymbol{\eta}\,
	d\boldsymbol{\zeta}.
\end{equation}

Integrating Eq.~(\ref{eq:Shakhov}) over the internal degrees of freedom yields the reduced kinetic equations:
\begin{equation}
	\frac{\partial g}{\partial t}
	+
	\boldsymbol{\xi}\cdot\nabla_{\boldsymbol{x}}g
	=
	-\frac{1}{\tau}
	(g-g^{S}),
	\label{eq:gEq}
\end{equation}

\begin{equation}
	\frac{\partial h}{\partial t}
	+
	\boldsymbol{\xi}\cdot\nabla_{\boldsymbol{x}}h
	=
	-\frac{1}{\tau}
	(h-h^{S}).
	\label{eq:hEq}
\end{equation}
For a \(D\)-dimensional physical space, the corresponding reduced equilibrium distributions are
\begin{equation}
	g^{S}
	=
	g^{M}
	\left[
	1
	+
	(1-\mathrm{Pr})
	\frac{\boldsymbol{c}\cdot\boldsymbol{q}}
	{5pRT}
	\left(
	\frac{|\boldsymbol{c}|^{2}}
	{RT}
	-
	D
	-
	2
	\right)
	\right],
\end{equation}

\begin{equation}
	h^{S}
	=
	(K+3-D)
	g^{M}
	RT
	\left[
	1
	+
	(1-\mathrm{Pr})
	\frac{\boldsymbol{c}\cdot\boldsymbol{q}}
	{5pRT}
	\left(
	\frac{|\boldsymbol{c}|^{2}}
	{RT}
	-
	D
	-
	\frac{2K}{K+3-D}
	\right)
	\right],
\end{equation}
where the reduced Maxwellian distribution takes the form

\begin{equation}
	g^{M}
	=
	\frac{\rho}
	{(2\pi RT)^{D/2}}
	\exp
	\left(
	-\frac{|\boldsymbol{c}|^{2}}
	{2RT}
	\right).
\end{equation}

The conservative variables and heat flux are recovered from the reduced distribution functions through

\begin{equation}
	\boldsymbol{W}
	=
	\begin{pmatrix}
		\rho\\
		\rho\boldsymbol{u}\\
		\rho E
	\end{pmatrix}
	=
	\int
	\begin{pmatrix}
		g\\
		\boldsymbol{\xi}g\\
		\left(
		|\boldsymbol{\xi}|^{2}g+h
		\right)/2
	\end{pmatrix}
	d\boldsymbol{\xi},
\end{equation}
and
\begin{equation}
	\boldsymbol{q}
	=
	\frac12
	\int
	\boldsymbol{c}
	\left(
	|\boldsymbol{c}|^{2}g+h
	\right)
	d\boldsymbol{\xi}.
\end{equation}

The resulting reduced kinetic system preserves the original conservation properties at a substantially reduced computational cost. To facilitate numerical implementation and comparisons across different flow regimes, the reduced kinetic equations are nondimensionalized using the reference density $\rho_{ref}$, reference length $L_{ref}$, reference temperature $T_{ref}$, and the characteristic molecular speed \(C_{ref}=\sqrt{2RT_{ref}}\). The characteristic scales adopted in the nondimensionalization procedure are summarized in Table~\ref{tab:nondim_variables}. Unless otherwise specified, all variables appearing hereafter are dimensionless, and the superscript ``$^*$'' is omitted for brevity. Under this scaling, the reduced kinetic equations, Eqs.~(\ref{eq:gEq})--(\ref{eq:hEq}), retain the same form as their dimensional counterparts. The equilibrium distributions are rewritten as

\begin{equation}
	g^{S}
	=
	g^{M}
	\left[
	1
	+
	(1-\mathrm{Pr})
	\frac{2\boldsymbol{c}\cdot\boldsymbol{q}}
	{5pT}
	\left(
	\frac{2|\boldsymbol{c}|^{2}}{T}
	-
	D
	-
	2
	\right)
	\right],
\end{equation}

\begin{equation}
	h^{S}
	=
	g^{M}
	\frac{(K+3-D)T}{2}
	\left[
	1
	+
	(1-\mathrm{Pr})
	\frac{2\boldsymbol{c}\cdot\boldsymbol{q}}
	{5pT}
	\left(
	\frac{2|\boldsymbol{c}|^{2}}{T}
	-
	D
	-
	\frac{2K}{K+3-D}
	\right)
	\right],
\end{equation}
where the corresponding dimensionless Maxwellian distribution is
\begin{equation}
	g^{M}
	=
	\frac{\rho}
	{(\pi T)^{D/2}}
	\exp
	\left(
	-\frac{|\boldsymbol{c}|^{2}}
	{T}
	\right),
\end{equation}
and the equation of state becomes

\begin{equation}
	p
	=
	\frac{\rho T}{2}.
\end{equation}

For numerical implementation, the velocity-space moments are approximated by a discrete quadrature. Let
$\{\boldsymbol{\xi}_i,\omega_i\}_{i=1}^{N_v}$ denote the discrete velocity set and the corresponding quadrature weights. The conservative variables and heat flux are then evaluated as
\begin{equation}
	\boldsymbol{W}
	=
	\begin{pmatrix}
		\rho\\
		\rho\boldsymbol{u}\\
		\rho E
	\end{pmatrix}
	=
	\sum_{i=1}^{N_v}
	\omega_i
	\begin{pmatrix}
		g_i\\
		\boldsymbol{\xi}_i g_i\\
		\left(
		|\boldsymbol{\xi}_i|^2g_i+h_i
		\right)/2
	\end{pmatrix},
\end{equation}
and
\begin{equation}
	\boldsymbol{q}
	=
	\frac12
	\sum_{i=1}^{N_v}
	\omega_i
	\boldsymbol{c}_i
	\left(
	|\boldsymbol{c}_i|^2g_i+h_i
	\right),
\end{equation}
where $g_i=g(\boldsymbol{\xi}_i)$, $h_i=h(\boldsymbol{\xi}_i)$, and $N_v$ is the total number of discrete velocity points. The above formulation converts the continuous velocity moments into weighted summations over a finite set of discrete velocities. Consequently, the accuracy and efficiency of the kinetic solver are largely determined by the underlying quadrature rule, which motivates the developments presented in the following sections.

\begin{table}[tbhp]
	\centering
	\caption{Reference scales used for nondimensionalization.}
	\label{tab:nondim_variables}
	\begin{tabular}{ll}
		\hline
		Variable & Reference scale \\
		\hline
		Length \(x\) & \(L_{ref}\) \\
		Density \(\rho\) & \(\rho_{ref}\) \\
		Molecular and macroscopic velocities \(\boldsymbol{\xi},\boldsymbol{u}\) & \(C_{ref}\)\\
		Temperature \(T\) & \(T_{ref}\) \\
		Temporal variables \(t,\;\tau\) & \(L_{ref}/C_{ref}\)\\
		Pressure \(p\) & \(\rho_{ref}C_{ref}^{2}\) \\
		Specific total energy \(E\) & \(C_{ref}^{2}\) \\
		Heat flux \(\boldsymbol{q}\) & \(\rho_{ref}C_{ref}^{3}\) \\
		Reduced distribution \(g\) & \(\rho_{ref}/C_{ref}^{D}\) \\
		Reduced distribution \(h\) & \(\rho_{ref}C_{ref}^{\,2-D}\) \\
		\hline
	\end{tabular}
\end{table}

\subsection{Boundary conditions}

The reduced Shakhov system is closed with appropriate boundary conditions. In the present work, fully diffuse reflection is imposed at solid walls, whereas equilibrium distributions are prescribed at far-field boundaries. Under the diffuse reflection assumption, molecules impinging on a wall are re-emitted with velocities independent of their incident velocities, following a Maxwellian distribution determined by the prescribed wall temperature and velocity.

Introducing the vector form $\phi=(g,h)^T$, the diffuse reflection boundary condition can be written as
\begin{equation}
	\phi
	(\boldsymbol{x}_{w},\boldsymbol{\xi}_{i})
	=
	\begin{pmatrix}
		1 \\
		(K+3-D)RT_{w}
	\end{pmatrix}
	g^{M}
	(
	\boldsymbol{\xi}_{i},
	\rho_{w},
	\boldsymbol{u}_{w},
	T_{w}
	),
	\quad
	\boldsymbol{\xi}_{i}\cdot\boldsymbol{S}_{w}>0,
\end{equation}
where \(\boldsymbol{x}_{w}\) denotes the wall boundary point, \(\boldsymbol{S}_{w}\) is the outward surface vector, \(\rho_{w}\), \(\boldsymbol{u}_{w}\), and \(T_{w}\) denote the wall density, velocity, and temperature, respectively. The wall density \(\rho_w\) is determined by enforcing the impermeability condition, namely, zero net mass flux across the solid surface,
\begin{equation}
	\sum_{\boldsymbol{\xi}_{i}\cdot\boldsymbol{S}_{w}>0}
	\omega_i (\boldsymbol{\xi}_{i}\cdot\boldsymbol{n})
	g^{M}
	(
	\boldsymbol{\xi}_{i},
	\rho_{w},
	\boldsymbol{u}_{w},
	T_{w}
	)
	+
	\sum_{\boldsymbol{\xi}_{i}\cdot\boldsymbol{S}_{w}<0}
	\omega_i (\boldsymbol{\xi}_{i}\cdot\boldsymbol{n})
	g
	(
	\boldsymbol{x}_{w},
	\boldsymbol{\xi}_{i}
	)
	=
	0,
\end{equation}
where \(\boldsymbol{n}\) is the outward unit normal vector to the wall surface. From the above constraint, the wall density is obtained as
\begin{equation}
	\rho_{w}
	=
	-
	\frac{
		\displaystyle
		\sum_{\boldsymbol{\xi}_{i}\cdot\boldsymbol{S}_{w}<0}
		\omega_i (\boldsymbol{\xi}_{i}\cdot\boldsymbol{n})
		g
		(
		\boldsymbol{x}_{w},
		\boldsymbol{\xi}_{i}
		)
	}{
		\displaystyle
		\sum_{\boldsymbol{\xi}_{i}\cdot\boldsymbol{S}_{w}>0}
		\omega_i (\boldsymbol{\xi}_{i}\cdot\boldsymbol{n})
		g^{M}
		(
		\boldsymbol{\xi}_{i},
		1,
		\boldsymbol{u}_{w},
		T_{w}
		)
	}.
\end{equation}

At the far-field boundary, the incoming distribution function is prescribed by
\begin{equation}
	\phi
	(
	\boldsymbol{x}_{b},
	\boldsymbol{\xi}_{i}
	)
	=
	\phi^{S}
	(
	\rho_{\infty},
	\boldsymbol{u}_{\infty},
	T_{\infty},
	\boldsymbol{\xi}_{i}
	).
\end{equation}
where \(\rho_{\infty}\), \(\boldsymbol{u}_{\infty}\), and \(T_{\infty}\) are the prescribed freestream density, velocity, and temperature, respectively.

\section{Bell-Shaped Gaussian Quadrature Framework}
\label{sec3}

This section presents the proposed Bell-shaped Gaussian quadrature framework. The velocity-space integral is first decomposed into radial and directional components. A radial transformation, together with its associated bell-shaped weight function, is then introduced to recast the radial integral into the standard Gauss–Jacobi form, while the directional integral is treated using a suitable discretization scheme. The resulting formulations for different velocity-space dimensions are subsequently presented, followed by a discussion of the properties of the resulting discrete velocity sets.

\subsection{Unified Quadrature Formulation}
The proposed formulation is based on a generalized velocity-space weight function, which provides a common basis for constructing different quadrature rules. The generalized weight function and its
associated quadrature construction are first presented, followed by the recovery of several existing quadrature rules.

\subsubsection{Construction of the Quadrature Rule}
Consider the following weighted integral defined over a $D$-dimensional velocity space:
\begin{equation}
	I_D
	=
	\int_{\mathbb{R}^D}
	w(\chi_\xi)
	f(\boldsymbol{\xi})
	\,d\boldsymbol{\xi},
	\label{eq:ID1}
\end{equation}
where $\boldsymbol{\xi}$ denotes the molecular velocity and $w(\chi_\xi)$ is a radially symmetric weight function depending on the magnitude of $\boldsymbol{\xi}$. To establish a unified formulation applicable to different velocity-space dimensions, the dimensionless radial variable is introduced as
\begin{equation}
	\chi_\xi
	=
	\frac{|\boldsymbol{\xi}|^2}{\alpha T_m},
	\label{eq:chi_xi}
\end{equation}
with $T_m$ denoting the median temperature and $\alpha$ a tunable parameter controlling the radial scaling. The velocity vector can then be decomposed into its radial magnitude and direction as
\begin{equation}
	\boldsymbol{\xi}
	=
	\sqrt{\alpha T_m\chi_\xi}\,\boldsymbol{s},
	\qquad
	\boldsymbol{s}\in S^{D-1},
	\label{eq:velocity_decomposition}
\end{equation}
where $\sqrt{\alpha T_m\chi_\xi}=|\boldsymbol{\xi}|$ is the velocity magnitude and $\boldsymbol{s}$ is a unit directional vector defined on the unit $(D-1)$-sphere $S^{D-1}$. Under this transformation, the
velocity-space volume element becomes
\begin{equation}
	d\boldsymbol{\xi}
	=
	\frac{(\alpha T_m)^{D/2}}{2}
	\chi_\xi^{D/2-1}
	d\chi_\xi\,d\Omega_{D-1},
	\label{eq:volume_element}
\end{equation}
in which $d\Omega_{D-1}$ denotes the directional measure on $S^{D-1}$. Since the weight function depends only on $\chi_\xi$, Eq.~\eqref{eq:ID1} can consequently be decomposed as
\begin{equation}
	I_D
	=
	\frac{(\alpha T_m)^{D/2}}{2}
	\int_0^{\infty}
	\chi_\xi^{D/2-1}
	w(\chi_\xi)
	\left[
	\int_{S^{D-1}}
	f\!\left(
	\sqrt{\alpha T_m\chi_\xi}\,\boldsymbol{s}
	\right)
	d\Omega_{D-1}
	\right]
	d\chi_\xi .
	\label{eq:ID_radial}
\end{equation}

To transform the radial integration domain into a finite interval, a radial mapping is introduced:
\begin{equation}
	r
	=
	\mathcal{R}(\chi_\xi),
	\qquad
	\mathcal{R}:(0,\chi_{\max})\rightarrow(0,1),
	\label{eq:bell_mapping}
\end{equation}
where $\chi_{\max}$ denotes the upper bound of the radial domain. The mapping can be applied directly to the unbounded interval by taking $\chi_{\max}\rightarrow\infty$, or to a truncated radial domain with a
finite $\chi_{\max}$. The mapping is assumed to be monotonically increasing and invertible over the prescribed interval, with $\mathcal{R}(0)=0$ and $\mathcal{R}(\chi_{\max})=1$, where the latter condition is understood as $\mathcal{R}(\infty)=1$ when $\chi_{\max}\rightarrow\infty$. Accordingly,
\begin{equation}
	\chi_\xi
	=
	\mathcal{R}^{-1}(r),
	\qquad
	d\chi_\xi
	=
	\frac{dr}
	{\mathcal{R}'\!\left(\mathcal{R}^{-1}(r)\right)}.
	\label{eq:bell_inverse}
\end{equation}
Since $\mathcal{R}(\chi_\xi)$ is monotonically increasing, with $\mathcal{R}'(\chi_\xi)>0$, the integration interval is mapped from $[0,\chi_{\max}]$ to $[0,1]$, yielding
\begin{equation}
	I_D
	=
	\frac{(\alpha T_m)^{D/2}}{2}
	\int_0^1
	\frac{
		\left[\mathcal{R}^{-1}(r)\right]^{D/2-1}}
	{
		\mathcal{R}'\!\left(\mathcal{R}^{-1}(r)\right)
	}w\!\left(\mathcal{R}^{-1}(r)\right)
	\left[
	\int_{S^{D-1}}
	f\!\left(
	\sqrt{\alpha T_m \mathcal{R}^{-1}(r)}\,\boldsymbol{s}
	\right)
	d\Omega_{D-1}
	\right]
	\,dr .
	\label{eq:ID2}
\end{equation}

The transformed radial integral generally involves both the inverse mapping and its derivative and therefore does not directly correspond to a standard Gaussian quadrature. To eliminate the additional factors introduced by the variable transformation, a radial weight function matched to the bell-shaped mapping is constructed as
\begin{equation}
	w(\chi_\xi)
	=
	\left[1-\mathcal{R}(\chi_\xi)\right]^\beta
	\mathcal{R}'(\chi_\xi)
	\left(
	\frac{\mathcal{R}(\chi_\xi)}
	{\chi_\xi}
	\right)^{D/2-1},
	\label{eq:bell_weight}
\end{equation}
where $\beta>-1$ is an additional tunable parameter controlling the Gauss--Jacobi weight function. Substituting Eq.~\eqref{eq:bell_weight} into Eq.~\eqref{eq:ID2}, and using
$\mathcal{R}(\chi_\xi)=r$ and $\chi_\xi=\mathcal{R}^{-1}(r)$, gives
\begin{equation}
	I_D
	=
	\frac{(\alpha T_m)^{D/2}}{2}
	\int_0^1
	r^{D/2-1}(1-r)^\beta
	\left[
	\int_{S^{D-1}}
	f\!\left(
	\sqrt{\alpha T_m \mathcal{R}^{-1}(r)}\,\boldsymbol{s}
	\right)
	d\Omega_{D-1}
	\right]
	\,dr .
	\label{eq:ID3}
\end{equation}

Equation~\eqref{eq:ID3} shows that the radial integration has been transformed into an integral over the finite interval $[0,1]$ with the Jacobi weight $r^{D/2-1}(1-r)^\beta$. It can therefore be approximated by the Gauss--Jacobi quadrature rule:
\begin{equation}
	\int_0^1
	r^{D/2-1}(1-r)^\beta
	G(r)\,dr
	\approx
	\sum_{i=1}^{N_r}
	\omega_i^{(r)}
	G(r_i),
	\label{eq:GJ}
\end{equation}
where $r_i$ and $\omega_i^{(r)}$ denote the Gauss--Jacobi nodes and weights, respectively. The corresponding discrete radial velocities are then given by
\begin{equation}
	\xi_i
	=
	\sqrt{
		\alpha T_m \mathcal{R}^{-1}(r_i)
	},
	\qquad
	i=1,\ldots,N_r.
	\label{eq:radial_velocity}
\end{equation}

The remaining directional integration is defined on the unit sphere $S^{D-1}$ and can be discretized independently using a dimension-dependent directional quadrature:
\begin{equation}
	\int_{S^{D-1}}
	H(\boldsymbol{s})
	\,d\Omega_{D-1}
	\approx
	\sum_{j=1}^{N_\Omega}
	\omega_j^{(\Omega)}
	H(\boldsymbol{s}_j),
	\label{eq:directional_quadrature}
\end{equation}
in which $\boldsymbol{s}_j$ and $\omega_j^{(\Omega)}$ are the directional quadrature nodes and weights, respectively. Combining Eqs.~\eqref{eq:GJ} and \eqref{eq:directional_quadrature} yields the unified discrete representation
\begin{equation}
	I_D
	\approx
	\frac{(\alpha T_m)^{D/2}}{2}
	\sum_{i=1}^{N_r}
	\sum_{j=1}^{N_\Omega}
	\omega_i^{(r)}
	\omega_j^{(\Omega)}
	f\!\left(
	\xi_i\boldsymbol{s}_j
	\right).
	\label{eq:unified_DV}
\end{equation}

Equation~\eqref{eq:unified_DV} constitutes the core formulation of the proposed bell-shaped Gaussian quadrature framework. The radial discretization is determined solely by the radial mapping $\mathcal{R}(\chi_\xi)$ and the associated Gauss--Jacobi quadrature, whereas the directional discretization is independent of the radial construction and can be selected according to the dimension and symmetry of the velocity space. This separation provides a unified procedure for constructing discrete velocity sets in different dimensions while retaining the flexibility to combine the radial quadrature with different directional quadrature rules.

\subsubsection{Recovery of existing quadratures}
\label{sec:classical_recovery}

\begin{table}[htbp]
	\centering
	\caption{Recovery of existing quadratures within the proposed bell-shaped Gaussian quadrature framework.}
	\label{tab:classical}
	\begin{tabular}{@{}ccccc@{}}
		\toprule
		$\mathcal{D}_{\boldsymbol{\xi}}$
		&
		$\mathcal{R}(\chi_{\xi})$
		&
		Parameter choice
		&
		Weight function
		&
		Quadrature rule
		\\
		\midrule
		$\mathbb{R}^{2}$
		&
		$1-\exp(-\chi_{\xi})$
		&
		$\alpha=\beta+1, D=2$
		&
		$\exp(-|\xi|^{2}/T_{m})$
		&
		PGQ \cite{PGQ}
		\\
		$\mathbb{R}^{D}$
		&
		$\dfrac{2}{\pi}\arctan(\frac{\pi}{2}\chi_{\xi})$
		&
		$\alpha,\beta$
		&
		ATGJ weight
		&
		ATGJ \cite{ATGJ}
		\\
		$\mathbb{R}^{D}$
		&
		$\tanh(\chi_{\xi})$
		&
		$\alpha,\beta$
		&
		GGJQ weight
		&
		GGJQ \cite{GGJQ}
		\\
		$[-1,1]$
		&
		$\chi_{\xi}$
		&
		$\alpha T_{m}=1$
		&
		$(1-\xi^{2})^{\beta}$
		&
		symmetric Gauss--Jacobi
		\\
		$[-1,1]$
		&
		$\chi_{\xi}$
		&
		$\alpha T_{m}=1,\ \beta=0$
		&
		$1$
		&
		Gauss--Legendre
		\\
		$[-1,1]$
		&
		$\chi_{\xi}$
		&
		$\alpha T_{m}=1,\ \beta=-\dfrac{1}{2}$
		&
		$(1-\xi^{2})^{-1/2}$
		&
		Gauss--Chebyshev
		\\
		\bottomrule
	\end{tabular}
\end{table}

According to Eqs.~\eqref{eq:chi_xi} and \eqref{eq:bell_weight}, the proposed bell-shaped Gaussian quadrature framework is defined by the unified velocity-space weight function
\begin{equation}
	w(\chi_{\xi})
	=
	[1-\mathcal{R}(\chi_{\xi})]^{\beta} \mathcal{R}'(\chi_{\xi})
	\left(
	\frac{\mathcal{R}(\chi_{\xi})}{\chi_{\xi}}
	\right)^{D/2-1},
	\qquad
	\chi_{\xi}
	=
	\frac{|\xi|^{2}}{\alpha T_{m}}.
	\label{eq:unified_bell_weight}
\end{equation}
Different choices of $\mathcal{R}(\chi_{\xi})$ give rise to different weight functions. The previously developed Parametric Gaussian quadratures (PGQ), Arctangent-based Gauss--Jacobi (ATGJ) quadrature, and Generalised Gauss--Jacobi quadrature (GGJQ) \cite{PGQ, ATGJ, GGJQ} are recovered as special cases corresponding to specific choices of $\mathcal{R}(\chi_{\xi})$.

For $D=2$ on the unbounded velocity domain $\mathbb{R}^D$, the third factor in Eq.~\eqref{eq:unified_bell_weight} becomes unity. By setting $\mathcal{R}(\chi_{\xi})=1-\exp(-\chi_{\xi})$ and $\alpha=\beta+1$, Eq.~\eqref{eq:unified_bell_weight} then reduces to
\begin{equation}
	w(\chi_{\xi})
	=
	\exp[-(\beta+1)\chi_{\xi}]
	=
	\exp\left(-\frac{|\xi|^{2}}{T_{m}}\right),
	\label{eq:exp_bell_recovery}
\end{equation}
which recovers the Gaussian-type weight used in PGQ. The same construction also recovers ATGJ and GGJQ by
taking
\begin{equation}
	\mathcal{R}(\chi_{\xi})	=\frac{2}{\pi}\arctan(\frac{\pi}{2}\chi_{\xi})
\end{equation}
and
\begin{equation}
	\mathcal{R}(\chi_{\xi})	=\tanh(\chi_{\xi}),
\end{equation}
respectively. Thus, PGQ, ATGJ, and GGJQ correspond to different choices of the radial function within the same weight construction.

The same formulation also extends naturally to bounded velocity domains and compactly supported weights. Let $\alpha T_m=1$ and consider the linear function
\begin{equation}
	\mathcal{R}(\chi_{\xi})=\chi_{\xi}.
	\label{eq:linear_bell}
\end{equation}
For this choice, $\mathcal{R}'(\chi_{\xi})=1$ and
$\mathcal{R}(\chi_{\xi})/\chi_{\xi}=1$, Eq.~\eqref{eq:unified_bell_weight}
therefore reduces to
\begin{equation}
	w(\chi_{\xi})=(1-\chi_{\xi})^{\beta}.
\end{equation}
With $\alpha T_{m}=1$, the weight becomes
\begin{equation}
	w(\chi_{\xi})=(1-\xi^{2})^{\beta},
	\qquad
	-1\leq\xi\leq1,
	\label{eq:jacobi_weight}
\end{equation}
which is the symmetric Gauss--Jacobi weight on the standard interval $[-1,1]$. In particular, $\beta=0$ gives $w(\xi)=1$ and hence the Gauss--Legendre rule, whereas $\beta=-1/2$ gives
$w(\xi)=(1-\xi^{2})^{-1/2}$ and hence the Gauss--Chebyshev rule. These classical cases, together with the above unbounded-domain quadratures, are summarized in Table~\ref{tab:classical}, showing that the proposed framework covers Gaussian-type weights with both finite and infinite support within a common formulation.

It should be noted that, beyond recovering the weight functions and quadratures discussed above, the proposed framework also provides a systematic way to construct new weight functions through different choices of the radial mapping $\mathcal{R}(\chi_\xi)$. Several representative mappings are listed in Table~\ref{tab:other_bell}, covering both finite and infinite intervals and exhibiting different asymptotic decay behaviors.

\begin{table}[t]
	\centering
	\caption{Representative monotonically increasing radial mappings.}
	\label{tab:other_bell}
	\begin{tabular}{lccc}
		\toprule
		$\mathcal{R}(\chi_\xi)$ & $\mathcal{R}'(\chi_\xi)$ & Interval & $1-\mathcal{R}(\chi_\xi)$ asymptotics \\
		\midrule
		
		$\displaystyle
		\mathrm{erf}\!\left(
		\frac{\sqrt{\pi}}{2}\chi_\xi
		\right)$
		&
		$\displaystyle
		\exp\!\left(
		-\frac{\pi\chi_\xi^2}{4}
		\right)$
		&
		$[0,\infty)$
		&
		$\displaystyle
		O\!\left(
		\chi_\xi^{-1}
		e^{-\pi\chi_\xi^2/4}
		\right)$
		\\[6pt]
		
		$\displaystyle
		\frac{\chi_\xi}
		{\sqrt{1+\chi_\xi^2}}$
		&
		$\displaystyle
		\frac{1}
		{(1+\chi_\xi^2)^{3/2}}$
		&
		$[0,\infty)$
		&
		$\displaystyle
		O(\chi_\xi^{-2})$
		\\[6pt]
		
		$\displaystyle
		\frac{2}{\pi}
		\mathrm{gd}\!\left(
		\frac{\pi}{2}\chi_\xi
		\right)$
		&
		$\displaystyle
		\mathrm{sech}\!\left(
		\frac{\pi}{2}\chi_\xi
		\right)$
		&
		$[0,\infty)$
		&
		$\displaystyle
		O\!\left(
		e^{-\pi\chi_\xi/2}
		\right)$
		\\[6pt]
		
		$\displaystyle
		\sin\chi_\xi$
		&
		$\displaystyle
		\cos\chi_\xi$
		&
		$\displaystyle
		[0,\frac{\pi}{2}]$
		&
		$\displaystyle
		O\!\left[
		\left(
		\frac{\pi}{2}-\chi_\xi
		\right)^2
		\right]$
		\\[6pt]
		
		$\displaystyle
		\left(
		\frac{\chi_\xi}{\chi_{\max}}
		\right)^{\chi_{\max}}$
		&
		$\displaystyle
		\left(
		\frac{\chi_\xi}{\chi_{\max}}
		\right)^{\chi_{\max}-1}$
		&
		$[0,\chi_{\max}]$
		&
		$\displaystyle	O(\chi_{\max}-\chi)	$
		\\
		\bottomrule
	\end{tabular}
	
	\vspace{2pt}
	\begin{minipage}{0.7\linewidth}
		\footnotesize
		Here, $\mathrm{gd}(\cdot)$ denotes the Gudermannian function.
	\end{minipage}
\end{table}

\subsection{Dimensional Formulations}

For a given velocity-space dimension $D$, the radial weight in Eq.~\eqref{eq:ID3} is fixed, while the directional discretization is selected according to the corresponding directional space. The resulting
formulations for $D=1$, $2$, and $3$ are summarized in Table~\ref{tab:dimensional_formulations}, with details given below.

\begin{table}[t]
	\centering
	\caption{Radial and directional discretizations for different velocity-space dimensions.}
	\label{tab:dimensional_formulations}
	\begin{tabular}{lccc}
		\toprule
		Dimension
		& $D=1$
		& $D=2$
		& $D=3$ \\
		\midrule
		
		Radial nodes $\xi_i$
		& $\sqrt{\alpha T_m \mathcal{R}^{-1}(r_i^{(-1/2,\beta)})}$
		& $\sqrt{\alpha T_m \mathcal{R}^{-1}(r_i^{(0,\beta)})}$
		& $\sqrt{\alpha T_m \mathcal{R}^{-1}(r_i^{(1/2,\beta)})}$
		\\[6pt]
		
		Directional nodes
		& $s_j=\pm1$
		& $\boldsymbol{s}_j=(\cos\theta_j,\sin\theta_j)$
		& $\boldsymbol{s}_j=(s_{j,1},s_{j,2},s_{j,3})$
		\\[4pt]
		
		Discrete velocities
		& $\xi_{i,j}=\xi_i s_j$
		& $\boldsymbol{\xi}_{i,j}=\xi_i\boldsymbol{s}_{ij}$
		& $\boldsymbol{\xi}_{i,j}=\xi_i\boldsymbol{s}_{j}$
		\\[6pt]
		
		Discrete weights
		& $\displaystyle
		W_{i,j}^{(1)}
		=\frac{\sqrt{\alpha T_m}}{2}\omega_i^{(r)}$
		& $\displaystyle
		W_{i,j}^{(2)}
		=\frac{\alpha T_m}{2}
		\omega_i^{(r)}\omega_{ij}^{(\theta)}$
		& $\displaystyle
		W_{i,j}^{(3)}
		=\frac{(\alpha T_m)^{3/2}}{2}
		\omega_i^{(r)}\omega_j^{(\Omega)}$
		\\
		\bottomrule
	\end{tabular}
	
	\vspace{3pt}
	\begin{minipage}{0.95\linewidth}
		\footnotesize
		Here, $r_i^{(D/2-1,\beta)}$ denotes the $i$th root of the Gauss--Jacobi polynomial associated with the weight	$r^{D/2-1}(1-r)^\beta$ on $[0,1]$, and $\omega_i^{(r)}$ is the corresponding radial quadrature weight. The radial velocity magnitude	is defined as
		$\xi_i=\sqrt{\alpha T_m \mathcal{R}^{-1}(r_i)}$.
		The Bell-shaped weight function used in the present framework is
		$w(\chi_\xi)
		=
		\left[1-\mathcal{R}(\chi_\xi)\right]^\beta
		\mathcal{R}'(\chi_\xi)
		\left[
		(\mathcal{R}(\chi_\xi))/\chi_\xi
		\right]^{D/2-1}$, $\chi_\xi=|\boldsymbol{\xi}|^2/(\alpha T_m)$.
	\end{minipage}
\end{table}

\subsubsection{ One-dimensional formulation}

For $D=1$, the directional space reduces to the zero-dimensional unit sphere, \(S^0=\{-1,+1\}\).
Equation~\eqref{eq:ID3} therefore becomes
\begin{equation}
	I_1
	=
	\frac{\sqrt{\alpha T_m}}{2}
	\int_0^1
	r^{-1/2}(1-r)^\beta
	\sum_{s_j=\pm1}
	f\!\left(
	s_j\sqrt{\alpha T_m \mathcal{R}^{-1}(r)}
	\right)
	\,dr.
	\label{eq:ID1_1D}
\end{equation}
The radial integral is discretized using the Gauss--Jacobi rule in Eq.~\eqref{eq:GJ}, with the weight function $r^{-1/2}(1-r)^\beta$. Since the directional space consists of only two antipodal points, the directional integration is represented exactly by summation over $s_j=\pm1$. The resulting discrete quadrature is
\begin{equation}
	I_1
	\approx
	\sum_{i=1}^{N_r}
	\sum_{s_j=\pm1}
	W_{i,j}^{(1)} f(\xi_{i,j}),
	\label{eq:ID1_1D_sum}
\end{equation}
where the discrete velocities and their corresponding weights are
\begin{equation}
	\xi_{i,\pm}
	=
	\pm\sqrt{\alpha T_m \mathcal{R}^{-1}(r_i)},
	\qquad
	W_{i,\pm}^{(1)}
	=
	\frac{\sqrt{\alpha T_m}}{2}\,
	\omega_i^{(r)},
	\qquad
	i=1,\ldots,N_r.
	\label{eq:DVW_1D}
\end{equation}
where $r_i=r_i^{(-1/2,\beta)}$ and $\omega_i^{(r)}$ denote the $i$th Gauss--Jacobi node and its corresponding weight associated with the radial weight $r^{-1/2}(1-r)^\beta$, respectively. Thus, each radial quadrature node generates two antipodal discrete velocities with identical weights, resulting in a total of $2N_r$ discrete velocities that are exactly symmetric about the velocity origin.

\subsubsection{Two-dimensional formulation}

For $D=2$, the directional space is the unit circle $S^1$, parameterized by $\boldsymbol{s}(\theta)=(\cos\theta,\sin\theta)$ with $\theta\in[0,2\pi)$. Since $d\Omega_1=d\theta$, Eq.~\eqref{eq:ID3} becomes
\begin{equation}
	I_2
	=
	\frac{\alpha T_m}{2}
	\int_0^1
	(1-r)^\beta
	\left[
	\int_0^{2\pi}
	f\!\left(
	\sqrt{\alpha T_m \mathcal{R}^{-1}(r)}
	\,\boldsymbol{s}(\theta)
	\right)
	\,d\theta
	\right]
	\,dr.
	\label{eq:ID1_2D}
\end{equation}
Using the Gauss--Jacobi rule for the radial integration and the trapezoidal rule for the angular integration gives
\begin{equation}
	I_2
	\approx
	\frac{\alpha T_m}{2}
	\sum_{i=1}^{N_r}
	\sum_{j=1}^{N_{\theta,i}}
	\omega_i^{(r)}
	\omega_{ij}^{(\theta)}
	f\!\left(
	\sqrt{\alpha T_m \mathcal{R}^{-1}(r_i)}
	\,\boldsymbol{s}_{ij}
	\right).
	\label{eq:ID1_2D_sum}
\end{equation}
Accordingly, the discrete velocities and their corresponding weights are
\begin{equation}
	\boldsymbol{\xi}_{ij}
	=
	\sqrt{\alpha T_m \mathcal{R}^{-1}(r_i)}
	\,\boldsymbol{s}_{ij},
	\qquad
	W_{ij}^{(2)}
	=
	\frac{\alpha T_m}{2}\,
	\omega_i^{(r)}
	\omega_{ij}^{(\theta)}.
	\label{eq:DVW_2D}
\end{equation}
Here, $\boldsymbol{s}_{ij}=(\cos\theta_{ij},\sin\theta_{ij})$ is the unit directional vector associated with the $j$th angular node on the $i$th radial shell. The quantities $r_i$ and $\omega_i^{(r)}$ denote the Gauss--Jacobi nodes and weights associated with the radial weight $(1-r)^\beta$, while $\theta_{ij}$ and $\omega_{ij}^{(\theta)}$ are the corresponding angular nodes and weights.

The radial and angular discretizations are constructed independently in this framework. In particular, the number and angular positions of the directional nodes may be varied from one radial shell to another. Let
$N_{\theta,i}$ denote the number of angular nodes on the $i$th shell and $\Delta\theta_i$ the corresponding angular offset. The periodic trapezoidal rule then gives
\begin{equation}
	\theta_{ij}
	=
	\frac{2\pi(j-1)}{N_{\theta,i}}
	+
	\Delta\theta_i,
	\qquad
	\omega_{ij}^{(\theta)}
	=
	\frac{2\pi}{N_{\theta,i}},
	\qquad
	j=1,\ldots,N_{\theta,i}.
	\label{eq:2D_angular_rule}
\end{equation}

\begin{figure*}[!th]
	\centering
	\subfigure{
		\includegraphics[width=0.22\textwidth]{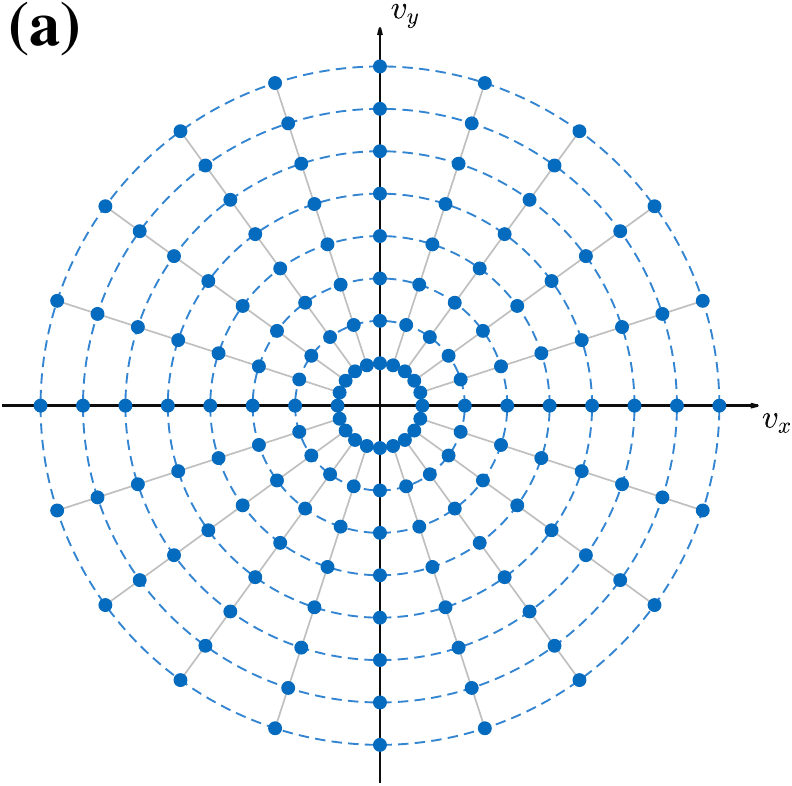}}
	\subfigure{
		\includegraphics[width=0.22\textwidth]{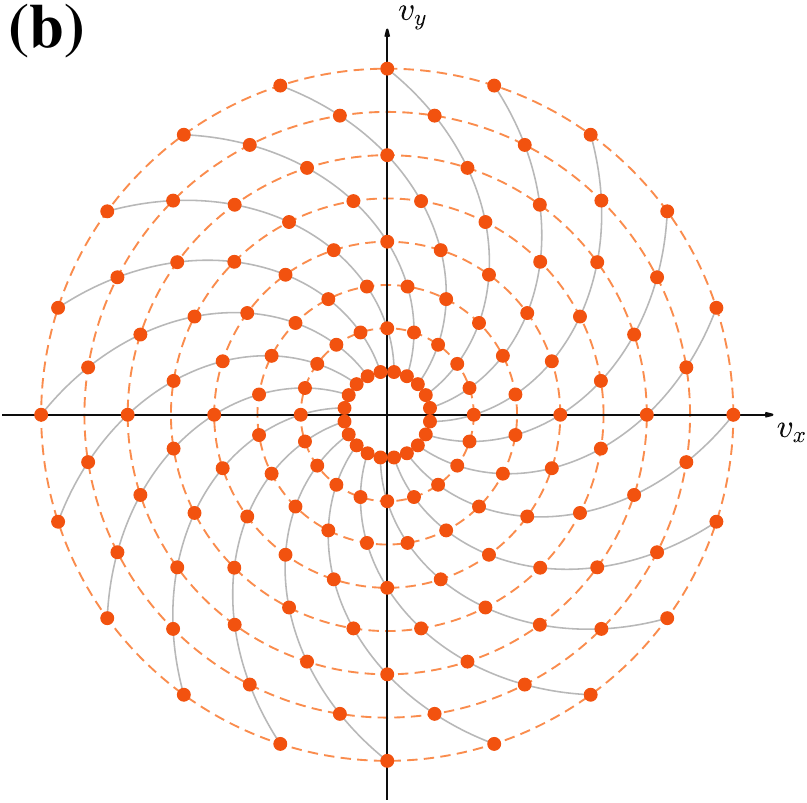}}
	\subfigure{
		\includegraphics[width=0.22\textwidth]{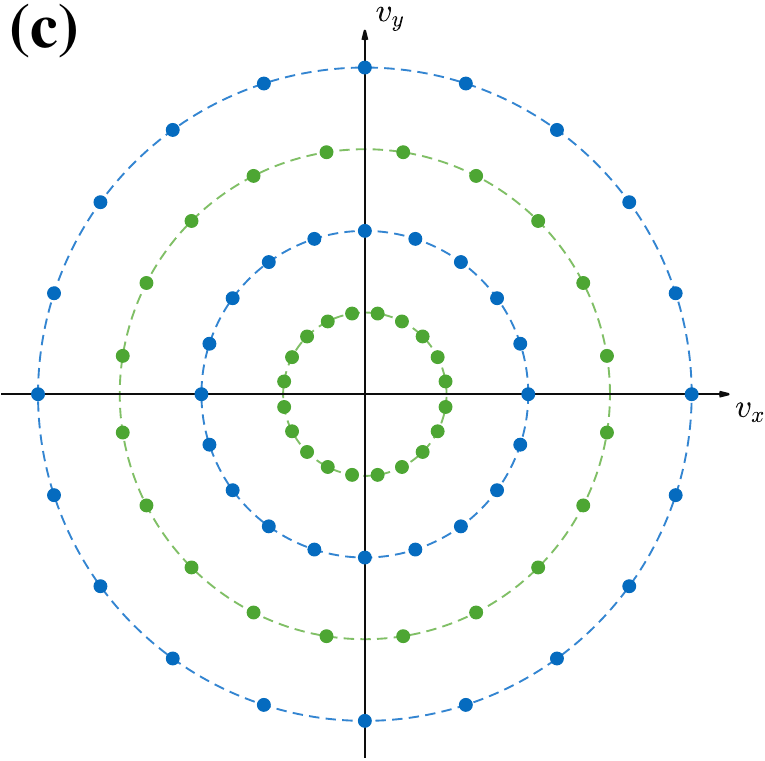}}
	\subfigure{
		\includegraphics[width=0.22\textwidth]{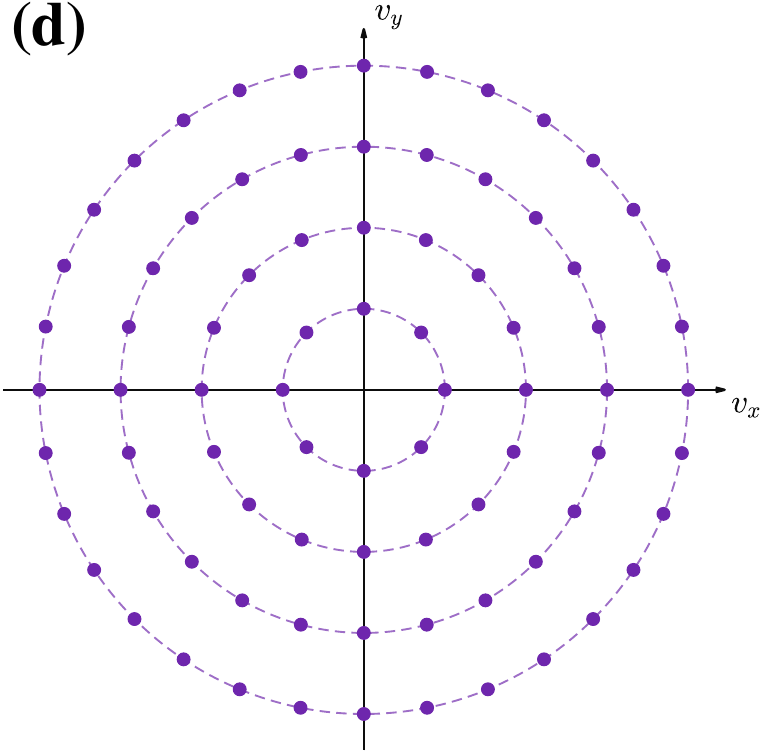}}
	\caption{Schematic illustrations of the two-dimensional velocity-space configurations:
		(a) aligned, (b) shifted, (c) staggered, and (d) variable-point arrangements.}
	\label{Fig_2D_configurations}
\end{figure*}

Different choices of $N_{\theta,i}$ and $\Delta\theta_i$ lead to different angular arrangements, four representative examples of which are shown in Figure.~\ref{Fig_2D_configurations}. The \emph{aligned}
configuration uses the same angular locations on all radial shells. In the \emph{shifted} configuration, the same number of angular nodes is retained on each shell, but their angular locations are shifted by a
prescribed offset. The \emph{staggered} configuration places the nodes on neighboring shells in an interlaced manner. The \emph{variable-point} configuration, in contrast, allows the number of
angular nodes to vary between radial shells, thereby providing a radially varying angular resolution.

This construction thus allows the radial resolution, angular resolution, and angular arrangement to be adjusted independently. Different two-dimensional discrete velocity sets can therefore be generated from
the same radial Gauss--Jacobi discretization by changing only the directional nodes and their associated weights.

\subsubsection{ Three-dimensional formulation}

For $D=3$, the directional space is the unit sphere $S^2$, with $\boldsymbol{s}\in S^2$ and $|\boldsymbol{s}|=1$. Equation~\eqref{eq:ID3} then takes the form
\begin{equation}
	I_3
	=
	\frac{(\alpha T_m)^{3/2}}{2}
	\int_0^1
	r^{1/2}(1-r)^\beta
	\left[
	\int_{S^2}
	f\!\left(
	\sqrt{\alpha T_m \mathcal{R}^{-1}(r)}\,\boldsymbol{s}
	\right)
	d\Omega_2
	\right]dr .
	\label{eq:ID1_3D}
\end{equation}

As in the two-dimensional formulation, the radial and directional discretizations are constructed independently. For the $i$th radial quadrature node $r_i$, the corresponding velocity magnitude is
$\xi_i=\sqrt{\alpha T_m \mathcal{R}^{-1}(r_i)}$. Projecting this magnitude onto the prescribed spherical directions gives
\begin{equation}
	\boldsymbol{\xi}_{ij}
	=
	\xi_i\boldsymbol{s}_j
	=
	\sqrt{\alpha T_m \mathcal{R}^{-1}(r_i)}\,\boldsymbol{s}_j .
	\label{eq:DV_3D}
\end{equation}
When the same spherical discretization is applied to all radial shells, the resulting discrete form of Eq.~\eqref{eq:ID1_3D} is
\begin{equation}
	I_3
	\approx
	\frac{(\alpha T_m)^{3/2}}{2}
	\sum_{i=1}^{N_r}
	\sum_{j=1}^{N_\Omega}
	\omega_i^{(r)}
	\omega_j^{(\Omega)}
	f\!\left(
	\xi_i\boldsymbol{s}_j
	\right),
	\label{eq:ID1_3D_sum}
\end{equation}
with the corresponding discrete velocity weight
\begin{equation}
	W_{ij}^{(3)}
	=
	\frac{(\alpha T_m)^{3/2}}{2}
	\omega_i^{(r)}
	\omega_j^{(\Omega)} .
	\label{eq:DVW_3D}
\end{equation}
Here, $r_i$ and $\omega_i^{(r)}$ are the Gauss--Jacobi nodes and weights associated with the radial weight $r^{1/2}(1-r)^\beta$, while $\boldsymbol{s}_j$ and $\omega_j^{(\Omega)}$ denote the spherical nodes and their corresponding directional weights.

\begin{figure*}[!th]
	\centering
	\subfigure{
		\includegraphics[width=0.22\textwidth]{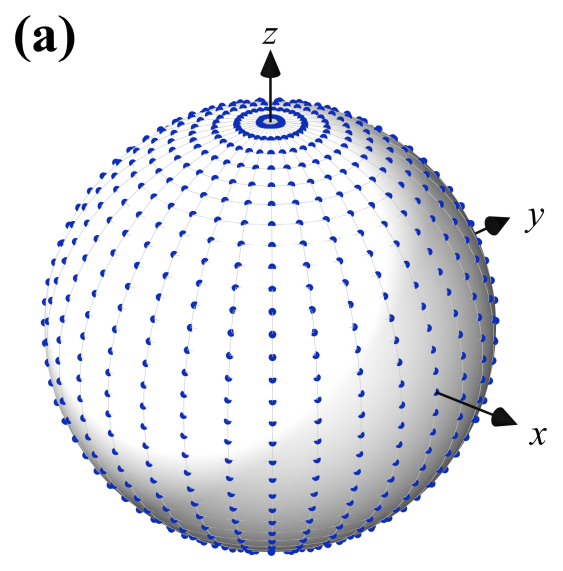}
	}
	\subfigure{
		\includegraphics[width=0.22\textwidth]{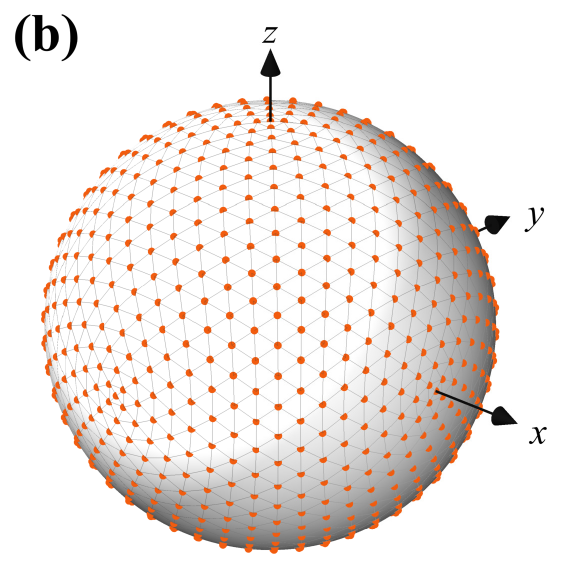}
	}
	\subfigure{
		\includegraphics[width=0.22\textwidth]{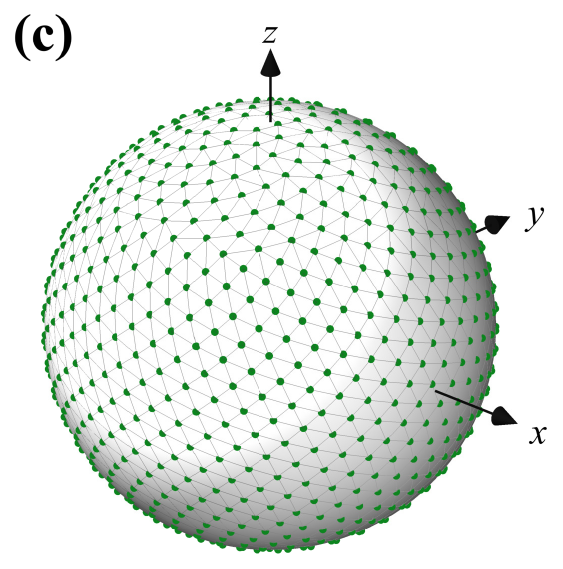}
	}
	\subfigure{
		\includegraphics[width=0.22\textwidth]{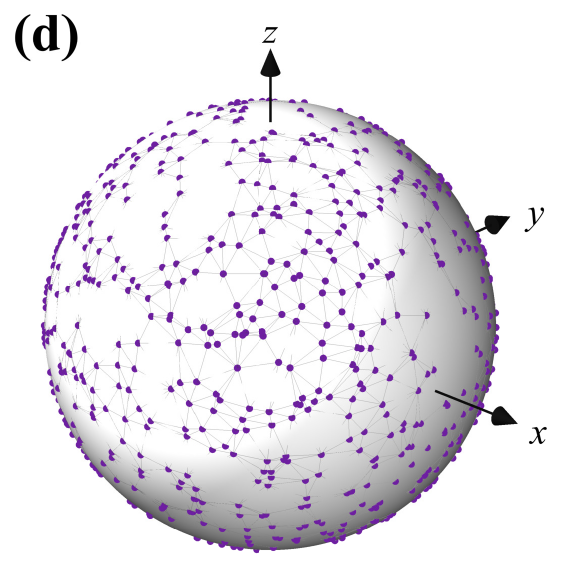}
	}
	\caption{Schematic illustrations of the spherical directional discretizations:
		(a) Lat-Lon tensor-product,
		(b) Lebedev,
		(c) Fibonacci,
		and (d) Monte Carlo distributions.}
	\label{Fig_3D_configurations}
\end{figure*}

The directional integration can be discretized using different spherical constructions without changing the radial quadrature. Four representative approaches are considered in this work,
namely the latitude--longitude (Lat-Lon) tensor-product, Lebedev, Fibonacci, and Monte Carlo discretizations, as illustrated in Figure.~\ref{Fig_3D_configurations}.
Their corresponding directional nodes and weights are summarized in Table~\ref{tab:spherical_discret}. The Lat-Lon construction generates a structured tensor-product distribution from the polar and azimuthal angles, whereas the Lebedev rule employs highly symmetric nodes and weights designed for accurate spherical integration.
The Fibonacci construction provides a quasi-uniform deterministic distribution of directions, while Monte Carlo sampling generates directions through random sampling on the sphere.
Beyond these commonly used constructions, spherical $t$-designs provide equal-weight quadrature rules with exactness for spherical polynomials up to degree $t$ \cite{SphericalDesign2024}.
Recent work has also considered optimizing quadrature weights on fixed scattered point sets as an alternative to node construction \cite{OptimizationWeight2026}.

These directional constructions differ only in the representation of the spherical directional space and can be directly coupled with the same Bell-shaped radial quadrature.
The resulting velocity set therefore consists of concentric spherical shells, with the shell radii determined by the Gauss--Jacobi quadrature and the directional nodes and weights determined independently by the selected spherical discretization.
This separation provides a direct means of examining the influence of spherical directional discretization while keeping the radial discretization unchanged.

\begin{table*}[!th]
	\centering
	\caption{Directional nodes and weights of the representative spherical discretizations employed in the three-dimensional Bell-shaped quadrature framework.}
	\label{tab:spherical_discret}
	\begin{tabular}{lcc}
		\toprule
		Discretization
		& Directional nodes $\boldsymbol{s}_j$
		& Directional weights $\omega_j^{(\Omega)}$
		\\
		\midrule
		
		Lebedev
		&
		$\displaystyle
		\boldsymbol{s}_j^{(\mathrm{L})}
		=
		\left(
		s_{j,1}^{(\mathrm{L})},
		s_{j,2}^{(\mathrm{L})},
		s_{j,3}^{(\mathrm{L})}
		\right),
		\quad
		|\boldsymbol{s}_j^{(\mathrm{L})}|=1$
		&
		$\displaystyle
		\omega_j^{(\mathrm{L})}$
		\\[14pt]
		
		Lat-Lon
		&
		$\displaystyle
		\boldsymbol{s}_{jk}^{(\mathrm{LL})}
		=
		\left(
		\sqrt{1-\mu_j^2}\cos\phi_k,\,
		\sqrt{1-\mu_j^2}\sin\phi_k,\,
		\mu_j
		\right)$
		&
		$\displaystyle
		\frac{2\pi}{N_\phi}
		\omega_j^{(\mu)}$
		\\[14pt]
		
		Fibonacci
		&
		$\displaystyle
		\boldsymbol{s}_j^{(\mathrm{F})}
		=
		\left(
		\sqrt{1-z_j^2}\cos\phi_j,\,
		\sqrt{1-z_j^2}\sin\phi_j,\,
		z_j
		\right)$
		&
		$\displaystyle
		\frac{4\pi}{N_\Omega}$
		\\[14pt]
		
		Monte Carlo
		&
		$\displaystyle
		\boldsymbol{s}_j^{(\mathrm{MC})}
		=
		\left(
		\sqrt{1-\mu_j^2}\cos\phi_j,\,
		\sqrt{1-\mu_j^2}\sin\phi_j,\,
		\mu_j
		\right)$
		&
		$\displaystyle
		\frac{4\pi}{N_\Omega}$
		\\
		
		\bottomrule
	\end{tabular}
	
	\vspace{2mm}
	\begin{minipage}{0.95\textwidth}
		\small
		For the Lebedev quadrature, the nodes and weights are determined by octahedral symmetry and exact integration of spherical harmonics up to degree $t$:
		\(
		\sum_{j=1}^{N_\Omega}
		\omega_j^{(\mathrm{L})}
		Y_{\ell m}\!\left(\boldsymbol{s}_j^{(\mathrm{L})}\right)
		=
		\int_{\mathbb{S}^2}
		Y_{\ell m}(\boldsymbol{s})\,\mathrm{d}\Omega\),\(0\leq \ell\leq t\).
		For the Lat--Lon discretization,
		$\{\mu_j,\omega_j^{(\mu)}\}$ are the Gauss--Legendre nodes and weights,
		and $\phi_k=2\pi(k-1)/N_\phi$.
		For the Fibonacci distribution,
		$z_j=1-(2j-1)/N_\Omega$ and
		$\phi_j=(j-1)\pi(3-\sqrt{5})$.
		For Monte Carlo sampling,
		$\mu_j\sim\mathcal{U}(-1,1)$ and
		$\phi_j\sim\mathcal{U}(0,2\pi)$.
	\end{minipage}
\end{table*}

\subsection{Properties of the proposed quadrature}

\subsubsection{Asymptotic Weight Behavior}

The asymptotic behavior of the Bell-shaped weight provides a useful connection between the mapping parameters and the resulting distribution of discrete velocities.
In particular, the relative magnitude of $\alpha$ and $\beta$ determines the effective width of the weight and, consequently, the radial extent over which the discrete velocity set is distributed.
This relation provides a criterion for normalizing the radial mappings and selecting the scaling parameters.

\noindent\textbf{General asymptotic form.}
Let \(\mathcal{R}\) be a differentiable mapping satisfying \(\mathcal{R}(0)=0\), \(\mathcal{R}(\infty)=1\), and \(\mathcal{R}'(\chi_\xi)>0\). The associated radial weight function in Eq.~\eqref{eq:bell_weight} is
\begin{equation}
	w(\chi_\xi)
	=
	\left[1-\mathcal{R}(\chi_\xi)\right]^\beta
	\mathcal{R}'(\chi_\xi)
	\left[
	\frac{\mathcal{R}(\chi_\xi)}{\chi_\xi}
	\right]^{D/2-1}.
	\label{eq:bell_weight_asymptotic}
\end{equation}
To characterize its behavior near the origin, we first consider the local expansion of the mapping,
\begin{equation}
	\mathcal{R}(\chi_\xi)
	=
	\mathcal{R}'(0)\chi_\xi
	+
	O(\chi_\xi^2),
	\qquad
	\mathcal{R}'(0)>0.
	\label{eq:bell_local_expansion}
\end{equation}
Taking logarithms and using \(\ln(1-x)=-x+O(x^2)\) as \(x\to0\), we obtain
\begin{equation}
	\ln\left[1-\mathcal{R}(\chi_\xi)\right]
	=
	-\mathcal{R}'(0)\chi_\xi
	+
	O(\chi_\xi^2).
\end{equation}
Therefore,
\begin{equation}
	\left[1-\mathcal{R}(\chi_\xi)\right]^\beta
	=
	\exp\left\{
	-\beta\mathcal{R}'(0)\chi_\xi
	+
	O(\chi_\xi^2)
	\right\}.
\end{equation}
In the limit \(\chi_\xi\to0\), the second and third factors in Eq.~\eqref{eq:bell_weight_asymptotic} remain finite, since
\begin{equation}
	\mathcal{R}'(\chi_\xi)
	\to
	\mathcal{R}'(0),
	\qquad
	\frac{\mathcal{R}(\chi_\xi)}{\chi_\xi}
	\to
	\mathcal{R}'(0).
\end{equation}
Combining these results yields the leading-order behavior
\begin{equation}
	w(\chi_\xi)
	\sim
	\left[\mathcal{R}'(0)\right]^{D/2}
	\exp\left[
	-\beta\mathcal{R}'(0)\chi_\xi
	\right],
	\qquad
	\chi_\xi\to0.
\end{equation}
Using \(\chi_\xi=|\boldsymbol{\xi}|^2/(\alpha T_m)\), the corresponding velocity-space weight can therefore be written as
\begin{equation}
	w(\boldsymbol{\xi})
	\sim
	\left[\mathcal{R}'(0)\right]^{D/2}
	\exp\left[
	-\mathcal{R}'(0)\frac{\beta}{\alpha}
	\frac{|\boldsymbol{\xi}|^2}{T_m}
	\right].
	\label{eq:general_asymptotic_weight}
\end{equation}

This expression shows that the asymptotic weight is governed by two quantities: the local slope $\mathcal{R}'(0)$ of the Bell-shaped mapping and the ratio $\beta/\alpha$.
The former determines the characteristic scaling introduced by the mapping, whereas the latter controls the effective concentration of the weight in velocity space.

\noindent\textbf{Normalization of $\mathcal{R}(\chi_\xi)$.}
Equation~\eqref{eq:general_asymptotic_weight} indicates that different values of $\mathcal{R}'(0)$ lead to different limiting amplitudes and velocity scales even when the same $\alpha$ and $\beta$ are used.
Such a mapping-dependent scaling obscures the role of the adjustable parameters and complicates a consistent comparison among different mappings.
It is therefore natural to impose
\begin{equation}
	\mathcal{R}'(0)=1.
	\label{eq:bell_normalization}
\end{equation}
This normalization is readily satisfied in practice, as demonstrated by all mappings listed in Tables~\ref{tab:classical} and~\ref{tab:other_bell}.
All mappings satisfying this condition have the common local expansion
\begin{equation}
	\mathcal{R}(\chi_\xi)=\chi_\xi+O(\chi_\xi^2),
\end{equation}
and the asymptotic weight becomes
\begin{equation}
	w(\boldsymbol{\xi})
	\sim
	\exp\left(
	-\frac{|\boldsymbol{\xi}|^2}{\alpha T_m / \beta}
	\right).
	\label{eq:normalized_asymptotic_weight}
\end{equation}
The leading-order behavior is thus independent of the particular radial mapping.
The mapping itself only affects the higher order corrections and, therefore, the finite parameter distribution of the velocity nodes.

\noindent\textbf{Limiting regimes of $\alpha/\beta$.}
The normalized asymptotic form in Eq.~\eqref{eq:normalized_asymptotic_weight} reveals three distinct parameter regimes.
Their implications for the weight and the resulting velocity-space distribution are summarized in Table~\ref{tab:bell_asymptotic_regimes}.

A more direct interpretation can be obtained from the characteristic velocity scale of the asymptotic weight. Defining
\begin{equation}
	\xi_c
	\sim
	\sqrt{\frac{\alpha T_m}{\beta}},
	\label{eq:characteristic_velocity_scale}
\end{equation}
the region with appreciable weight is characterized by $|\boldsymbol{\xi}|=O(\xi_c)$.
Thus, for $\alpha/\beta>1$, $\xi_c$ is relatively large and the discrete velocities are distributed over a broad velocity range.
As $\alpha/\beta$ approaches unity, the characteristic velocity scale becomes comparable with $\sqrt{T_m}$ and the velocity distribution approaches the conventional Gaussian weighting.
For $\alpha/\beta<1$, $\xi_c$ decreases and the discrete velocity set becomes progressively concentrated toward the velocity origin.
In the limiting case $\alpha/\beta \rightarrow 0$, the characteristic scale tends to zero and the weight vanishes at any fixed nonzero velocity.

It is worth noting that this asymptotic scaling concerns the effective weight rather than the outermost discrete velocity itself.
The latter is determined by the Gauss--Jacobi nodes through
\begin{equation}
	\xi_i
	=
	\sqrt{\alpha T_m \mathcal{R}^{-1}(r_i)}.
	\label{eq:asymptotic_velocity_nodes}
\end{equation}
Therefore, $\alpha$ primarily sets the radial scale of the discrete velocity set, whereas $\beta$ modifies the distribution of the Gauss--Jacobi nodes through the radial weight.
Their ratio controls the effective concentration of the weighted velocity representation. This separation is important in practice: increasing $\alpha$ expands the velocity-space scale, while increasing $\beta$ strengthens the concentration toward the low-velocity region.
These considerations therefore provide a theoretical basis for selecting $\alpha$ and $\beta$ according to the nonequilibrium characteristics of the flow considered.

\begin{table}[t]
	\centering
	\caption{Asymptotic regimes of $\mathcal{R}(\chi_\xi)$ for different values of $\alpha/\beta$.}
	\label{tab:bell_asymptotic_regimes}
	\begin{tabular}{c c l}
		\toprule
		$\alpha/\beta$
		& $w(\boldsymbol{\xi})$
		& Discrete velocity distribution
		\\
		\midrule
		$\gg 1$
		&
		$w\rightarrow 1$
		&
		Broad
		\\[3pt]
		
		$=1$
		&
		$\displaystyle
		w\rightarrow
		\exp(-|\boldsymbol{\xi}|^2/T_m)$
		&
		Gaussian
		\\[3pt]
		
		$\ll 1$
		&
		$\displaystyle
		w\rightarrow0,\quad
		\boldsymbol{\xi}\neq\boldsymbol{0}$
		&
		Concentrated
		\\
		\bottomrule
	\end{tabular}
\end{table}

\subsubsection{Convergence Characteristics}

The convergence of the proposed quadrature can be considered separately for the radial and directional integrations.
The radial component is determined by the approximation properties of the Gauss--Jacobi rule, whereas the directional accuracy depends on the quadrature or sampling procedure adopted on $S^{D-1}$.

\noindent\textbf{Radial convergence.} For a fixed directional integration, the radial part of Eq.~\eqref{eq:ID3} can be expressed as
\begin{equation}
	I_r
	=
	\int_0^1
	r^{D/2-1}(1-r)^\beta
	G(r)\,\mathrm{d}r,
	\label{eq:radial_integral_convergence}
\end{equation}
where $G(r)$ denotes the transformed integrand resulting from the radial mapping.
The $N_r$-point Gauss--Jacobi quadrature $\mathcal{Q}_{N_r}$ associated with the weight
$r^{D/2-1}(1-r)^\beta$ is exact for all polynomials of degree up to $2N_r-1$.
Therefore, its convergence is governed by the polynomial approximability of $G(r)$ rather than by the Jacobi weight itself.
In particular, if $p\in\mathbb{P}_{2N_r-1}$ is any polynomial of degree at most $2N_r-1$, the polynomial component is integrated exactly, and the quadrature error depends only on the residual $G-p$.
Since the Gauss--Jacobi weights are positive and satisfy
\begin{equation}
	\sum_{i=1}^{N_r}\omega_i^{(r)}
	=
	\int_0^1
	r^{D/2-1}(1-r)^\beta\,\mathrm{d}r
	=
	B\left(\frac{D}{2},\beta+1\right),
\end{equation}
the quadrature error can be bounded by
\begin{equation}
	\left|
	I_r-\mathcal{Q}_{N_r}[G]
	\right|
	\leq
	2B\left(\frac{D}{2},\beta+1\right)
	\inf_{p\in\mathbb{P}_{2N_r-1}}
	\left\|
	G-p
	\right\|_{\infty}.
	\label{eq:radial_error_bound}
\end{equation}
where $B(\cdot,\cdot)$ denotes the beta function. Thus, the quadrature error is directly bounded by the best polynomial approximation error of the transformed integrand.

The regularity of $G(r)$ therefore determines the asymptotic convergence of the radial quadrature.
After an affine transformation of $r\in[0,1]$ to $[-1,1]$, suppose that $G(r)$ admits an analytic continuation to and inside a Bernstein ellipse $E_\rho$ with $\rho>1$.
Classical polynomial approximation theory then gives
\begin{equation}
	\inf_{p\in\mathbb{P}_n}
	\left\|
	G-p
	\right\|_{\infty}
	\leq
	C_\rho \rho^{-n},
	\qquad
	\rho>1,
	\label{eq:radial_poly_approx}
\end{equation}
where $C_\rho$ is independent of $n$.
Since the $N_r$-point Gauss--Jacobi rule has polynomial exactness degree $2N_r-1$, setting $n=2N_r-1$ yields
\begin{equation}
	\inf_{p\in\mathbb{P}_{2N_r-1}}
	\left\|
	G-p
	\right\|_{\infty}
	\leq
	C_\rho
	\rho^{-(2N_r-1)}
	=
	\widetilde{C}_\rho
	\rho^{-2N_r},
	\label{eq:radial_analytic_approx}
\end{equation}
where $\widetilde{C}_\rho=\rho C_\rho$.
Substitution into Eq.~\eqref{eq:radial_error_bound} then gives
\begin{equation}
	\left|
	I_r-\mathcal{Q}_{N_r}[G]
	\right|
	\leq
	C_r
	\rho^{-2N_r},
	\qquad
	\rho>1,
	\label{eq:radial_exponential_convergence}
\end{equation}
where $C_r$ is independent of $N_r$, demonstrating exponential convergence with respect to the number of radial quadrature points.

The above analysis shows that the radial convergence is governed by the regularity of the transformed integrand $G(r)$, rather than by the Bell-shaped weight alone.
Since the velocity variable is related to $r$ through the inverse mapping $\mathcal{R}^{-1}(r)$, as given by Eq.~\eqref{eq:asymptotic_velocity_nodes}, the regularity of $G(r)$ depends on both the distribution function $f(\boldsymbol{\xi})$ and the mapping properties.
If $G(r)$ is analytic in a neighborhood of $[0,1]$, its polynomial approximation converges geometrically, leading to the exponential convergence in Eq.~\eqref{eq:radial_exponential_convergence}.
Conversely, loss of analyticity or reduced regularity in $G(r)$ may result in a reduced convergence rate, generally from exponential to algebraic.

\noindent\textbf{Directional convergence.}
The directional convergence differs across velocity spaces.
For $D=2$, the directional integral is periodic and is discretized using the equally spaced trapezoidal rule.
For an analytic $2\pi$-periodic integrand, the trapezoidal rule exhibits exponential convergence,
\begin{equation}
	\left|
	I_\theta-\mathcal{Q}_{N_\theta}^{(\theta)}
	\right|
	=
	O\!\left(e^{-aN_\theta}\right),
	\qquad a>0,
	\label{eq:trapezoidal_convergence}
\end{equation}
where the convergence rate is determined by the analyticity of the integrand in the complex plane \cite{exponentially2014}.
For non-analytic integrands, the convergence is instead governed by the decay of their Fourier coefficients.

For $D=3$, the directional accuracy is governed by the distribution of quadrature nodes over the unit sphere and the approximation properties of the corresponding spherical integration rule.
The Lat--Lon discretization employs a tensor product of one-dimensional angular quadratures, providing a straightforward and systematic refinement strategy.
The Lebedev rule, in contrast, is constructed to integrate spherical harmonics exactly up to a prescribed degree, making it particularly effective for smooth angular distributions whose variation can be well represented by low-order spherical harmonics.
The specific correspondence between the number of Lebedev nodes $N_\Omega$ and the spherical harmonic exactness degree $t$ is provided in Table~\ref{tab:lebedev_nodes_order}.
The Fibonacci discretization generates quasi-uniform deterministic points on the sphere and does not impose a prescribed degree of spherical harmonic exactness.
Its accuracy therefore depends primarily on the uniformity of the point distribution and the regularity of the integrand.
Monte Carlo sampling uses randomly distributed points and is statistically unbiased for uniform spherical integration, but its root-mean-square error decreases only as $O(N_\Omega^{-1/2})$.
Thus, unlike deterministic spherical quadratures, increasing the number of Monte Carlo points yields only the slow statistical convergence characteristic of random sampling.

Accordingly, the overall velocity-space quadrature error results from the combined radial and directional discretization errors.
Since the radial and directional integrations are separated in the present formulation, their resolutions can be adjusted independently according to the regularity of the transformed radial integrand and the approximation properties of the selected directional quadrature or sampling procedures.

\begin{table}[!tbhp]
	\centering
	\caption{Number of Lebedev nodes $N_\Omega$ and corresponding spherical harmonic exactness degree $t$.}
	\label{tab:lebedev_nodes_order}
	\begin{tabular}{rrrrrrrrrrrr}
		\hline
		$N_\Omega$
		& 6 & 14 & 26 & 38 & 50 & 74 & 86 & 110 & 146 & 170 & 194 \\
		$t$
		& 3 & 5 & 7 & 9 & 11 & 13 & 15 & 17 & 19 & 21 & 23 \\
		\hline
		$N_\Omega$
		& 230 & 266 & 302 & 350 & 434 & 590 & 770 & 974 & 1202 & 1454 & 1730 \\
		$t$
		& 25 & 27 & 29 & 31 & 35 & 41 & 47 & 53 & 59 & 65 & 71 \\
		\hline
		$N_\Omega$
		& 2030 & 2354 & 2702 & 3074 & 3470 & 3890 & 4334 & 4802 & 5294 & 5810 & \\
		$t$
		& 77 & 83 & 89 & 95 & 101 & 107 & 113 & 119 & 125 & 131 & \\
		\hline
	\end{tabular}
\end{table}

\section{Numerical tests}
\label{sec4}

In this section, a series of benchmark problems are presented to systematically evaluate the proposed quadrature framework from three perspectives: the performance of different radial velocity distributions, the accuracy of different directional quadrature strategies, and the applicability to both structured and unstructured spatial meshes. Two-dimensional Rayleigh flow and thermally-driven cavity flow are first considered to examine the effects of radial quadratures and circumferential directional discretizations. Subsequently, three-dimensional lid-driven cavity flow and supersonic flow past a sphere are employed to validate different spherical quadrature rules and radial velocity distributions, while demonstrating the robustness and flexibility of the proposed framework for simulations on both structured and unstructured meshes. It is worth noting that the radial mappings considered in this work, including the \emph{exp}, \emph{erf}, \emph{tanh}, \emph{arctan}, and \emph{gd}-based mappings, are summarized in Tables~\ref{tab:classical} and~\ref{tab:other_bell}.

\subsection{Rayleigh flow}
\label{sec4.1}

Rayleigh flow is selected as the first benchmark to assess the proposed bell-shaped velocity quadrature framework for unsteady nonequilibrium gas flows. Particular attention is given to the velocity-space distributions generated by different radial mappings and their effects on numerical accuracy.

The problem considers an infinitely long plate immersed in argon gas initially at rest with a uniform temperature of $T_0=273~\mathrm{K}$. At $t=0$, the plate is impulsively set into motion at $U_w=10~\mathrm{m/s}$ and heated to $T_w=373~\mathrm{K}$. Five representative times, $t=0.01\tau_0$, $0.1\tau_0$, $\tau_0$, $10\tau_0$, and $100\tau_0$, are considered, corresponding to $\mathrm{Kn}=26.6 \sim 0.00266$ and covering regimes from highly rarefied nonequilibrium to near-equilibrium conditions. The computational domain has a width of $L=1~\mathrm{m}$ and is discretized into 128 uniform cells. The variable hard sphere (VHS) model with $\omega=0.81$ is employed, and the CFL number is fixed at 0.95.

The velocity-space resolution is first determined from the most demanding case, $t=0.01\tau_0$ ($\mathrm{Kn}=26.6$). The analytical free-molecular solution reported in Ref.~\cite{Rayleigh2002} is used as the reference, and the $L_2$ errors of the macroscopic variables are evaluated. Using the \emph{tanh} mapping as a representative case, radial resolutions of $N_r=8$, 16, and 32 are examined with $\alpha=35$, 10, and 6, respectively, and $\beta=\alpha$, yielding a comparable effective velocity range of $\xi\in[-4,4]$. As shown in Table~\ref{tab:rayleigh_error_tanh}, increasing $N_r$ from 8 to 16 substantially reduces the errors, whereas further refinement to $N_r=32$ produces only marginal improvement. A similar trend is observed for the angular resolution, with $N_\theta=32$ providing sufficient accuracy. Therefore, the $16\times32$ velocity grid is adopted as a good compromise between accuracy and computational cost.

\begin{figure*}[!th]
	\centering
	\subfigure{
		\includegraphics[width=0.32\textwidth]{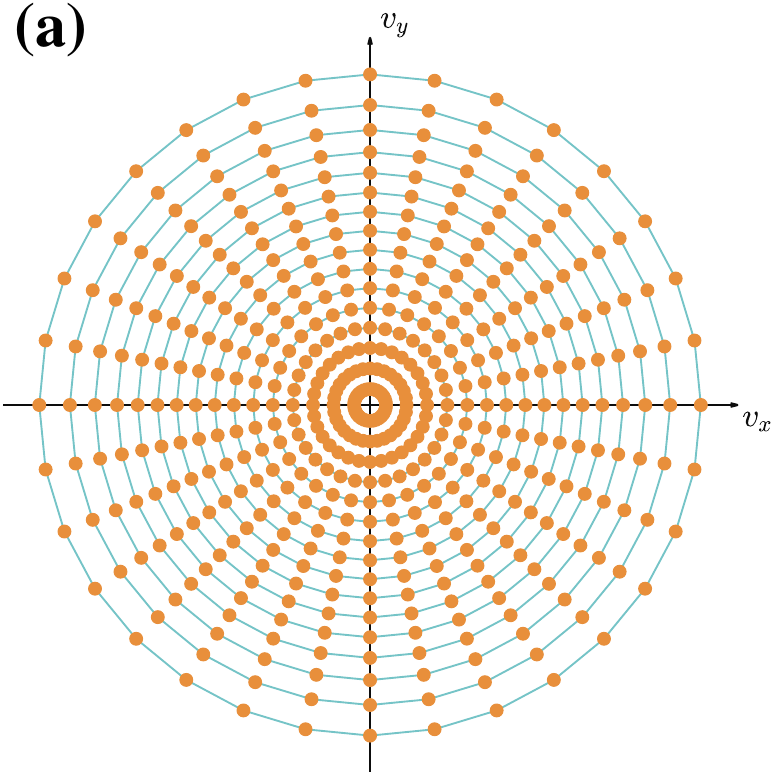}}
	\subfigure{
		\includegraphics[width=0.32\textwidth]{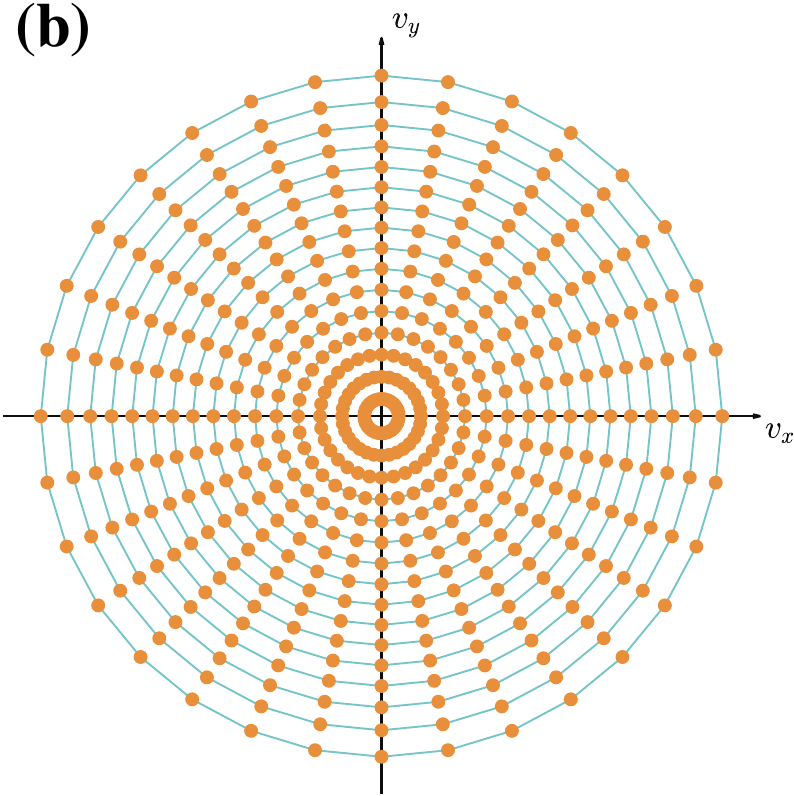}}
	\subfigure{
		\includegraphics[width=0.32\textwidth]{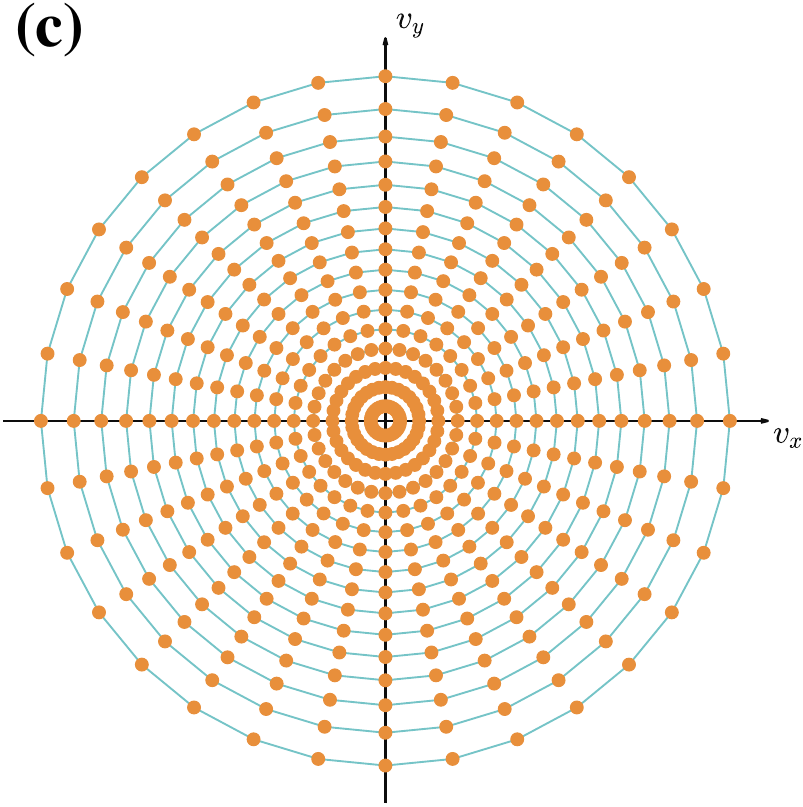}}
	\caption{Discrete velocity distributions for different radial mappings with $N_r \times N_\theta=16\times32$: (a) \emph{tanh}; (b) \emph{erf}; (c) \emph{arctan}.}
	\label{4.1Fig1_velocity_distribution}
\end{figure*}

Three radial mappings based on the \emph{tanh}, \emph{erf}, and \emph{arctan} functions are then compared. The parameters are chosen as $(\alpha,\beta)=(10,10)$, $(14,14)$, and $(70,200)$, respectively, to maintain the same effective velocity range $\xi\in[-4,4]$. As shown in Figure.~\ref{4.1Fig1_velocity_distribution}, all three mappings produce similar radial distributions, characterized by dense nodes near the center and progressively sparser nodes toward larger velocities. The \emph{arctan} mapping exhibits a slightly stronger central concentration. Correspondingly, the three mappings yield very similar numerical errors, with the \emph{arctan} mapping providing the highest accuracy, as reported in Table~\ref{tab:rayleigh_error}. This behavior indicates that the overall solution is relatively insensitive to the specific functional form of the radial mapping when the velocity range and resolution are properly controlled.

\begin{figure*}[!th]
	\centering
	\subfigure{
		\includegraphics[width=0.45\textwidth]{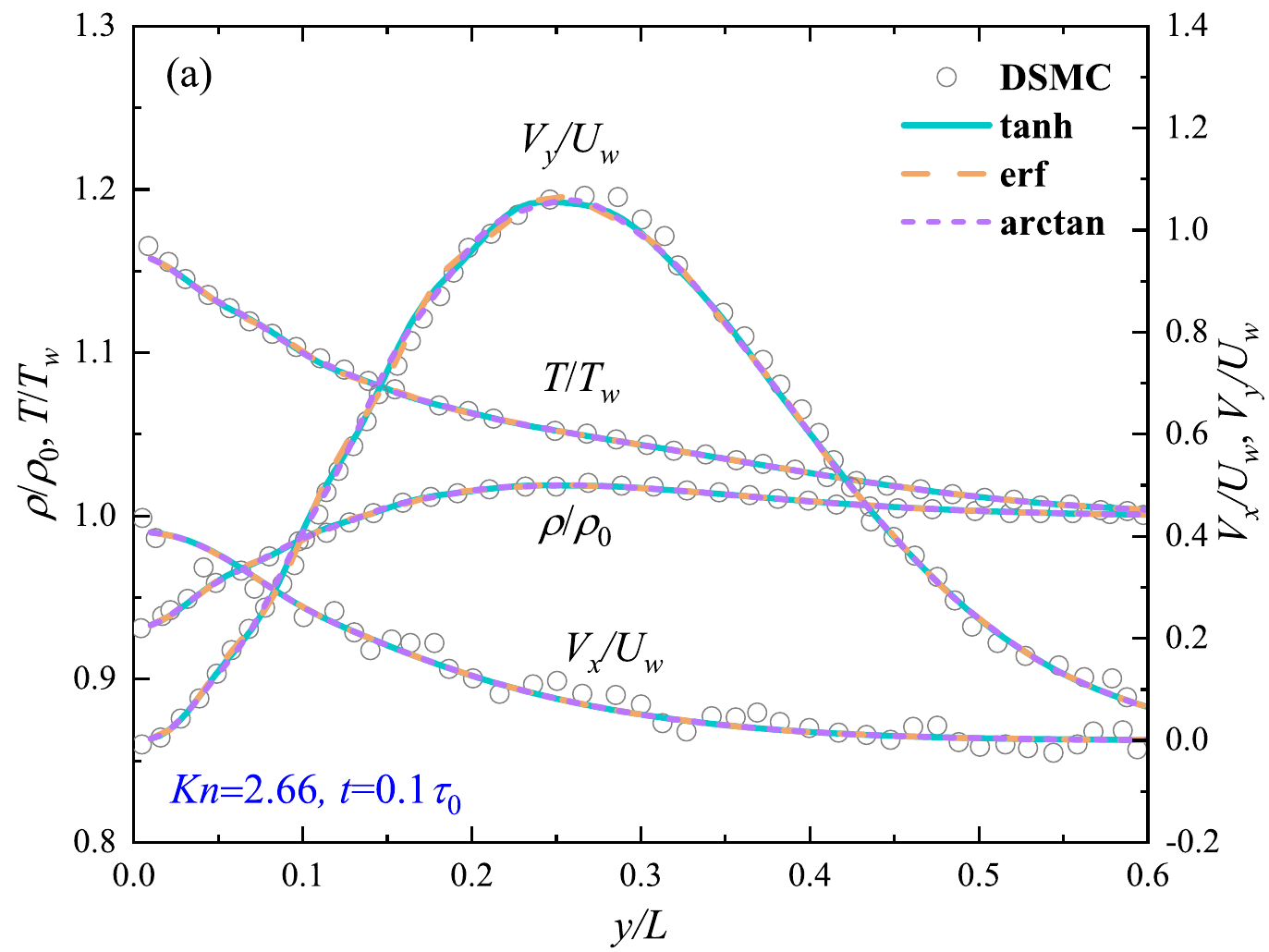}}\quad
	\subfigure{
		\includegraphics[width=0.45\textwidth]{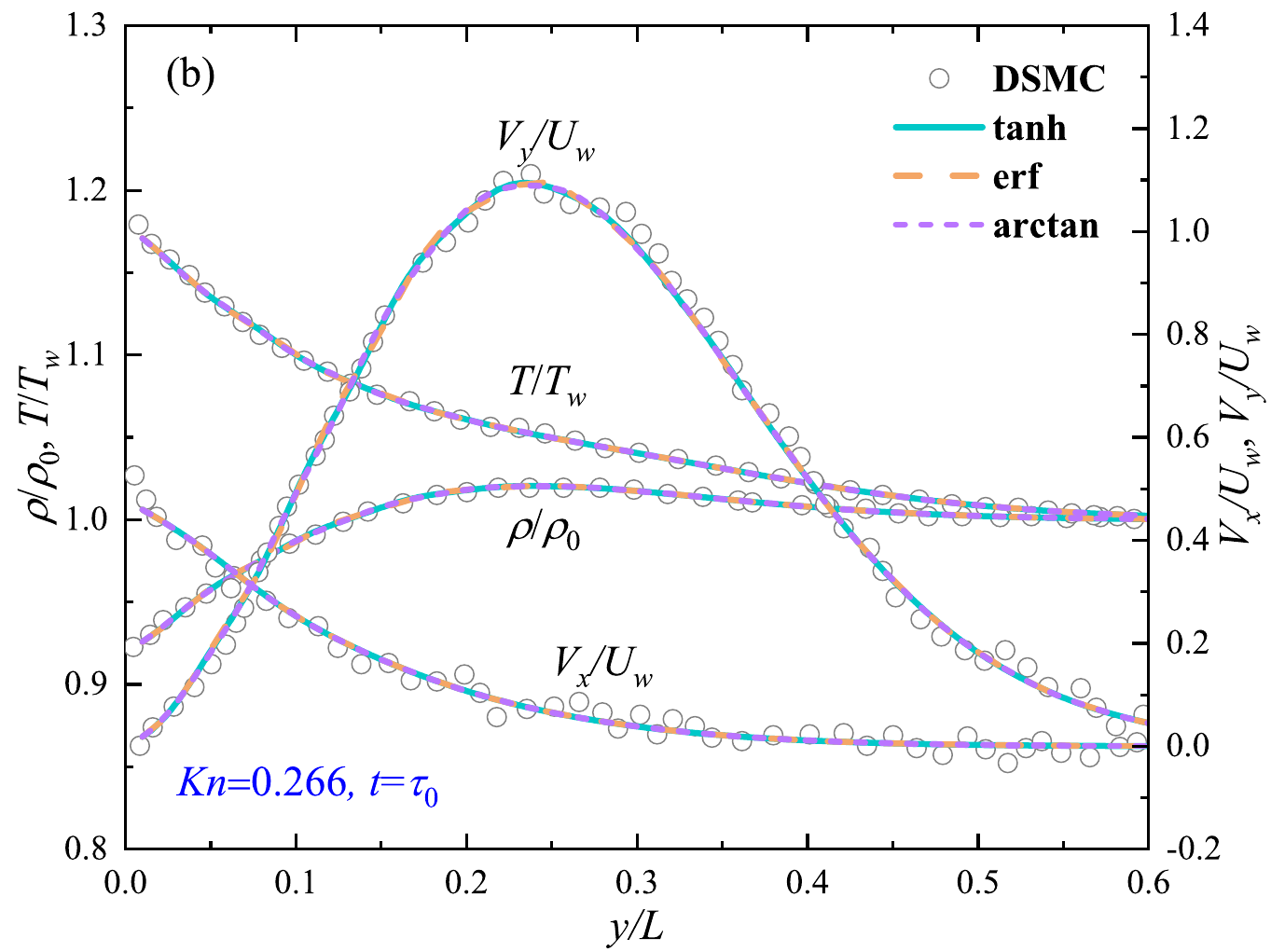}}
	\subfigure{
		\includegraphics[width=0.45\textwidth]{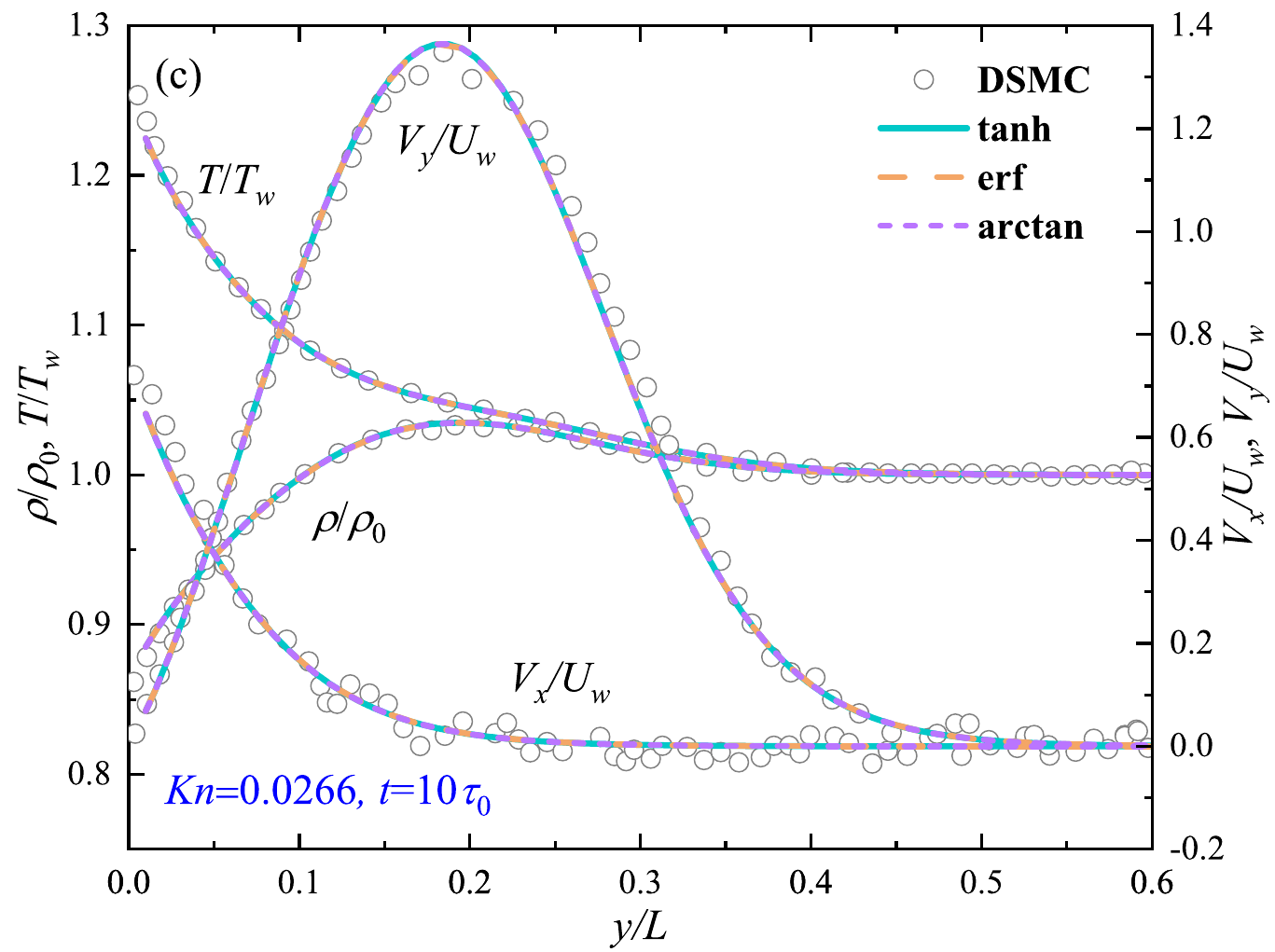}}\quad
	\subfigure{
		\includegraphics[width=0.45\textwidth]{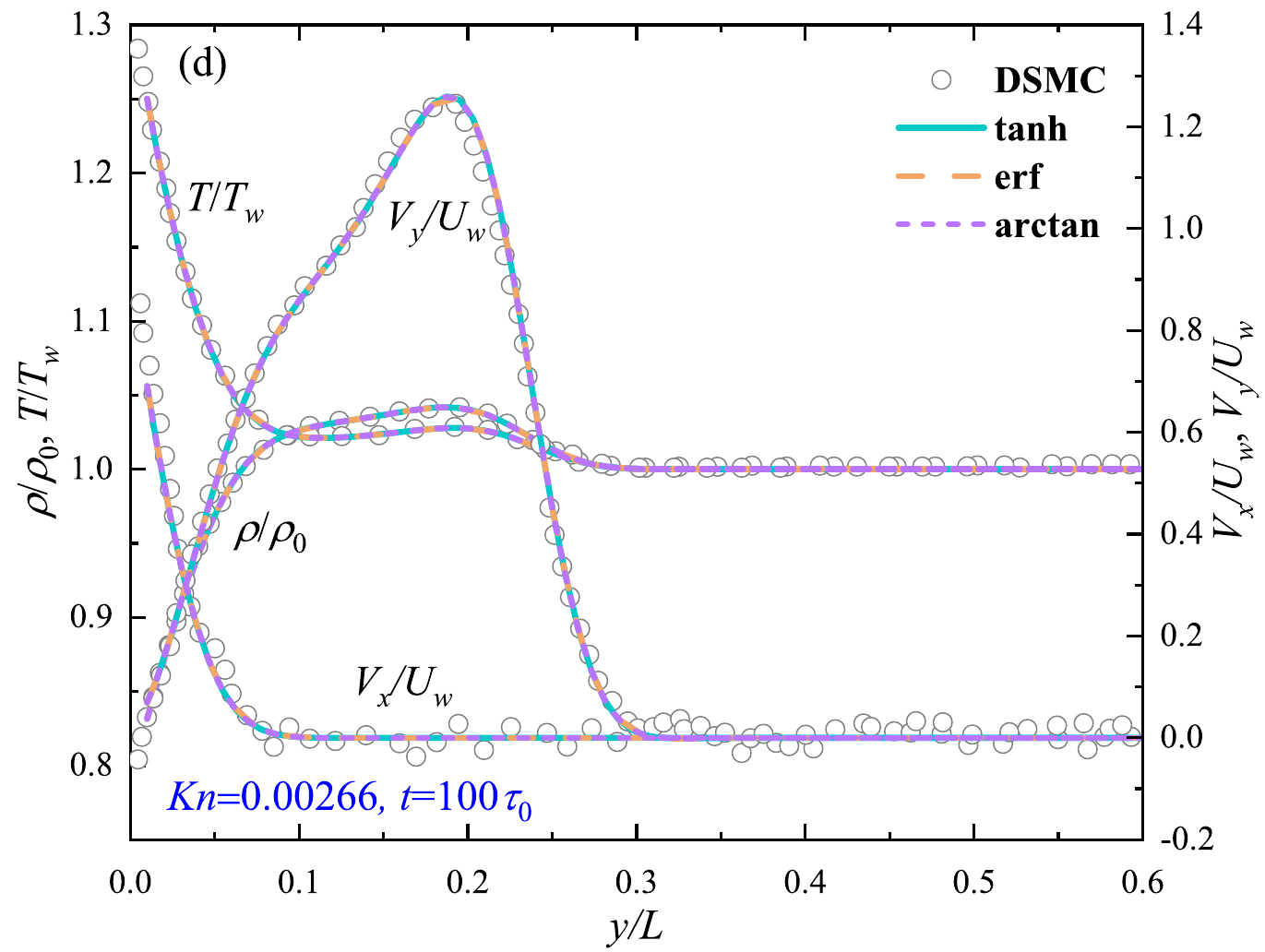}}
	\caption{\label{4.1Fig2_Rayleigh} \centering  Density, temperature, and velocity profiles for the Rayleigh flow obtained by DSMC and different quadrature methods.}	
\end{figure*}

Using the selected $16\times32$ velocity grid, Figure.~\ref{4.1Fig2_Rayleigh} compares the density, temperature, and velocity profiles obtained with the three mappings against DSMC results at $t=0.1\tau_0$, $\tau_0$, $10\tau_0$, and $100\tau_0$. At the early stage, where nonequilibrium effects are strongest, pronounced gradients develop in both the thermal and momentum boundary layers, yet all three mappings agree closely with the DSMC results. As the gas relaxes toward equilibrium, the macroscopic profiles become progressively smoother and the differences among the mappings further diminish.

Overall, the three radial mappings exhibit consistent velocity-space representations and comparable accuracy over a wide range of nonequilibrium conditions, demonstrating the robustness and generality of the proposed unified bell-shaped quadrature framework.

\begin{table}[htbp]
	\centering
	\caption{$L_2$ errors for Rayleigh flow at $t=0.01\tau_0$ under different discrete velocity grids.}
	\label{tab:rayleigh_error_tanh}
	\begin{tabular}{lcccccc}
		\toprule
		\multirow{2}{*}{Error}
		& \multicolumn{3}{c}{Varying $N_r$ ($N_{\theta}=32$)}
		& \multicolumn{3}{c}{Varying $N_{\theta}$ ($N_r=16$)} \\
		\cmidrule(r){2-4}
		\cmidrule(l){5-7}
		& $8\times32$
		& $16\times32$
		& $32\times32$
		& $16\times16$
		& $16\times32$
		& $16\times64$ \\
		\midrule
		$E(\rho)$
		& $1.80\times10^{-3}$
		& $4.18\times10^{-4}$
		& $2.69\times10^{-4}$
		& $9.09\times10^{-4}$
		& $4.18\times10^{-4}$
		& $3.11\times10^{-4}$ \\
		
		$E(T)$
		& $1.74\times10^{-3}$
		& $8.52\times10^{-4}$
		& $8.81\times10^{-4}$
		& $2.50\times10^{-3}$
		& $8.52\times10^{-4}$
		& $3.23\times10^{-4}$ \\
		
		$E(U)$
		& $6.21\times10^{-2}$
		& $1.62\times10^{-2}$
		& $1.58\times10^{-2}$
		& $3.80\times10^{-2}$
		& $1.62\times10^{-2}$
		& $1.22\times10^{-2}$ \\
		\bottomrule
	\end{tabular}
\end{table}

\begin{table}[htbp]
	\centering
	\caption{$L_2$ errors for Rayleigh flow at $t=0.01\tau_0$ under different radial mappings.}
	\label{tab:rayleigh_error}
	\begin{tabular}{lccc}
		\toprule
		{Error}
		& tanh
		& erf
		& arctan\\
		\midrule
		$E(\rho)$
		& $4.18\times10^{-4}$
		& $5.54\times10^{-4}$
		& $3.47\times10^{-4}$ \\
		
		$E(T)$
		& $8.52\times10^{-4}$
		& $8.67\times10^{-4}$
		& $8.43\times10^{-4}$\\
		
		$E(U)$
		& $1.62\times10^{-2}$
		& $1.77\times10^{-2}$
		& $1.59\times10^{-2}$\\
		\bottomrule
	\end{tabular}
\end{table}

\subsection{Thermally-Driven Cavity Flow}
\label{sec4.2}

\begin{figure*}[!th]
	\centering
	
	\subfigure[$T$, $\mathrm{Kn}=0.001$]{
		\includegraphics[width=0.32\textwidth]{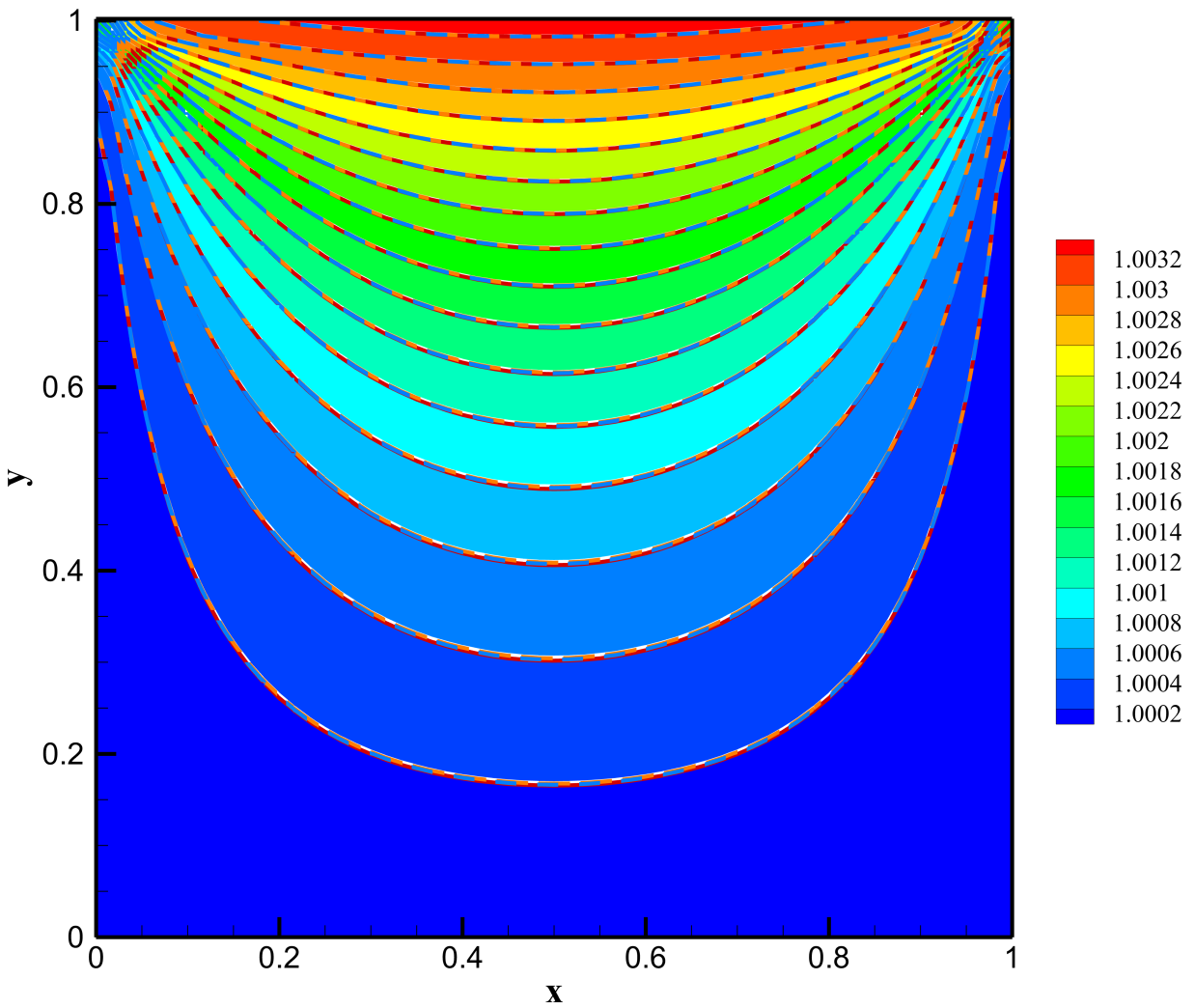}}
	\subfigure[$U_x$, $\mathrm{Kn}=0.001$]{
		\includegraphics[width=0.32\textwidth]{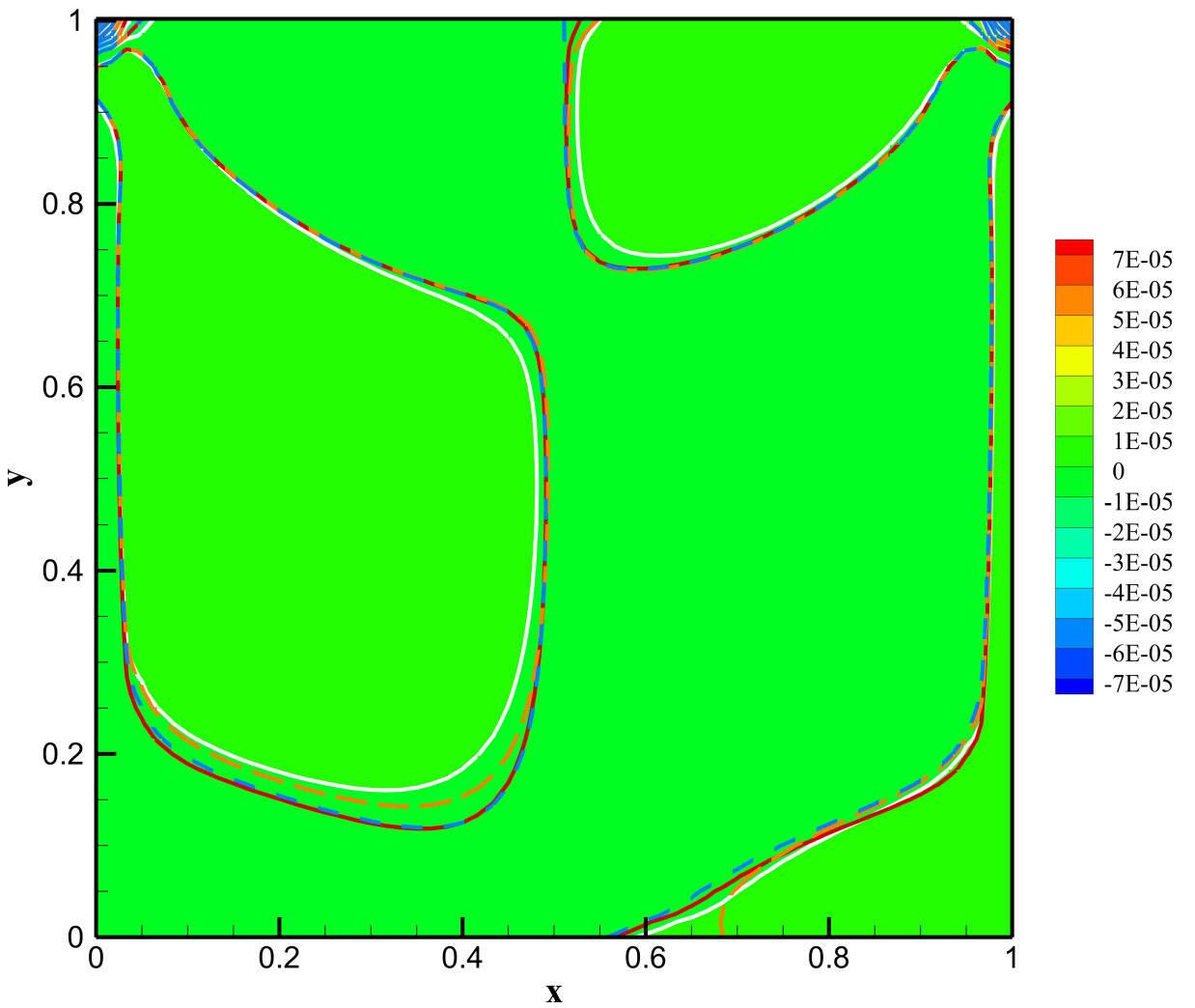}}
	\subfigure[$U_y$, $\mathrm{Kn}=0.001$]{
		\includegraphics[width=0.32\textwidth]{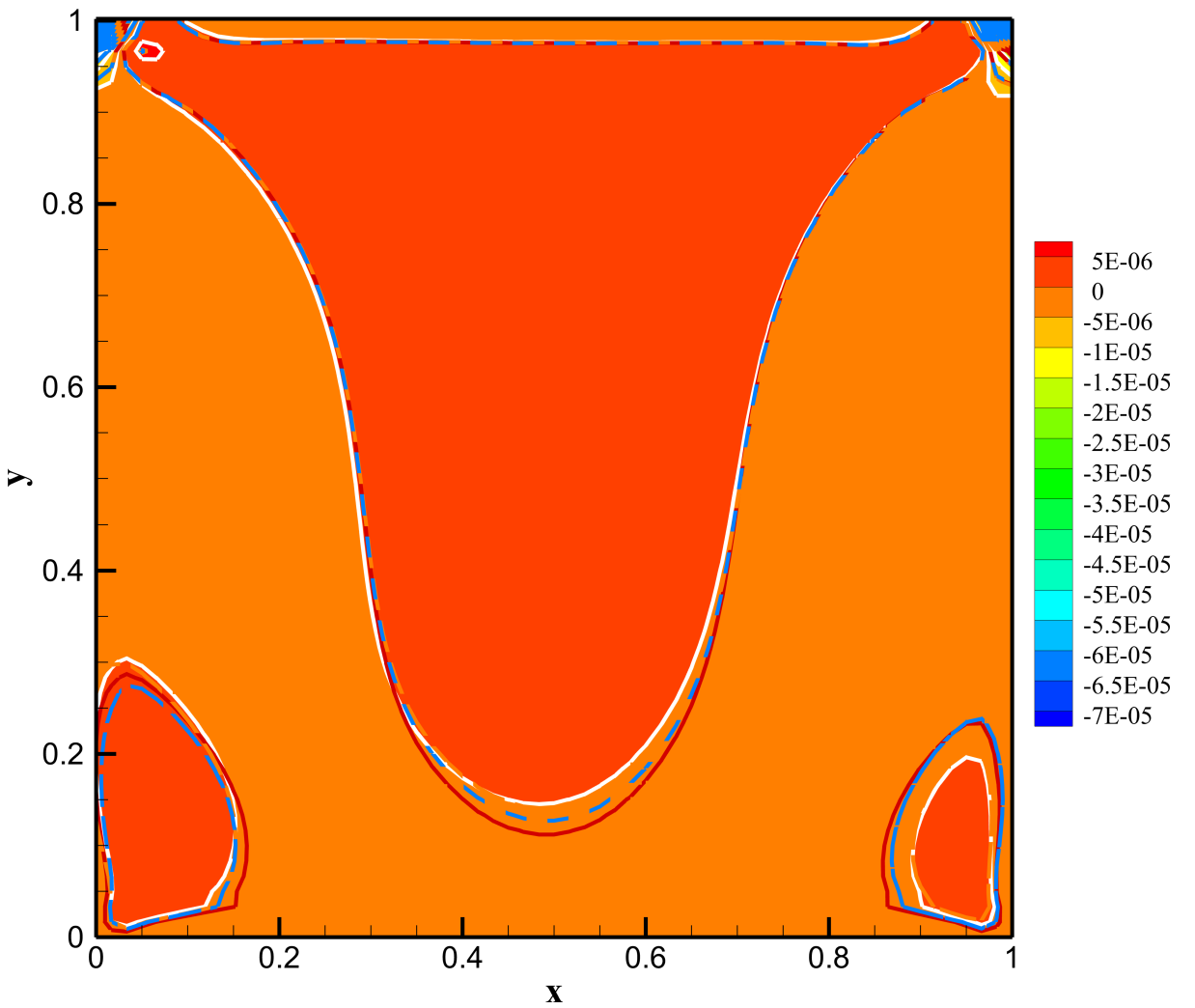}}
	
	\vspace{2mm}
	
	\subfigure[$T$, $\mathrm{Kn}=0.1$]{
		\includegraphics[width=0.32\textwidth]{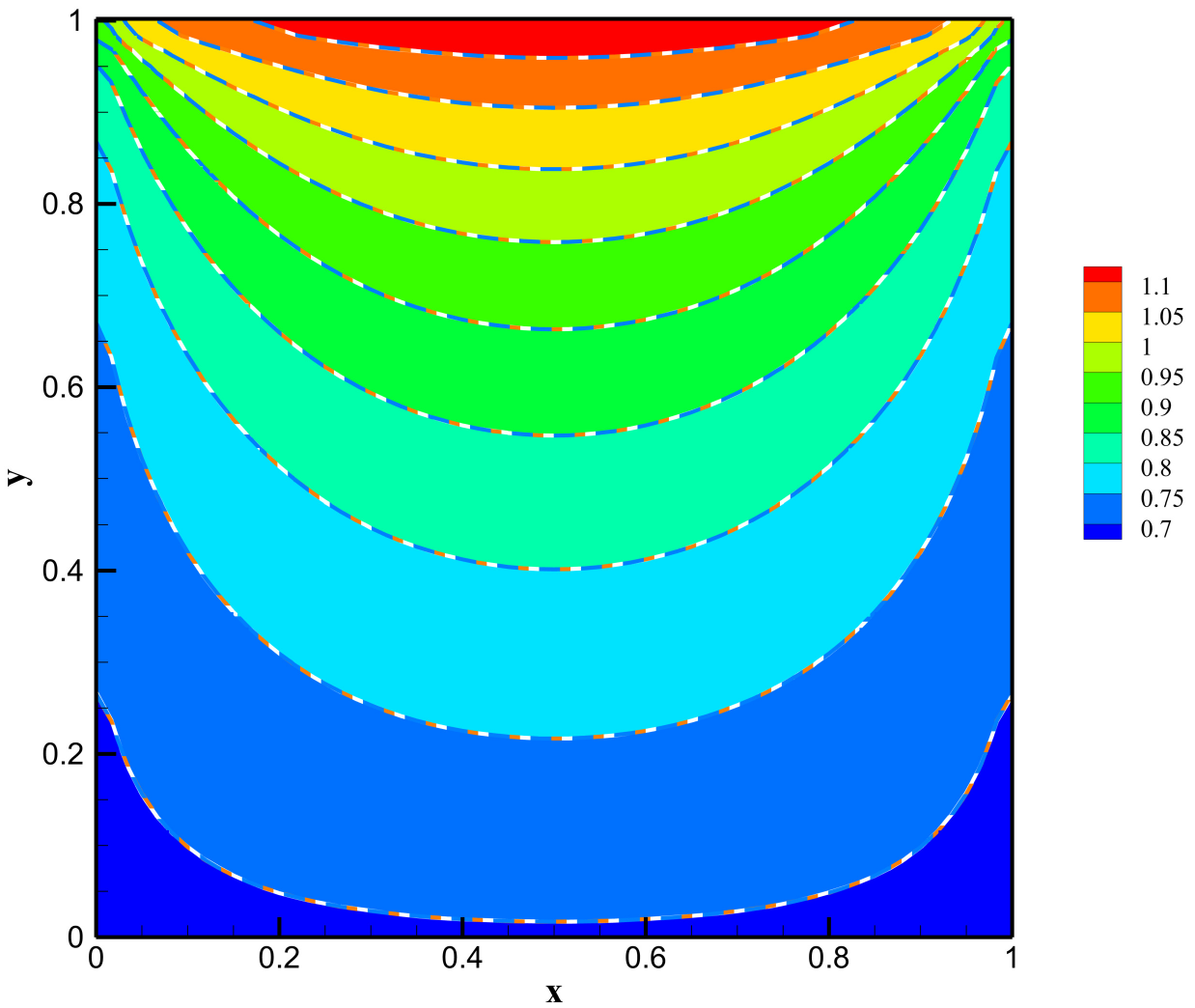}}
	\subfigure[$U_x$, $\mathrm{Kn}=0.1$]{
		\includegraphics[width=0.32\textwidth]{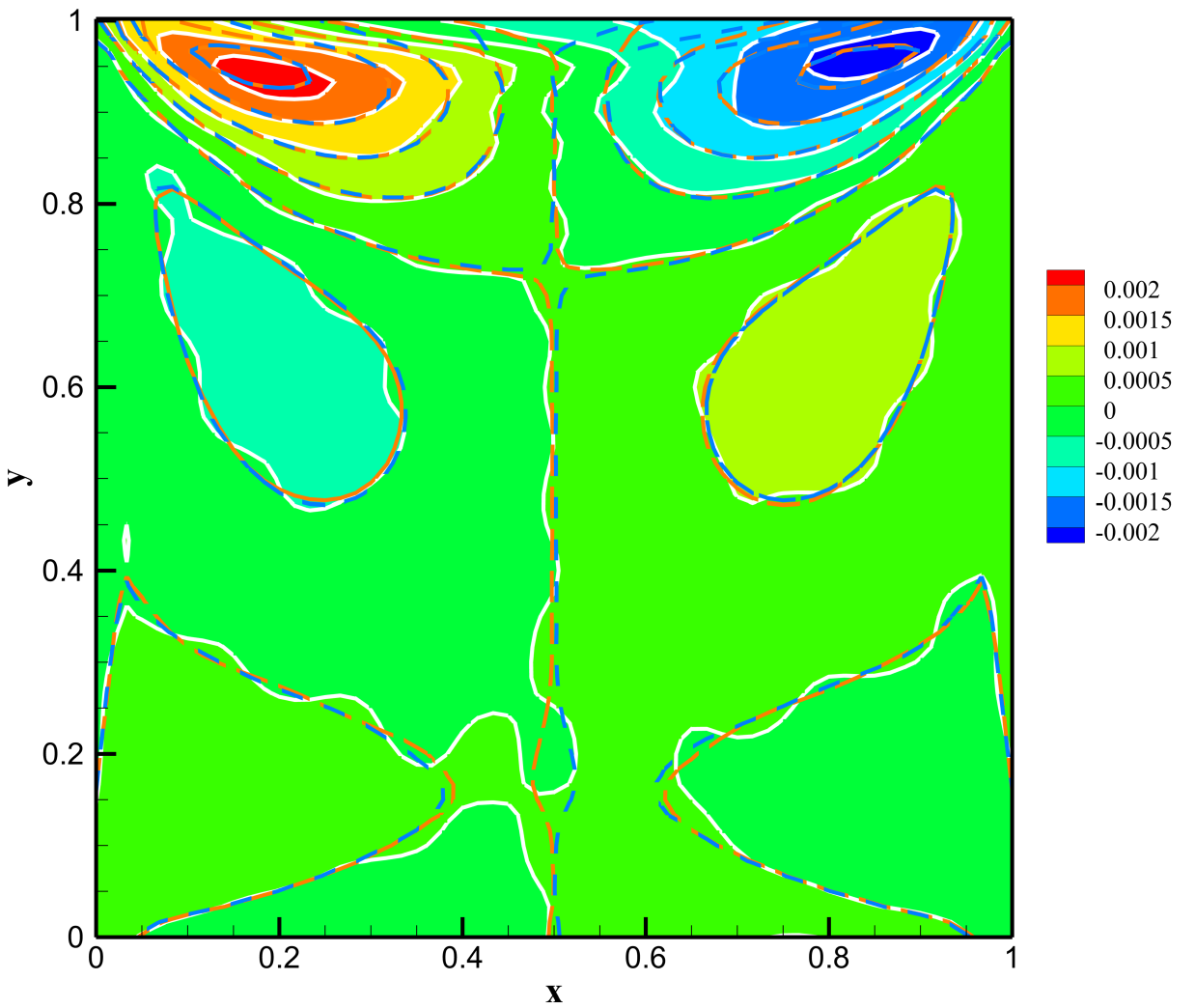}}
	\subfigure[$U_y$, $\mathrm{Kn}=0.1$]{
		\includegraphics[width=0.32\textwidth]{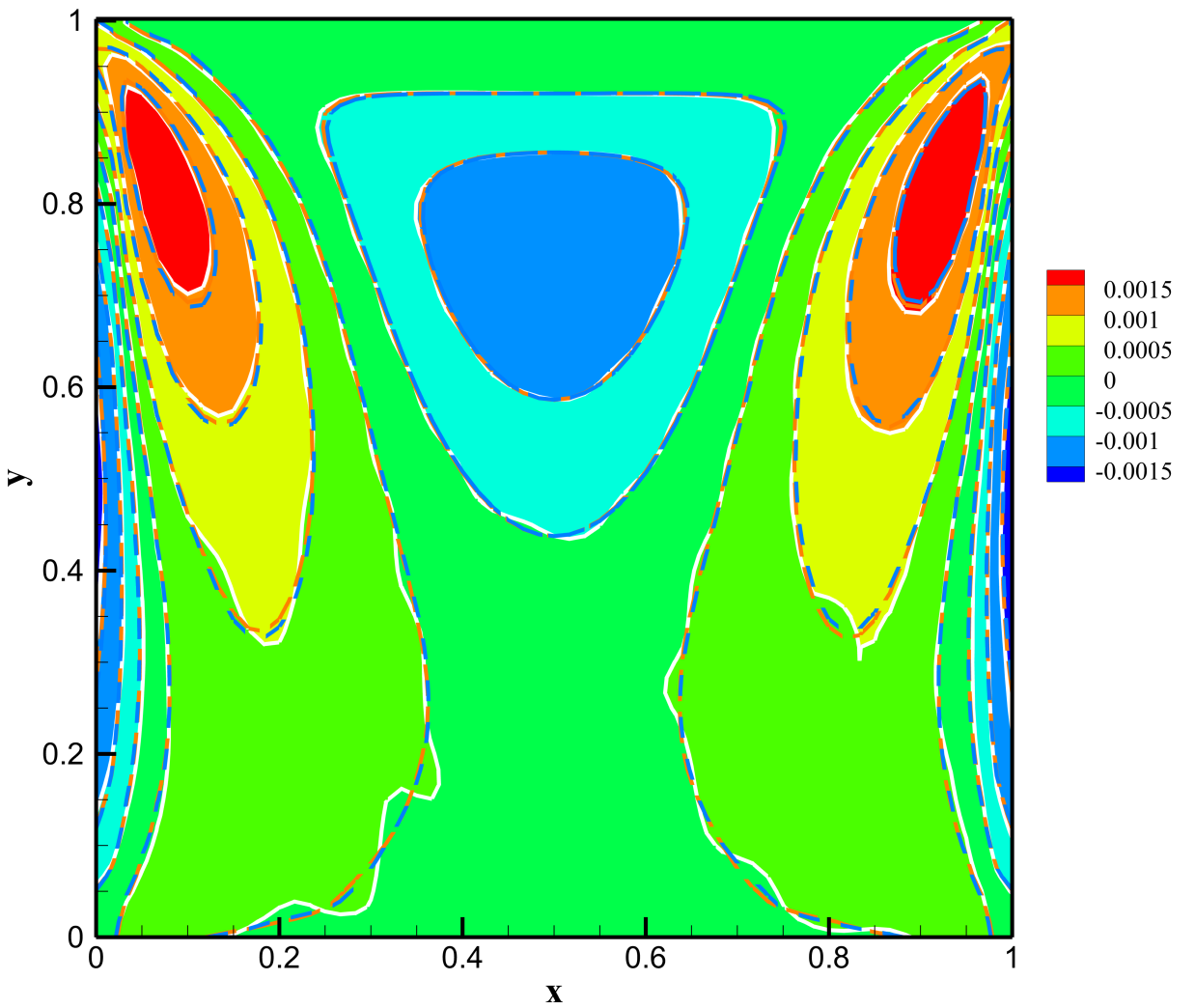}}
	
	\vspace{2mm}
	
	\subfigure[$T$, $\mathrm{Kn}=1$]{
		\includegraphics[width=0.32\textwidth]{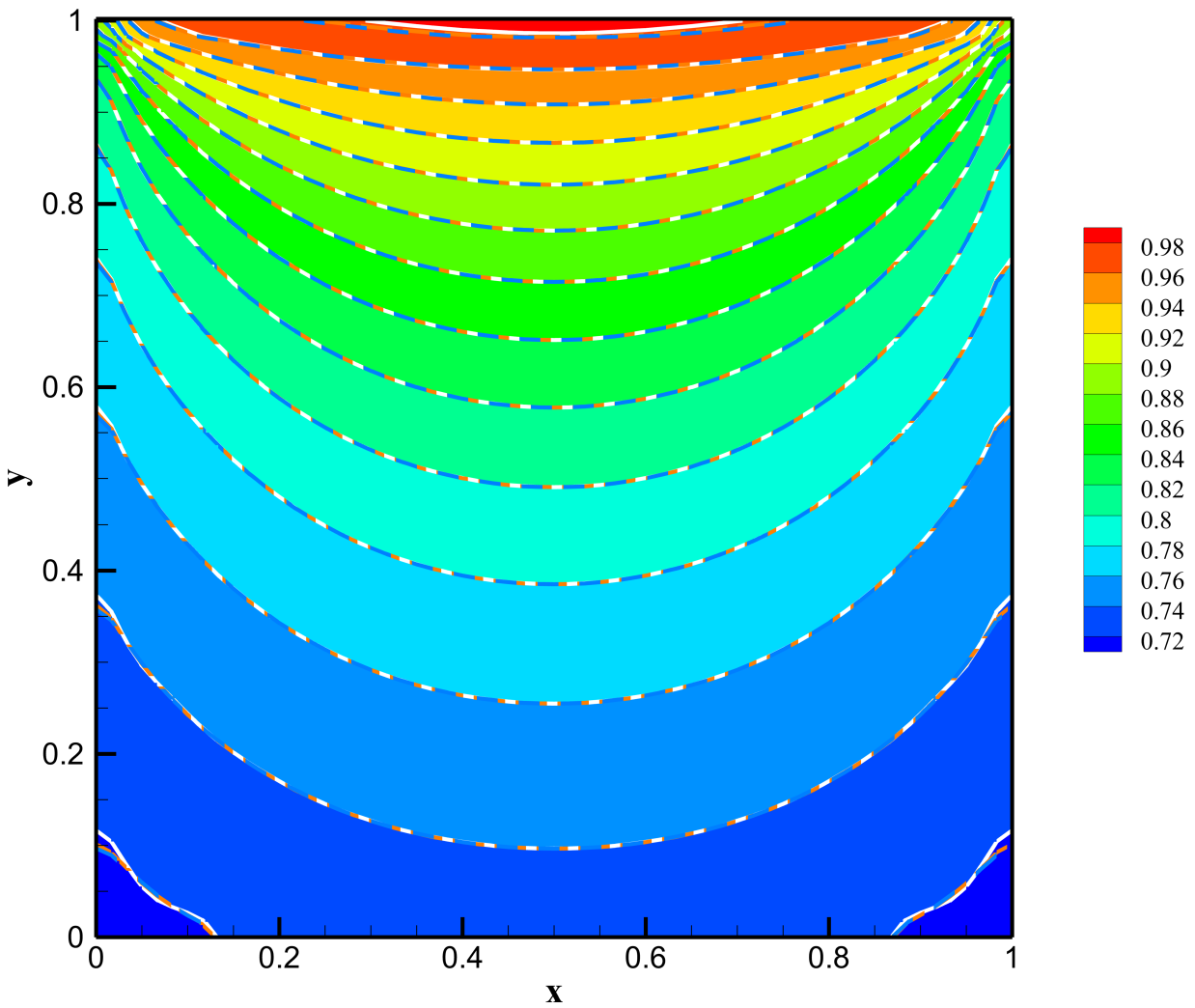}}
	\subfigure[$U_x$, $\mathrm{Kn}=1$]{
		\includegraphics[width=0.32\textwidth]{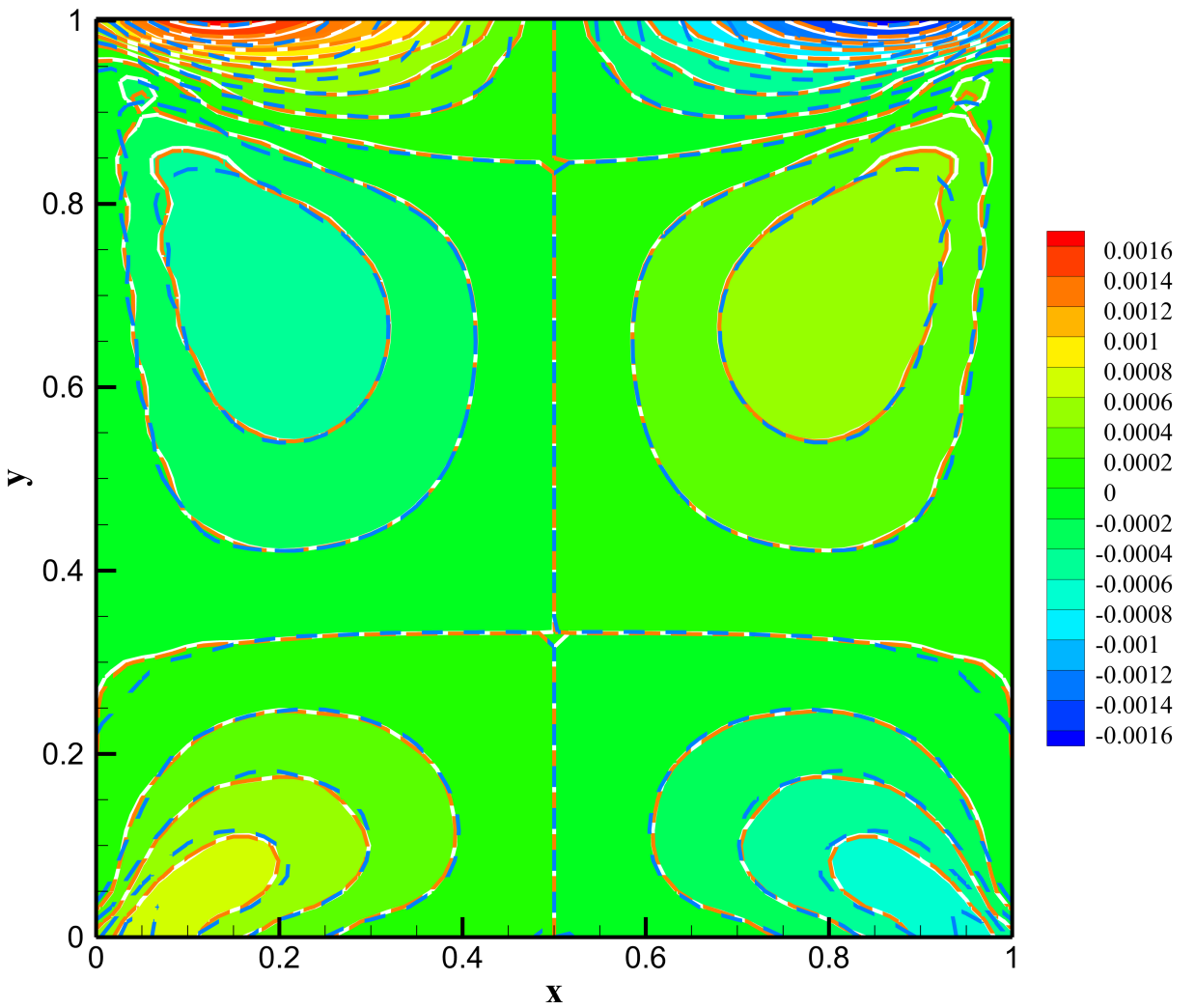}}
	\subfigure[$U_y$, $\mathrm{Kn}=1$]{
		\includegraphics[width=0.32\textwidth]{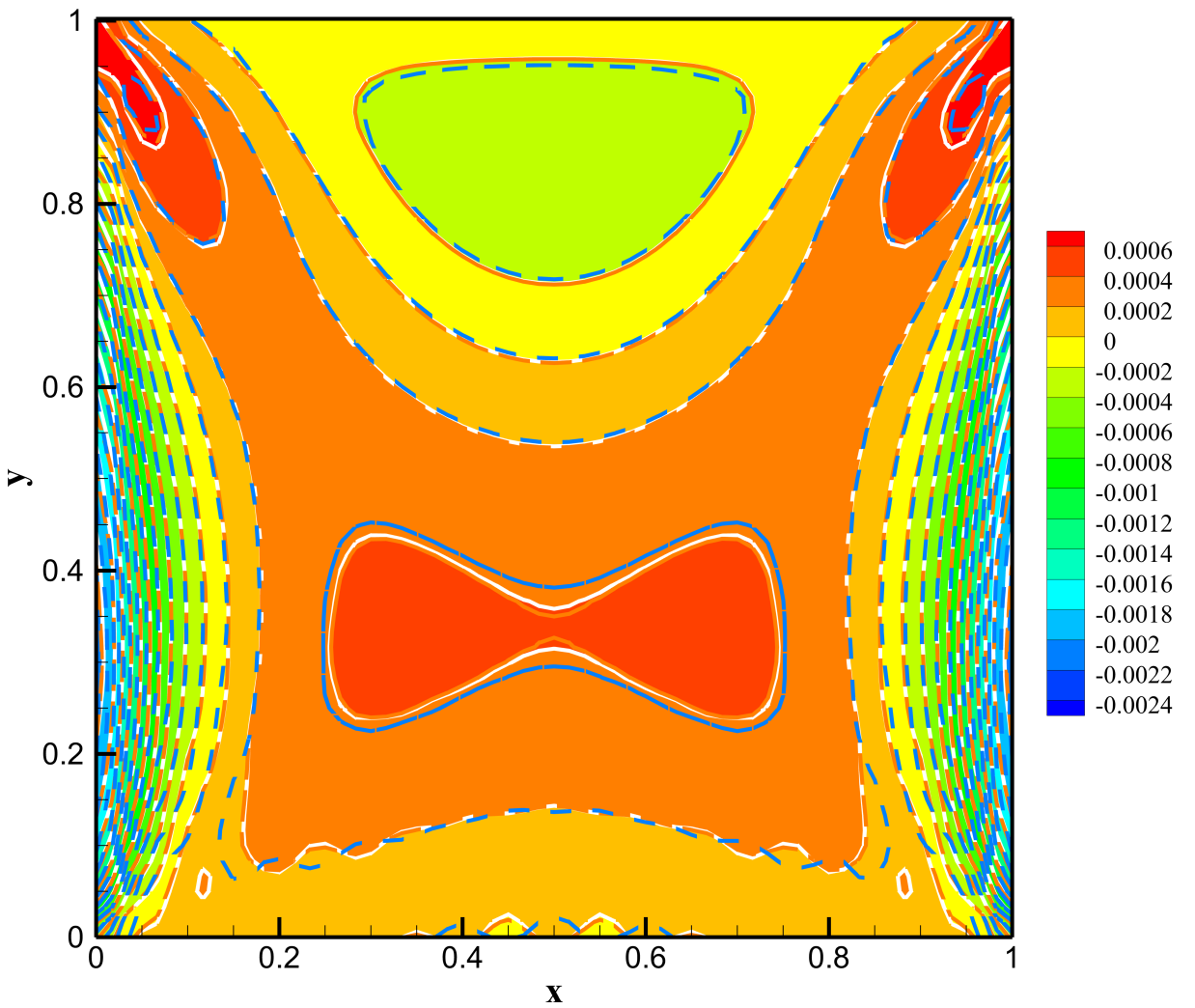}}
	
	\vspace{2mm}
	
	\subfigure[$T$, $\mathrm{Kn}=10$]{
		\includegraphics[width=0.32\textwidth]{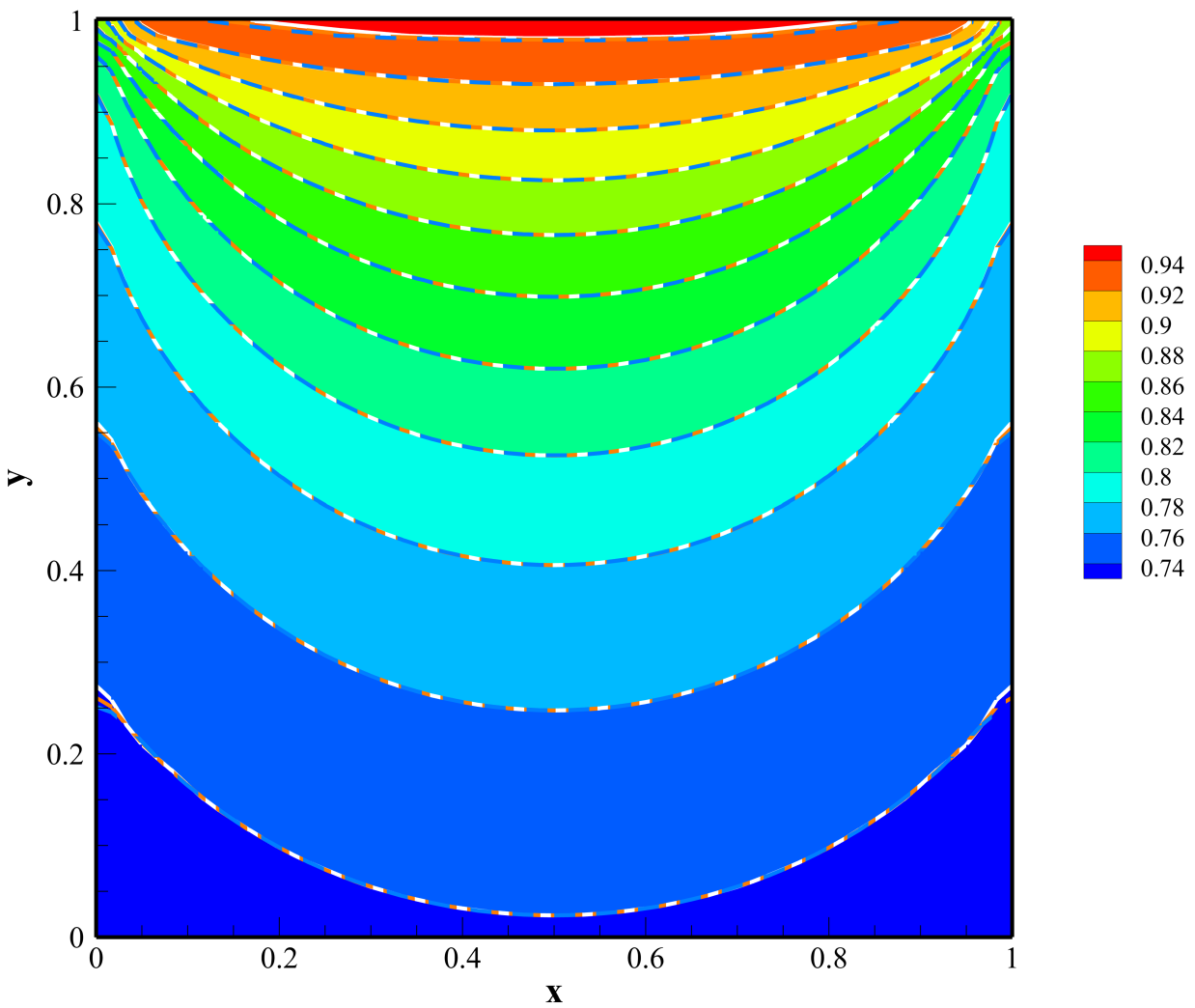}}
	\subfigure[$U_x$, $\mathrm{Kn}=10$]{
		\includegraphics[width=0.32\textwidth]{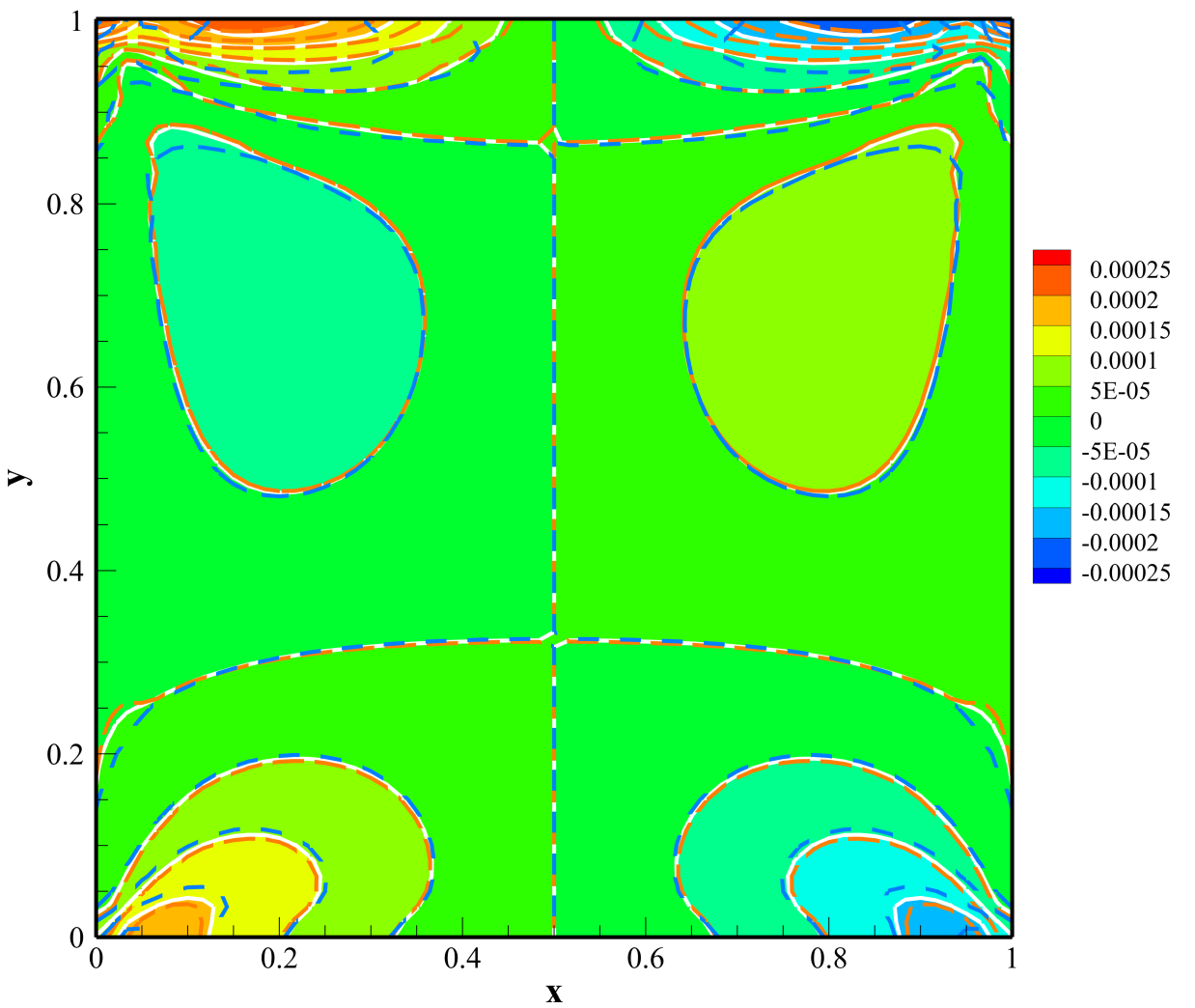}}
	\subfigure[$U_y$, $\mathrm{Kn}=10$]{
		\includegraphics[width=0.32\textwidth]{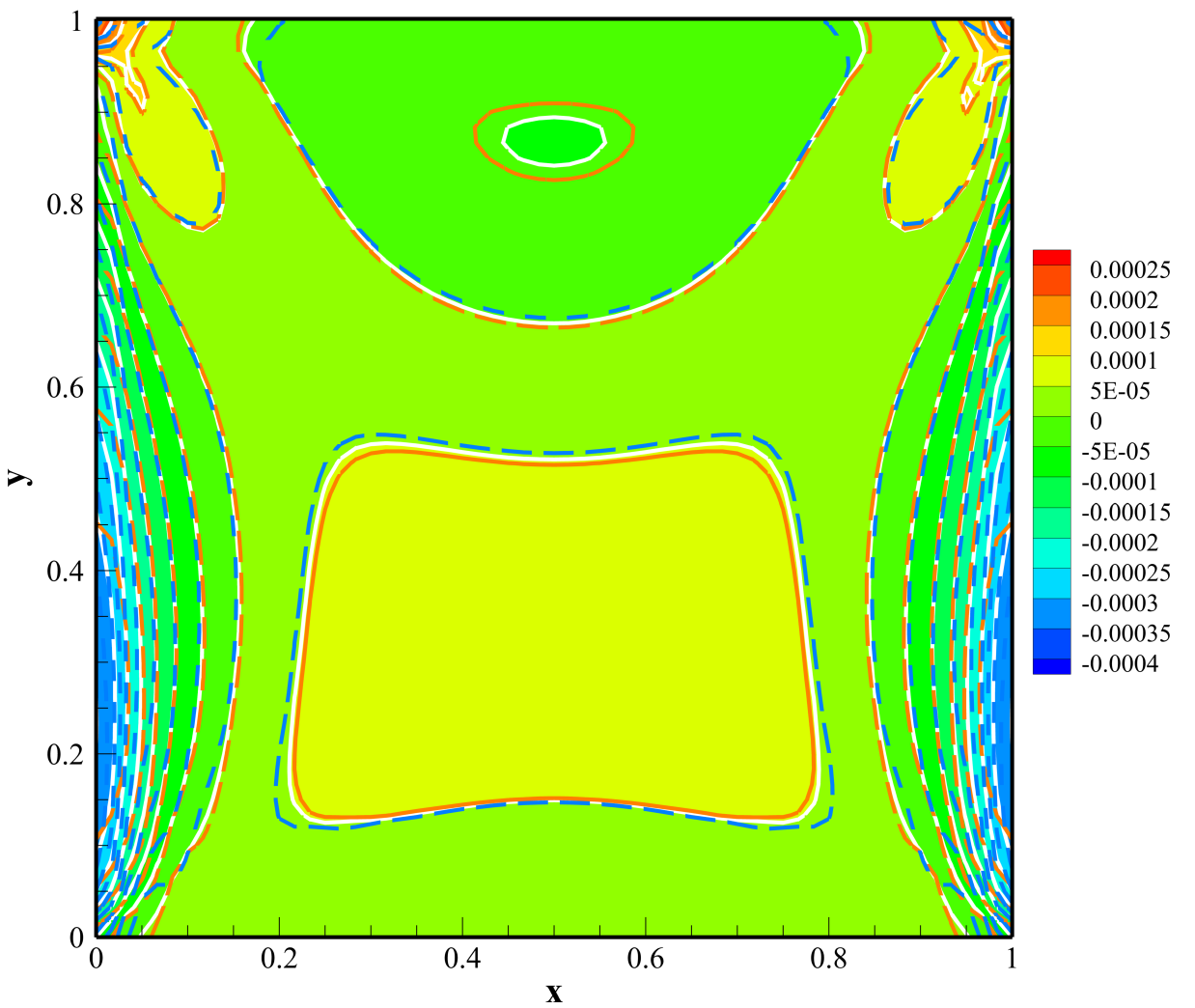}}
	
	\caption{Temperature and velocity contours of the thermally driven cavity flow.	
		(\emph{aligned}: coloured background with white solid lines; \emph{shifted}: orange long-dashed lines; \emph{staggered}: blue dashed lines). 	
		For $\mathrm{Kn}=0.001$, results from the dense \emph{aligned} configuration are also included (red solid line).
		(For interpretation of the references to colour in this figure legend, the reader is referred to the web version of this article).}
	\label{Fig4.1}
\end{figure*}

\begin{figure*}[!th]
	\centering
	\subfigure[]{
		\includegraphics[width=0.48\textwidth]{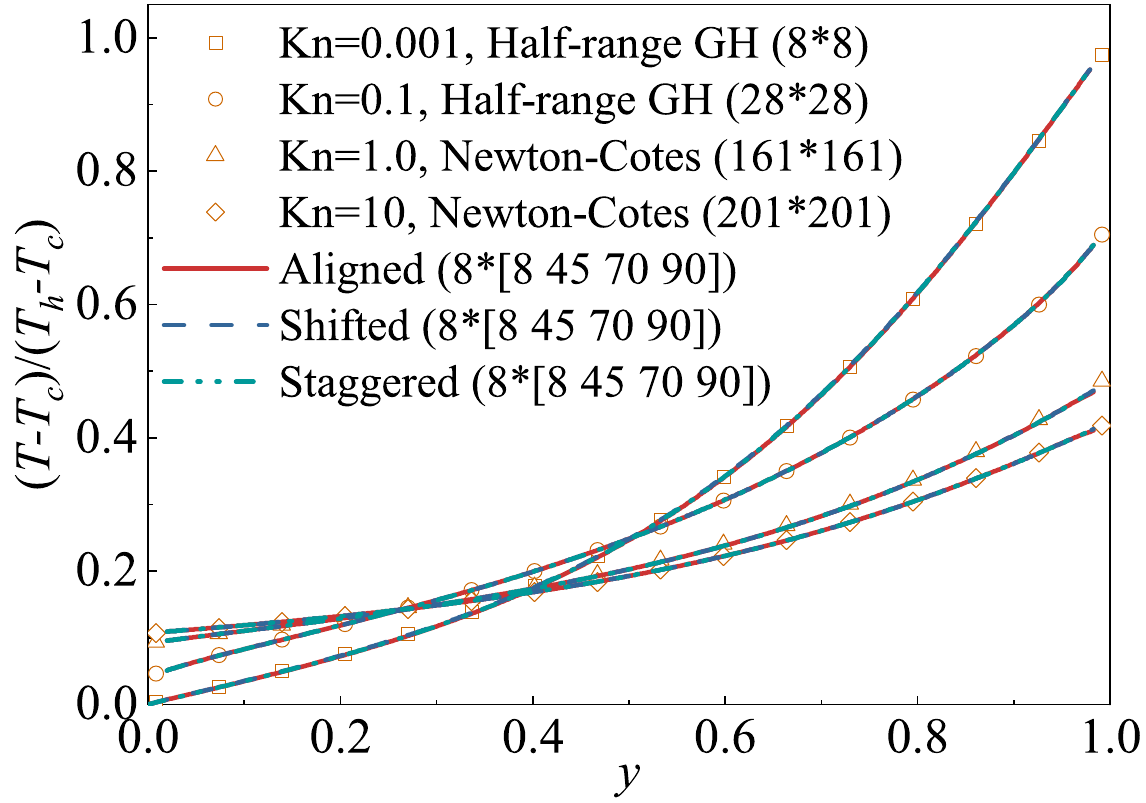}}
	\subfigure[]{
		\includegraphics[width=0.48\textwidth]{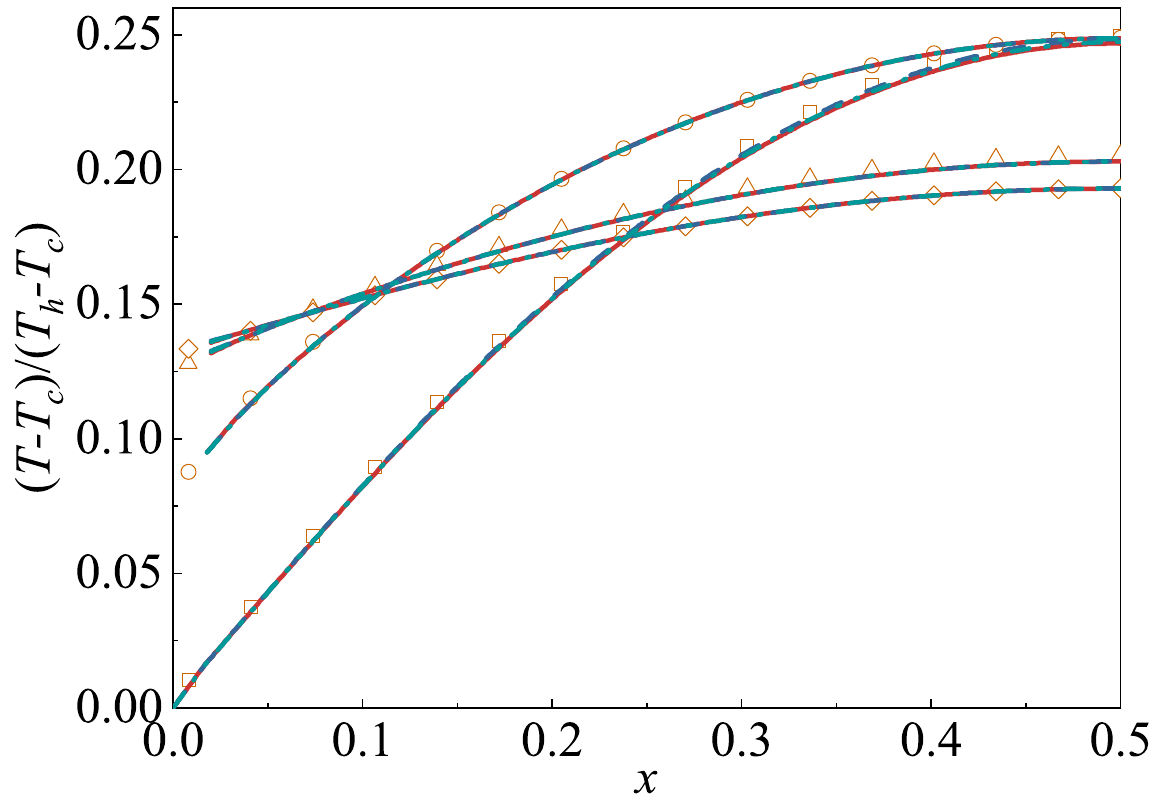}}
	\subfigure[]{
		\includegraphics[width=0.48\textwidth]{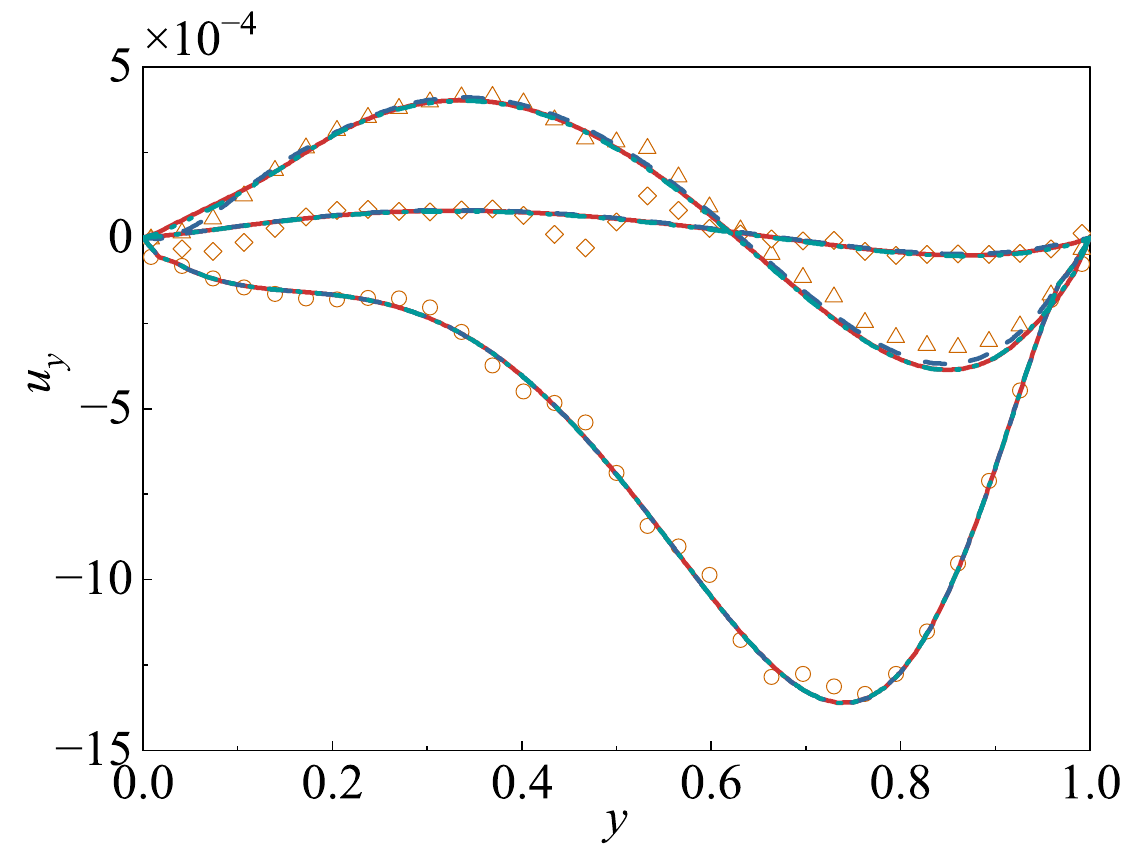}}
	\subfigure[]{
		\includegraphics[width=0.48\textwidth]{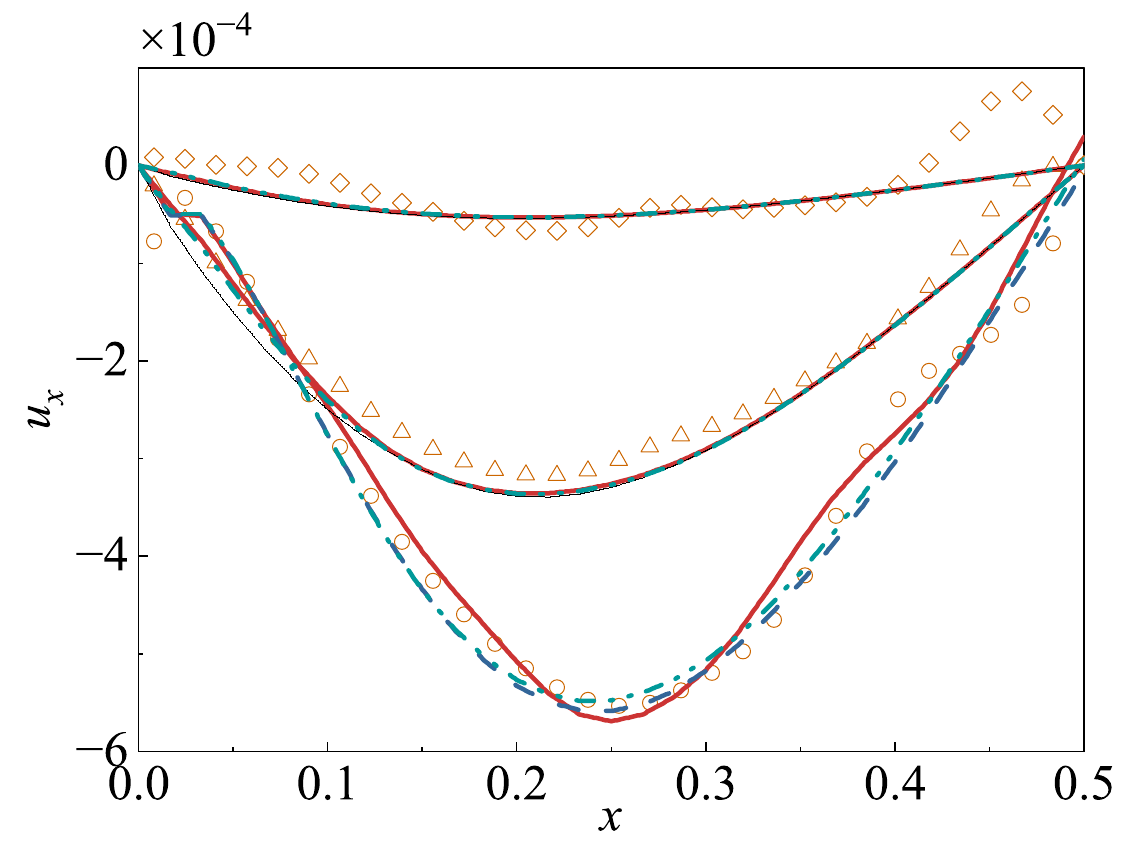}}
	\caption{\label{Fig4.2} Temperature and velocity profiles along the centerlines of the thermally driven cavity flow: (a) vertical temperature; (b) horizontal temperature; (c) vertical velocity; (d) horizontal velocity.}
\end{figure*}

The thermally driven cavity flow \cite{TDI2019} is considered to assess the influence of different azimuthal velocity distributions on the accuracy and robustness of the proposed velocity-space discretizations. The computational domain is a square cavity of side length $L=1$. The top wall is maintained at a temperature $T_h$, while the remaining three walls are kept at $T_c$, generating a thermally driven recirculating flow within the cavity. Fully diffuse reflection boundary conditions are imposed on all walls.

The Prandtl number are set to $Pr=2/3$, and the reference temperature is $T_{\mathrm{ref}}=300~\mathrm{K}$. Simulations are performed at $Kn=0.001$, 0.1, 1, and 10. For the near-continuum case of $Kn=0.001$, a weak thermal forcing is adopted, with $T_h=301~\mathrm{K}$ and $T_c=300~\mathrm{K}$, to minimize compressibility effects unrelated to the velocity-space discretization. For the more rarefied cases, $Kn=0.1$, 1, and 10, the temperature difference is increased to $T_h=400~\mathrm{K}$ and $T_c=200~\mathrm{K}$ to produce stronger nonequilibrium effects and thereby provide a more stringent assessment of the different velocity-space configurations.

The physical domain is discretized using a uniform $60\times60$ mesh. In velocity space, an exponential weight function is employed with $\alpha=\beta=500$, corresponding to the standard bell-shaped weight function. Eight radial nodes are distributed over $\xi\in[-5,5]$, while the numbers of azimuthal nodes are chosen as 8, 45, 70, and 90 for $Kn=0.001$, 0.1, 1, and 10, respectively. Three azimuthal arrangements are considered, namely, the \emph{aligned}, \emph{shifted}, and \emph{staggered} configurations.

Figure~\ref{Fig4.1} compare the temperature and velocity fields obtained with the three configurations at different Knudsen numbers. At $Kn=0.001$, the predicted temperature fields are nearly indistinguishable, whereas the velocity field remains more sensitive to the azimuthal arrangement. Notably, at the same $8\times8$ resolution, the \emph{shifted} and \emph{staggered} configurations yield results closer to those obtained using the denser aligned $8\times16$ discretization, with the \emph{staggered} configuration showing the closest agreement. At $Kn=0.1$, localized oscillations appear in the solution obtained with the aligned configuration, while both the shifted and staggered arrangements produce smoother fields and substantially reduce these oscillations. As the Knudsen number increases to $Kn=1$ and 10, the differences among the three configurations progressively diminish, and all configurations yield closely consistent temperature and velocity fields.

Overall, the \emph{shifted} and \emph{staggered} configurations provide improved performance relative to the \emph{aligned} arrangement, although the extent of this improvement depends on the flow regime. Their advantage is most pronounced in the near-continuum and transition regimes, where the distribution function varies relatively smoothly in velocity space and neighboring radial trajectories remain strongly correlated. Under these conditions, offsetting the azimuthal sampling locations improves velocity-space coverage and reduces redundant sampling. With increasing rarefaction, the velocity distribution becomes increasingly anisotropic and develops stronger local variations, placing greater demands on local velocity-space resolution. As a result, the benefit associated with the different azimuthal arrangements gradually decreases, and their numerical predictions become increasingly similar.

We further compare the temperature and velocity profiles along the horizontal and vertical centerlines with those obtained using the conventional half-range Gauss--Hermite (GH) and Newton--Cotes (NC)
quadratures, as shown in Fig.~\ref{Fig4.2}.
For low Knudsen numbers, the half-range GH quadrature is employed with \(8\times 8\) and
\(28\times 28\) velocity points for \(\mathrm{Kn}=0.001\) and \(0.1\), respectively.
At higher Knudsen numbers, where stronger nonequilibrium effects arise, the NC quadrature is used with
\(161\times 161\) and \(201\times 201\) velocity points for \(\mathrm{Kn}=1.0\) and \(10\), respectively, as the half-range GH quadrature becomes less effective in capturing these effects.

At \(\mathrm{Kn}=0.1\), the horizontal velocity profile obtained using the \emph{aligned} configuration exhibits some deviations from the other two configurations, whereas the \emph{shifted} and \emph{staggered} configurations yield nearly identical results.
This observation is consistent with the flow-field contours discussed above.
Apart from this difference, the temperature and velocity profiles obtained using all three configurations are in good agreement.
Notably, the flow velocities are as small as \(10^{-4}\), making their accurate resolution particularly challenging for conventional half-range GH and NC quadratures.
The corresponding profiles exhibit noticeable noise and oscillations, whereas the present framework
produces smooth velocity distributions. At higher Knudsen numbers, the present method achieves similarly smooth profiles using only approximately \(\mathbf{2\%}\) of the velocity points required by the NC
quadrature.
These results demonstrate the efficiency and robustness of the proposed framework in resolving both weak flow fields and strong nonequilibrium effects.

\subsection{3D Lid-Driven Cavity Flow}
\label{sec4.3}

\begin{figure*}[!th]
	\centering
	\subfigure[]{
		\includegraphics[width=0.32\textwidth]{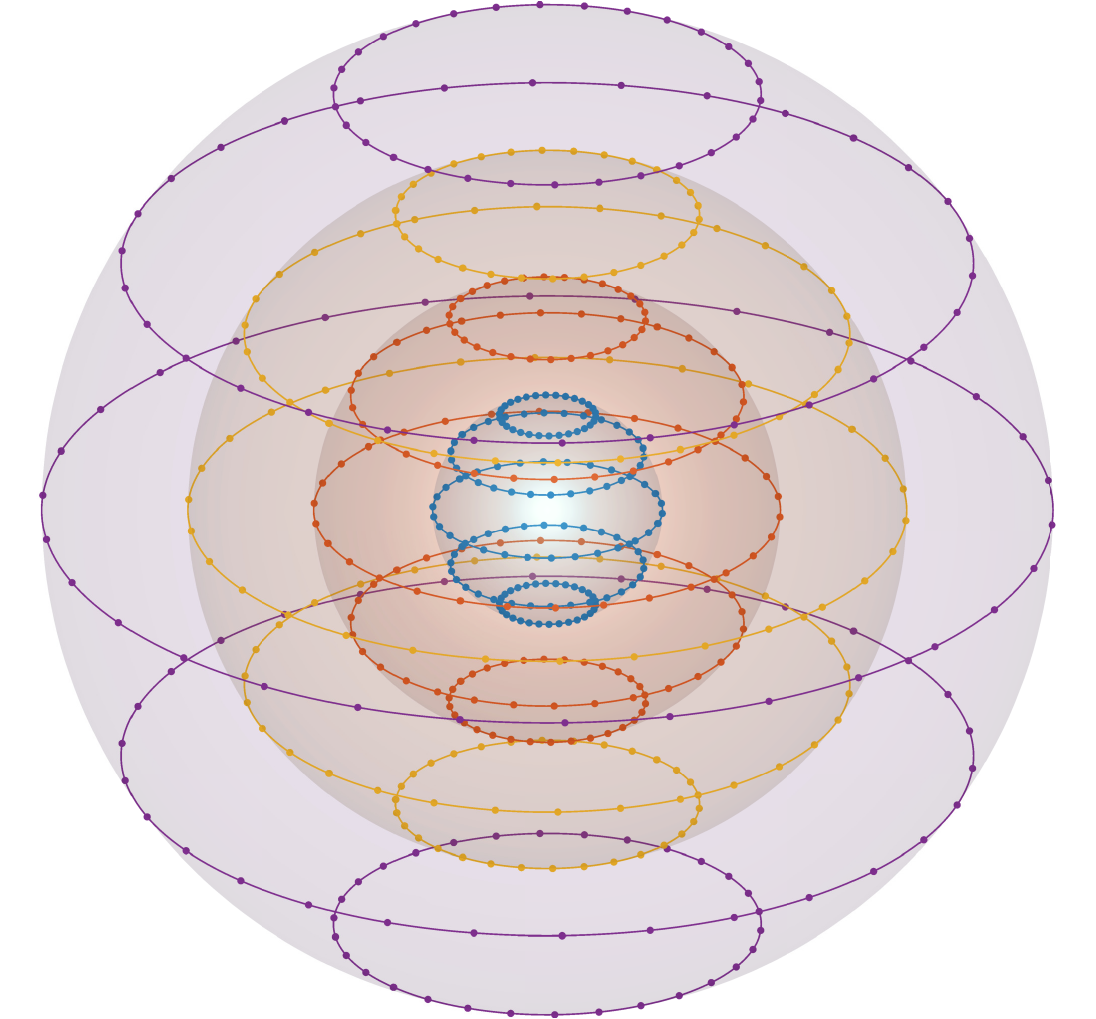}}
	\subfigure[]{
		\includegraphics[width=0.32\textwidth]{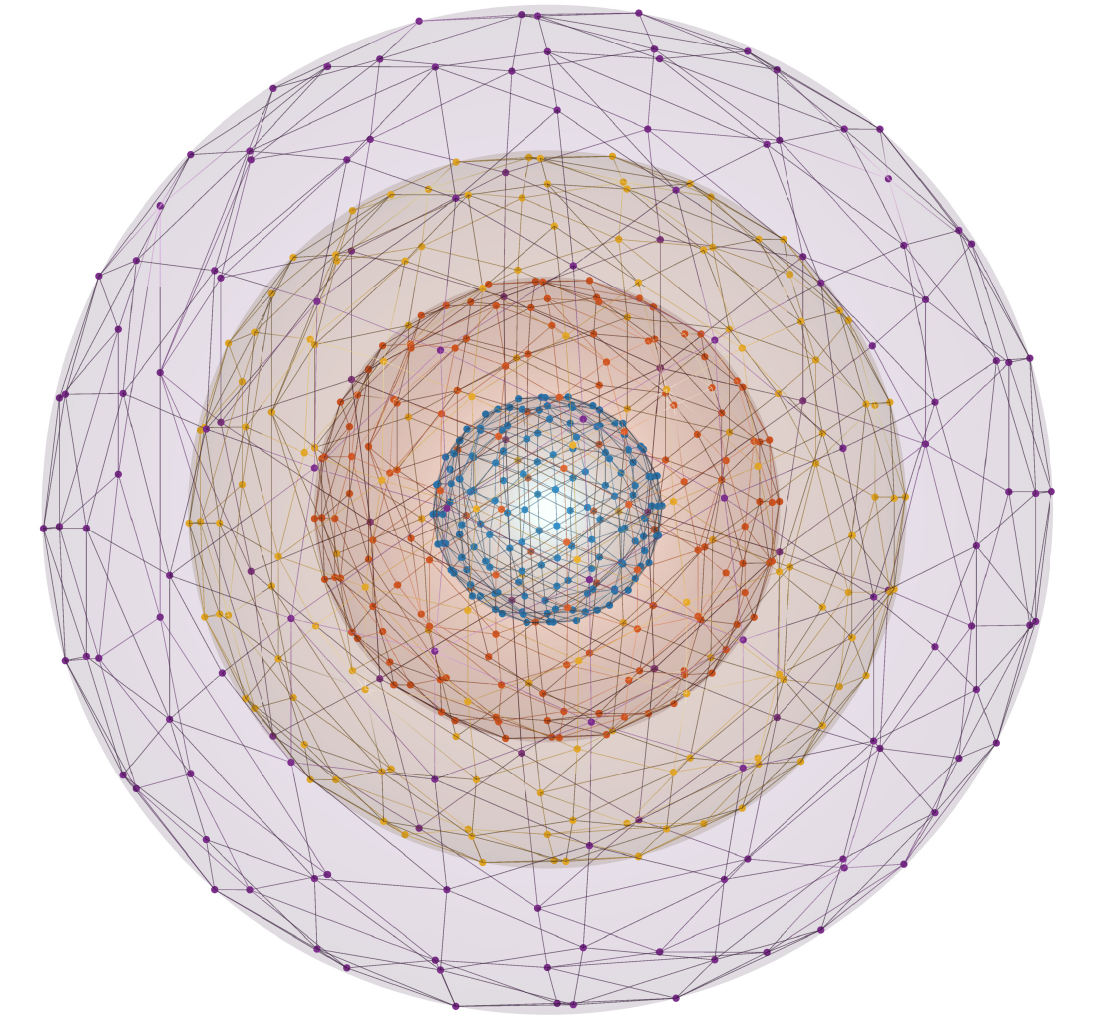}}
	\subfigure[]{
		\includegraphics[width=0.32\textwidth]{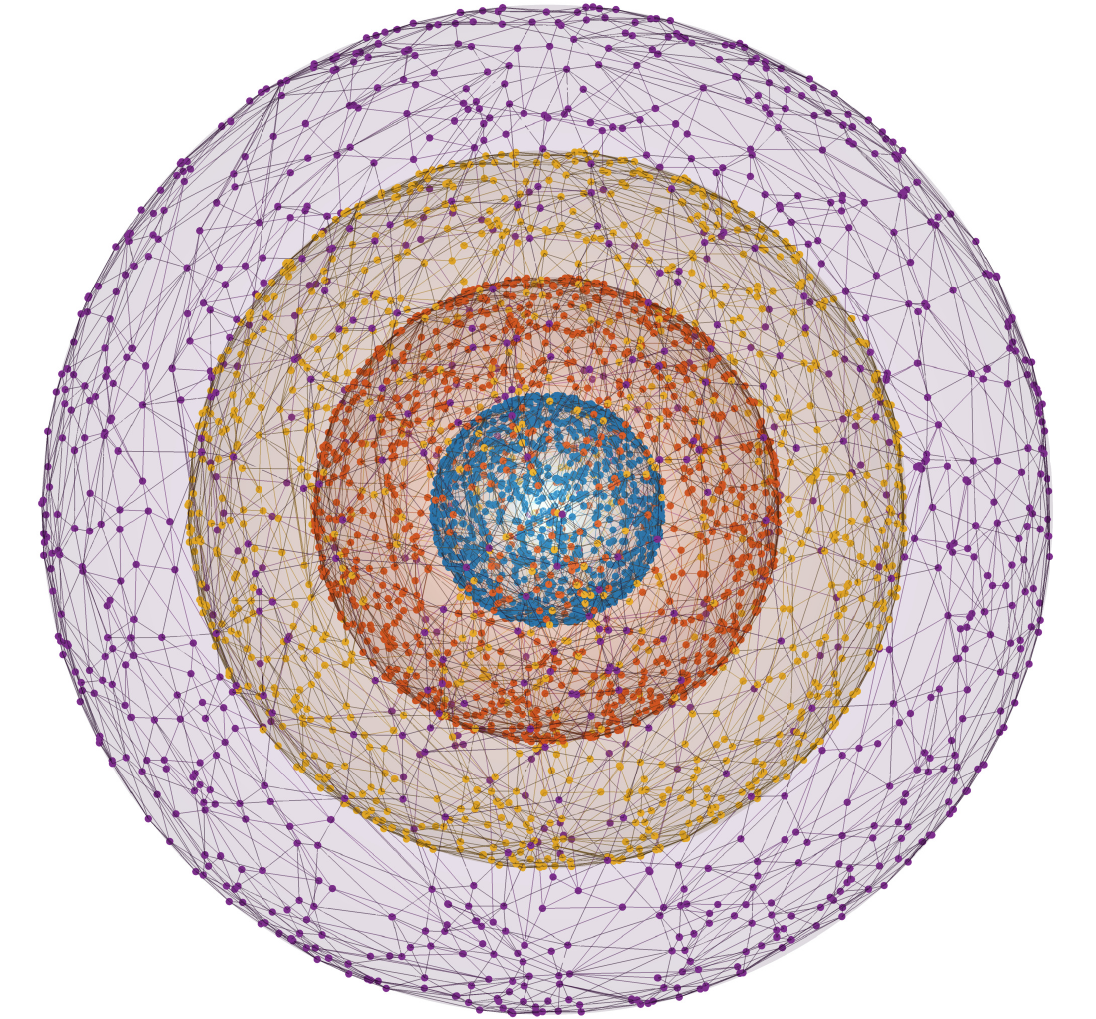}}
	\caption{Discrete velocity distributions of the 3D lid-driven cavity flow for $Kn=1$ and $10$; (a) Lat-Lon tensor-product,	(b) Lebedev, and (c) Monte Carlo distributions.}
	\label{4.3Fig1}
\end{figure*}

\begin{figure*}[!th]
	\centering
	\subfigure{
		\includegraphics[width=0.48\textwidth]{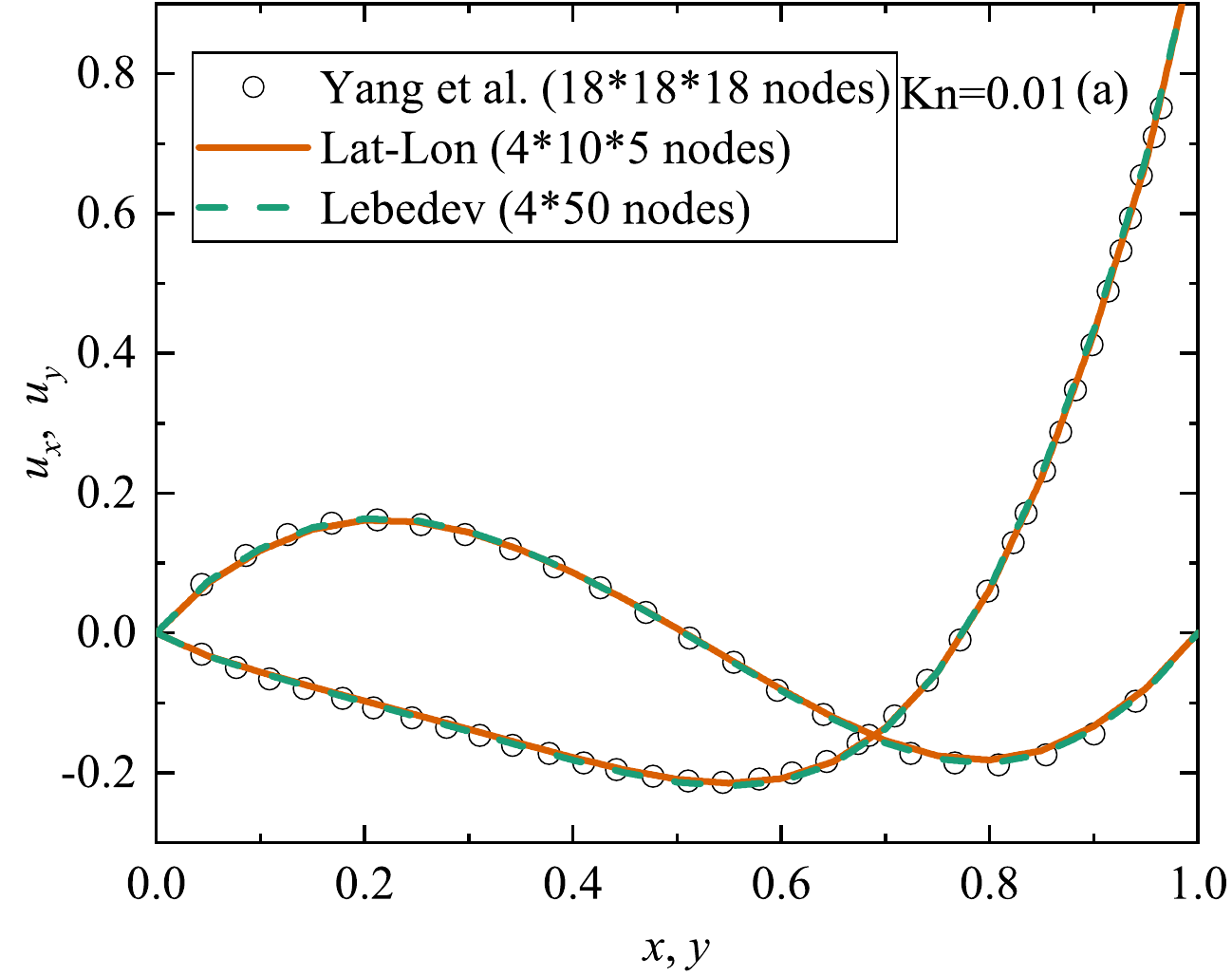}}
	\subfigure{
		\includegraphics[width=0.48\textwidth]{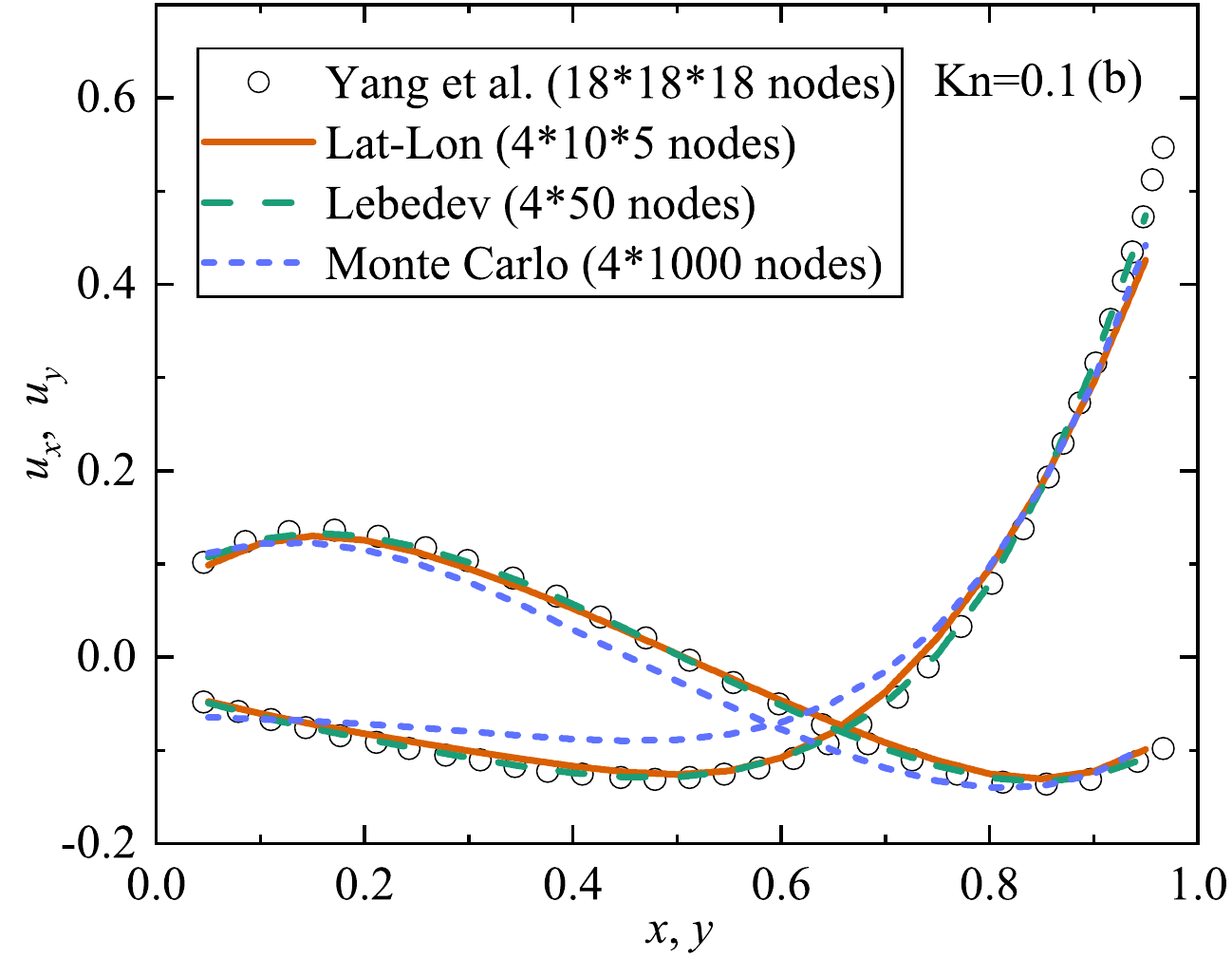}}
	\subfigure{
		\includegraphics[width=0.48\textwidth]{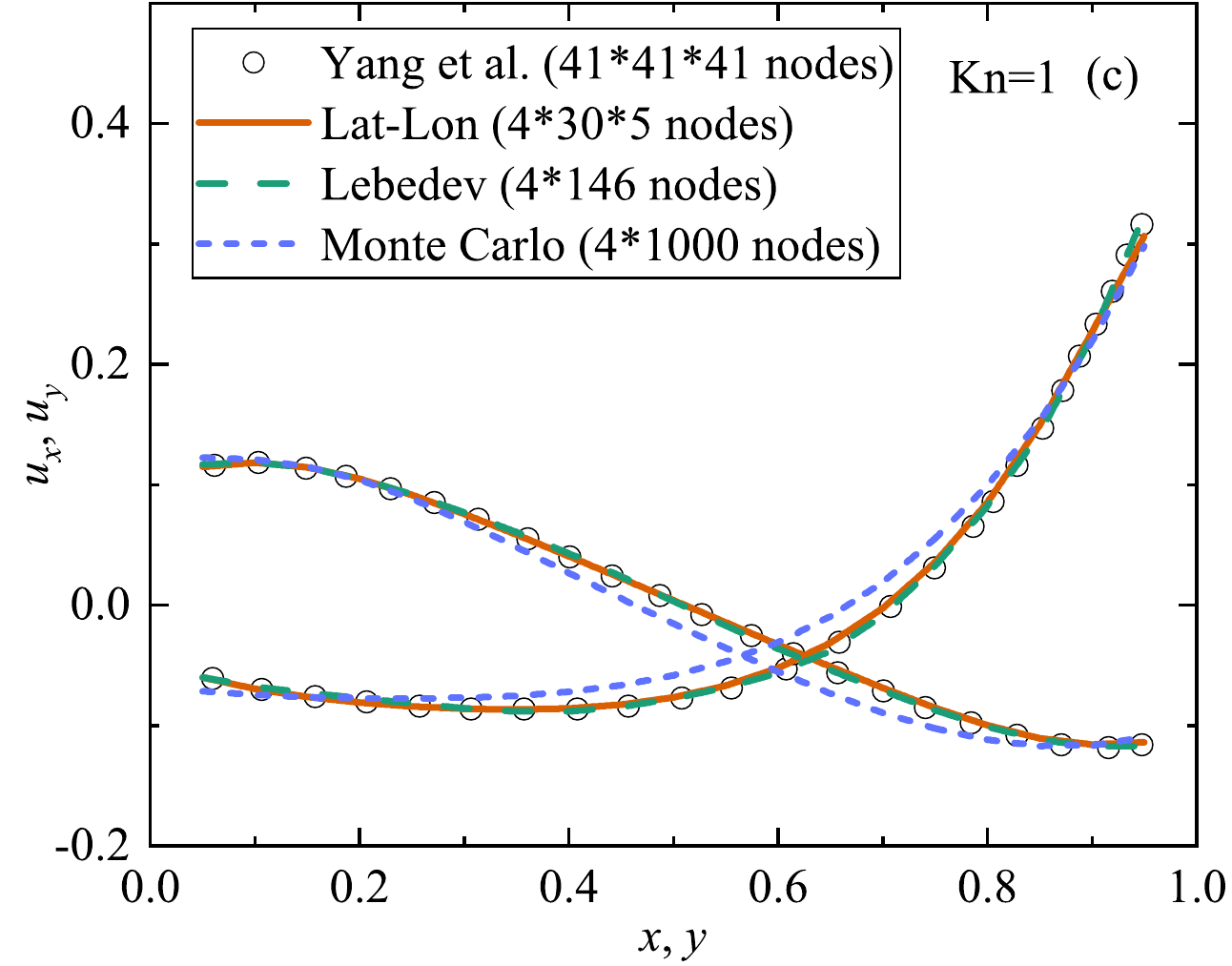}}
	\subfigure{
		\includegraphics[width=0.48\textwidth]{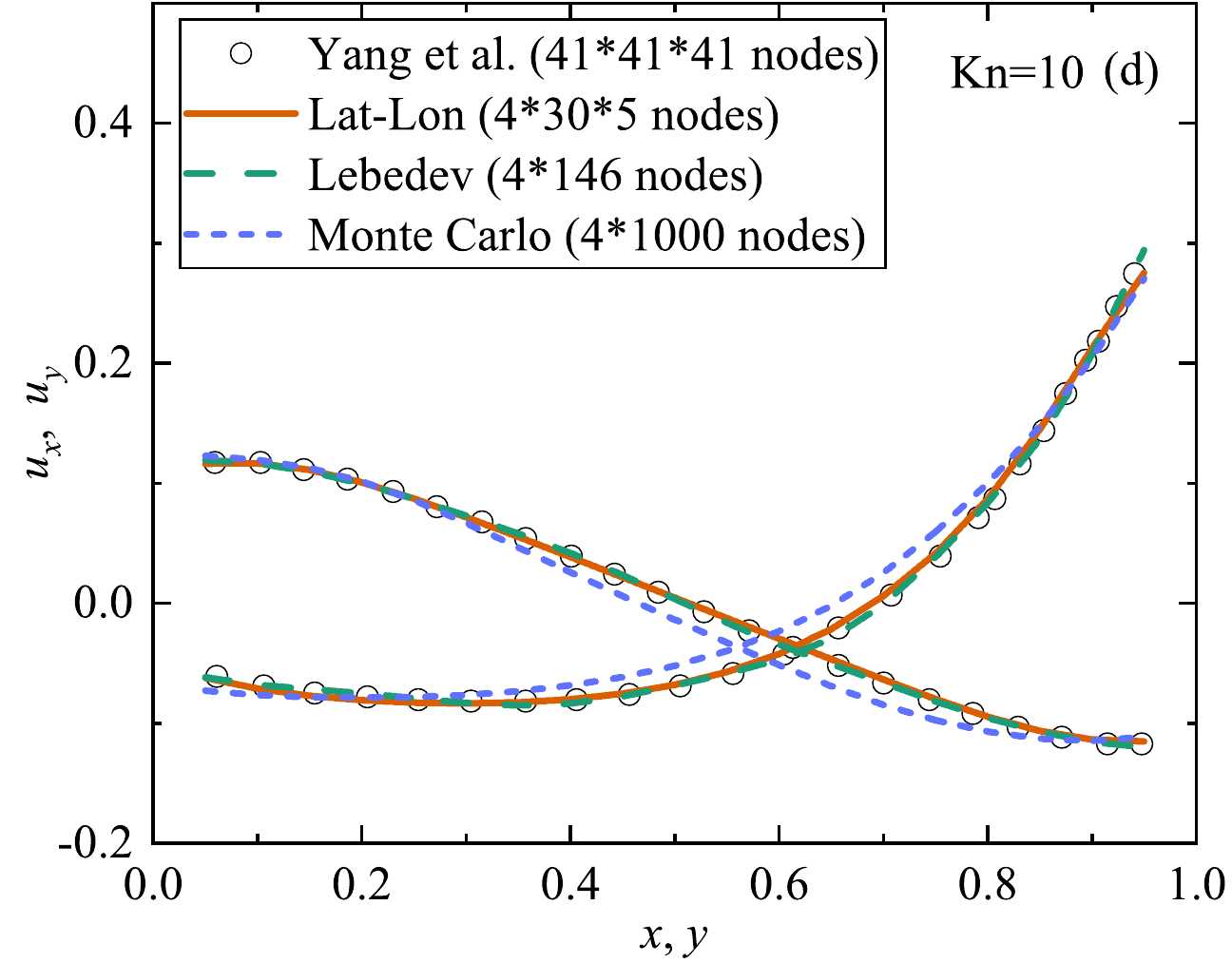}}
	\caption{\label{4.3Fig2} Comparison of velocity profiles for the three-dimensional lid-driven cavity flow.}
\end{figure*}

\begin{figure*}[!th]
	\centering
	
	\subfigure[$\mathrm{Kn}=0.1$, Lat--Lon]{
		\includegraphics[width=0.32\textwidth]{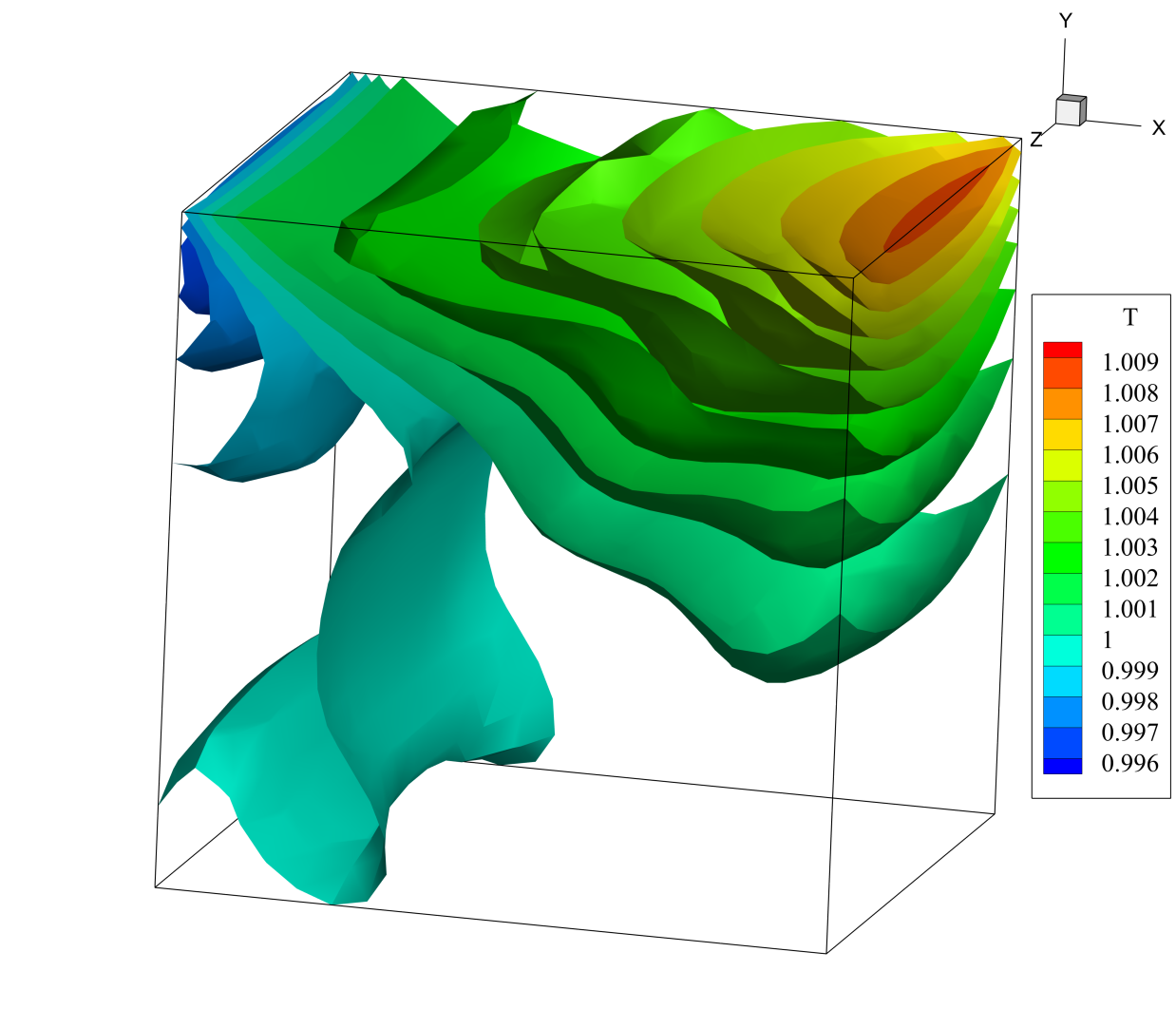}}
	\subfigure[$\mathrm{Kn}=0.1$, Lebedev]{
		\includegraphics[width=0.32\textwidth]{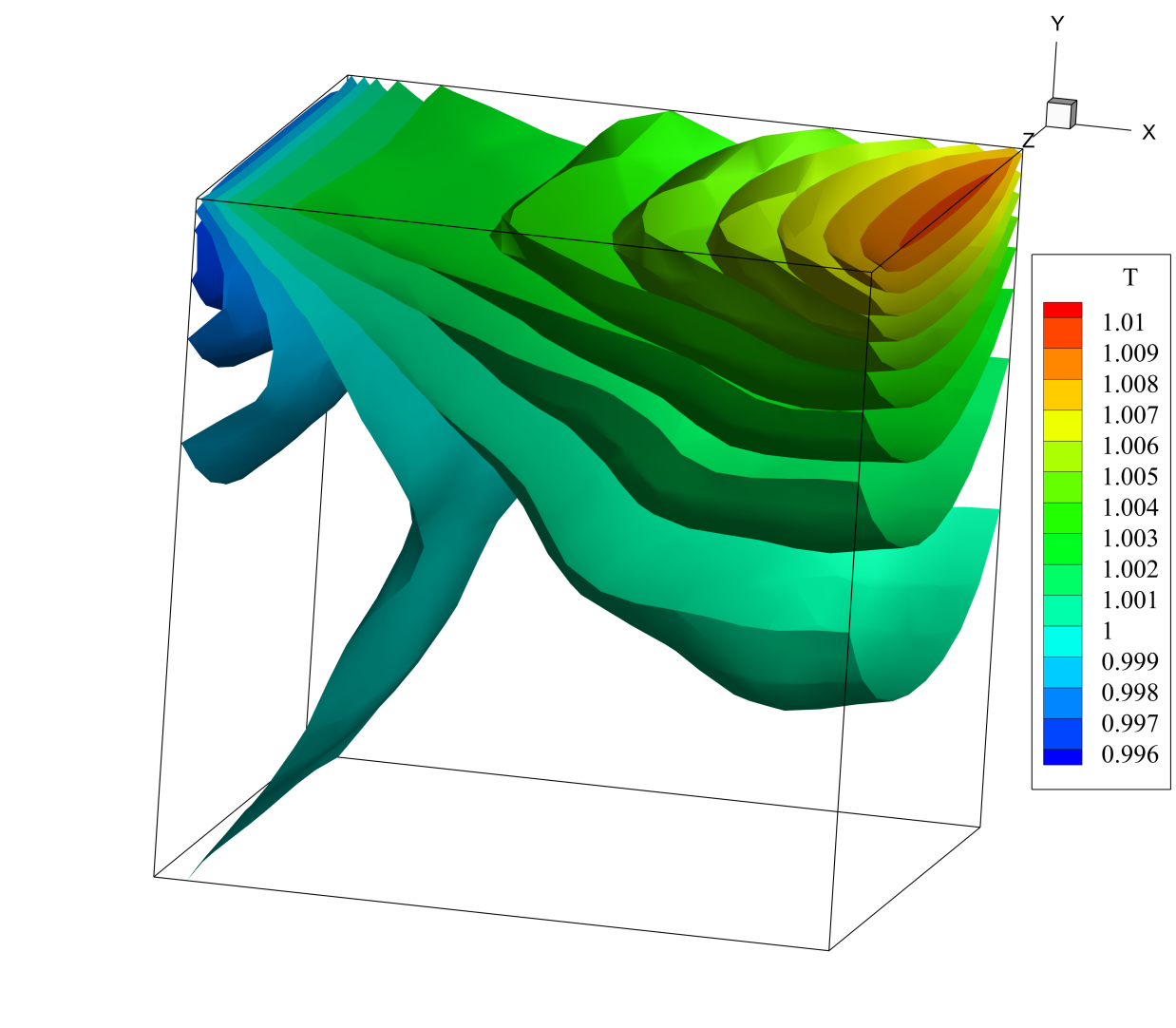}}
	\subfigure[$\mathrm{Kn}=0.1$, Monte Carlo]{
		\includegraphics[width=0.32\textwidth]{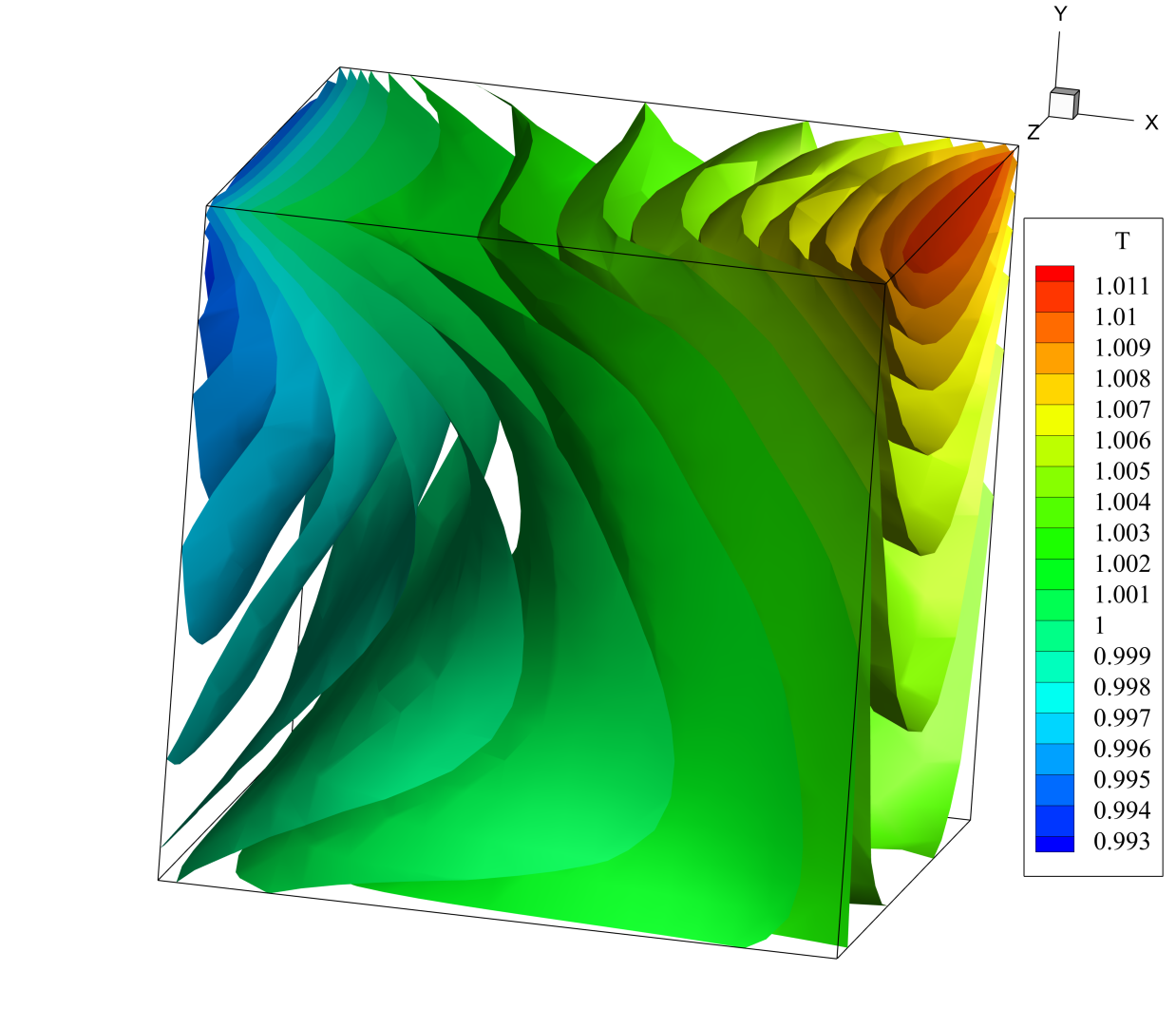}}
	
	\vspace{2mm}
	
	\subfigure[$\mathrm{Kn}=1.0$, Lat--Lon]{
		\includegraphics[width=0.32\textwidth]{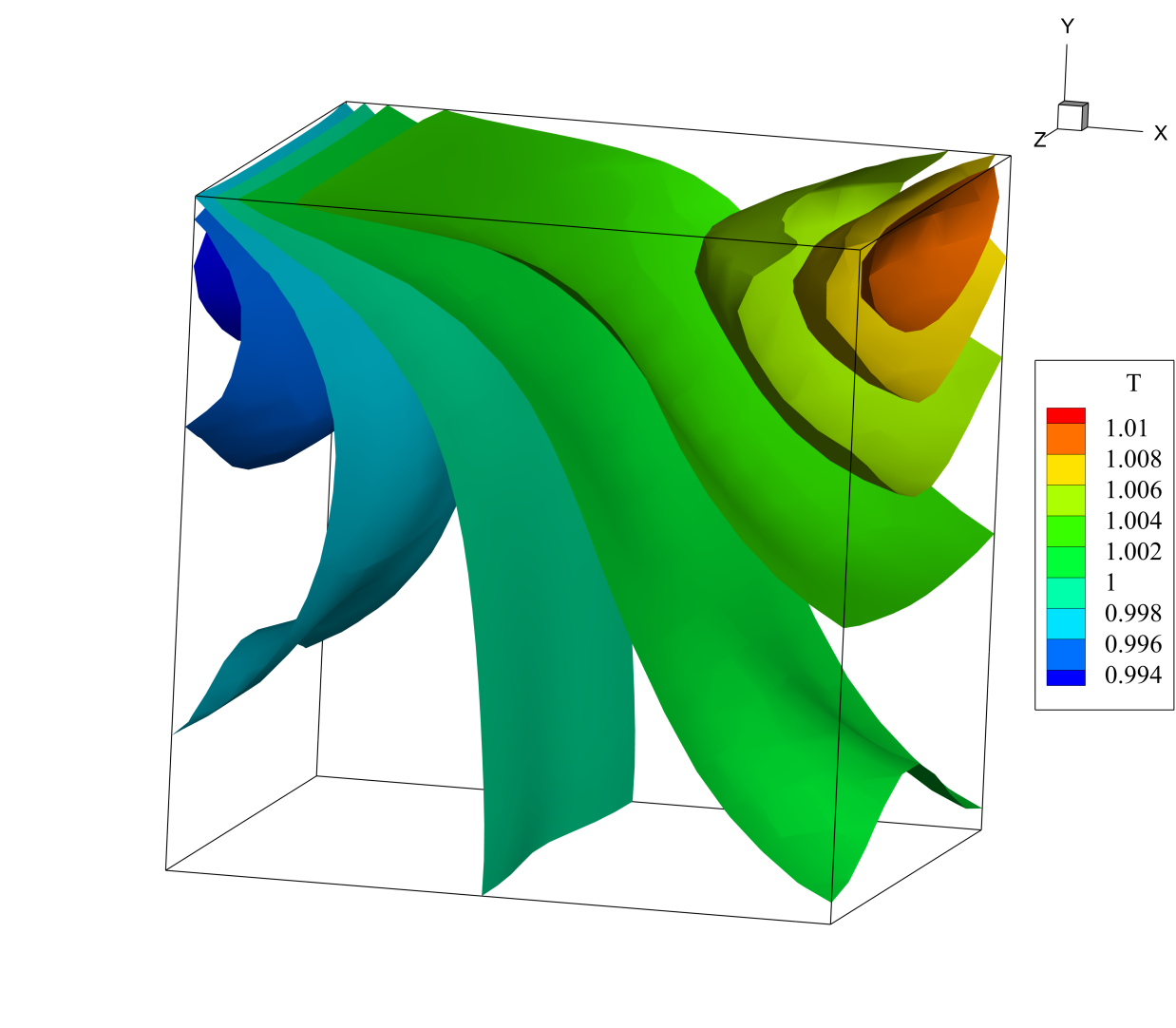}}
	\subfigure[$\mathrm{Kn}=1.0$, Lebedev]{
		\includegraphics[width=0.32\textwidth]{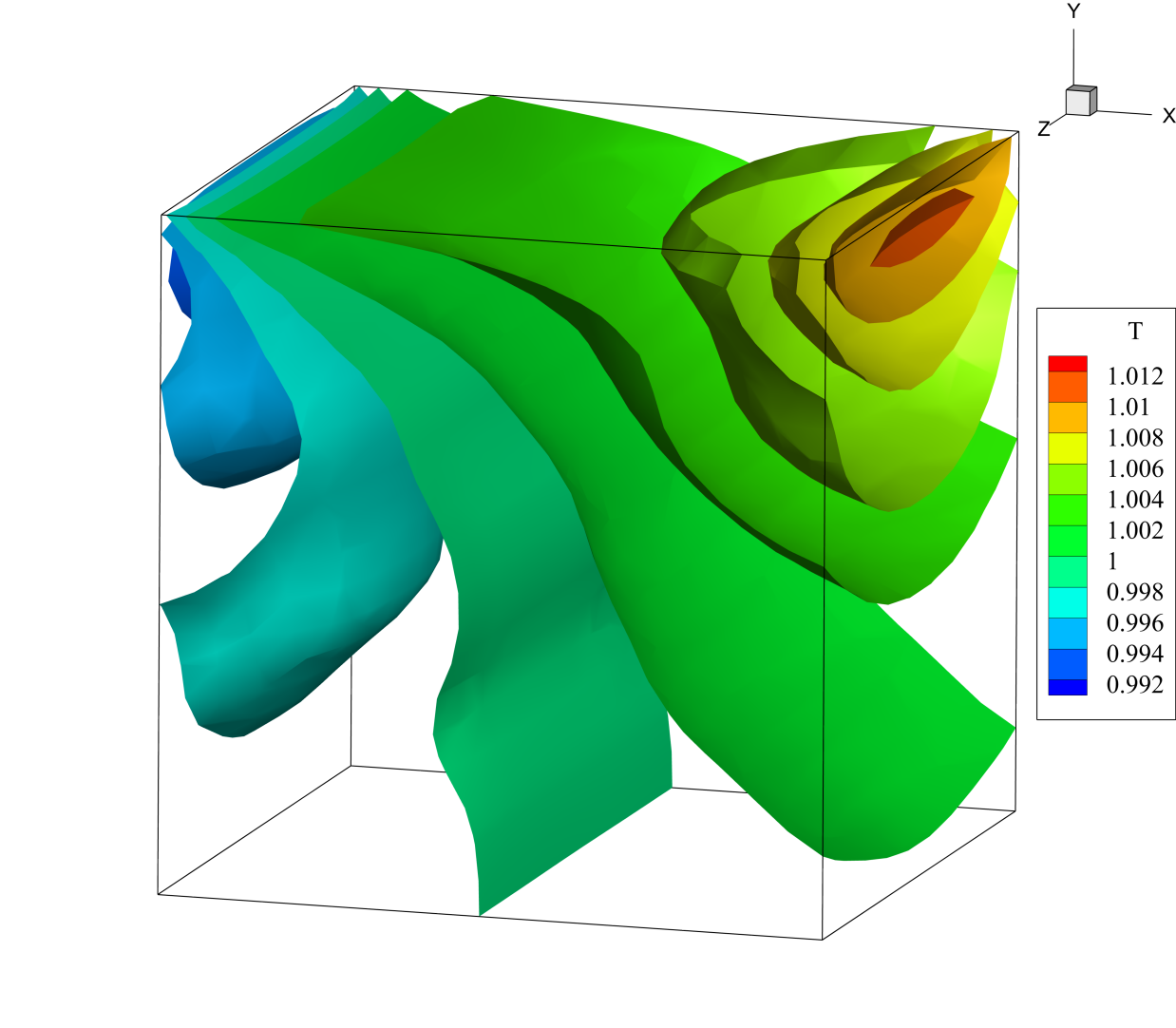}}
	\subfigure[$\mathrm{Kn}=1.0$, Monte Carlo]{
		\includegraphics[width=0.32\textwidth]{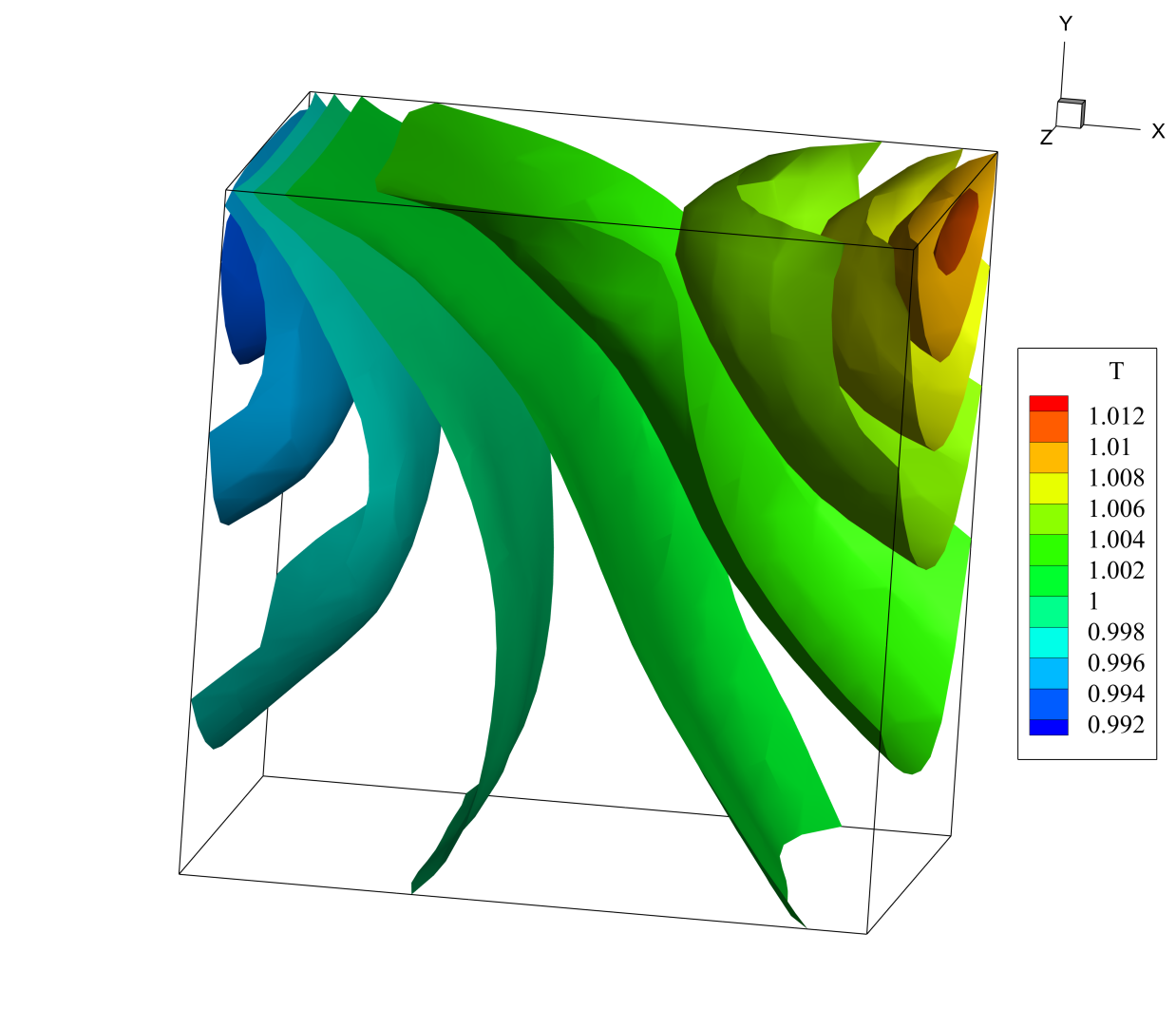}}
	
	\vspace{2mm}
	
	\subfigure[$\mathrm{Kn}=10$, Lat--Lon]{
		\includegraphics[width=0.32\textwidth]{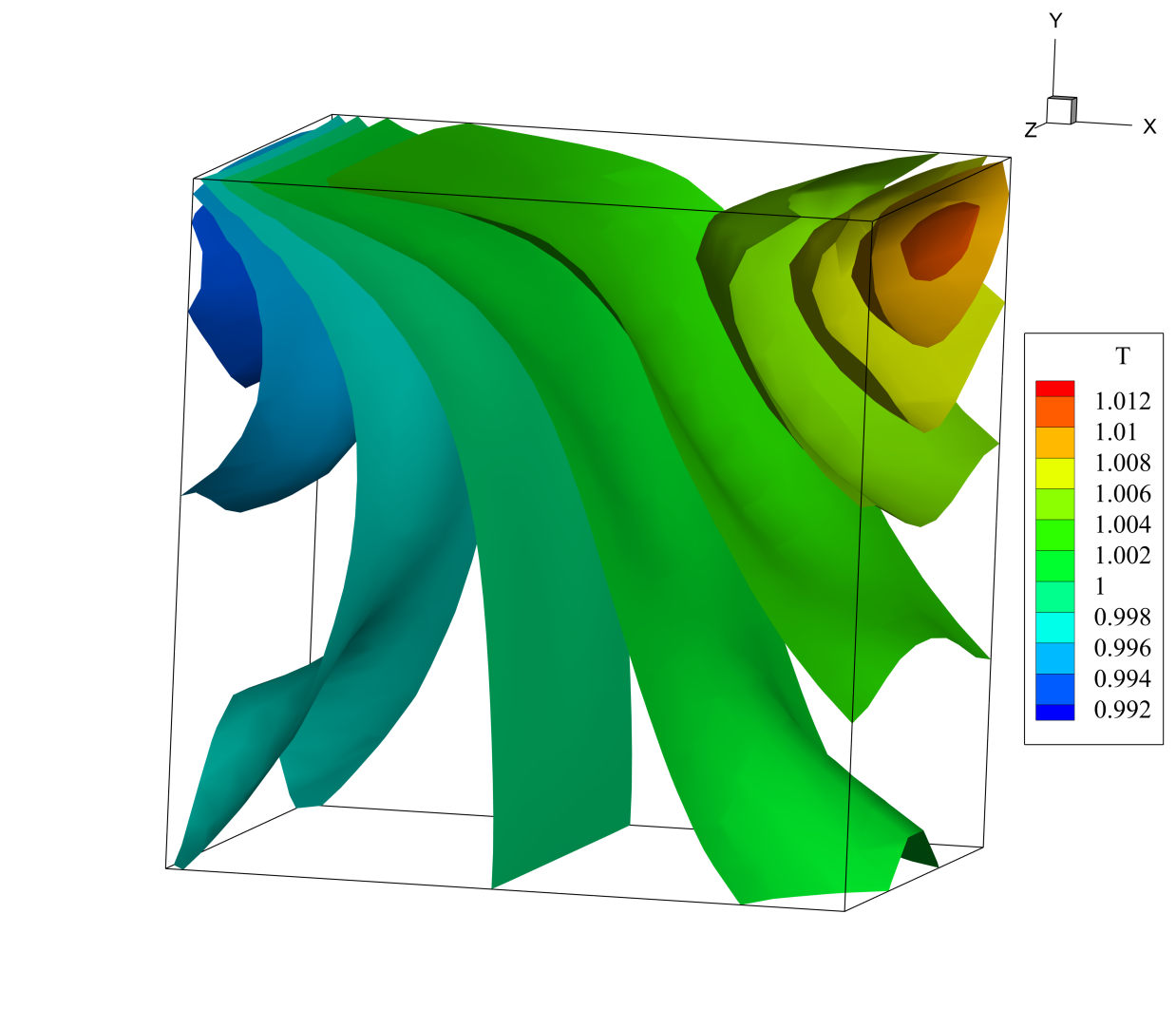}}
	\subfigure[$\mathrm{Kn}=10$, Lebedev]{
		\includegraphics[width=0.32\textwidth]{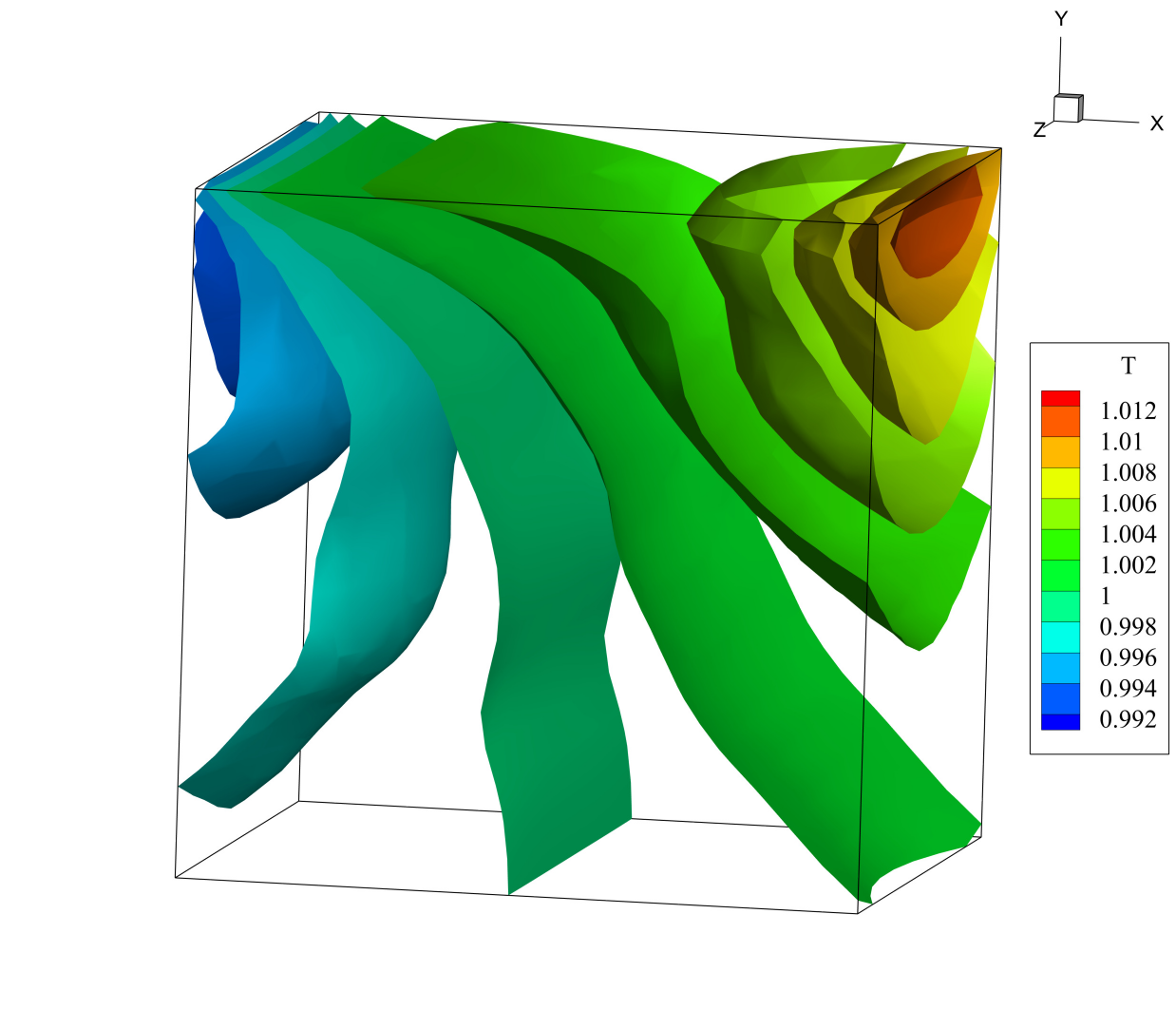}}
	\subfigure[$\mathrm{Kn}=10$, Monte Carlo]{
		\includegraphics[width=0.32\textwidth]{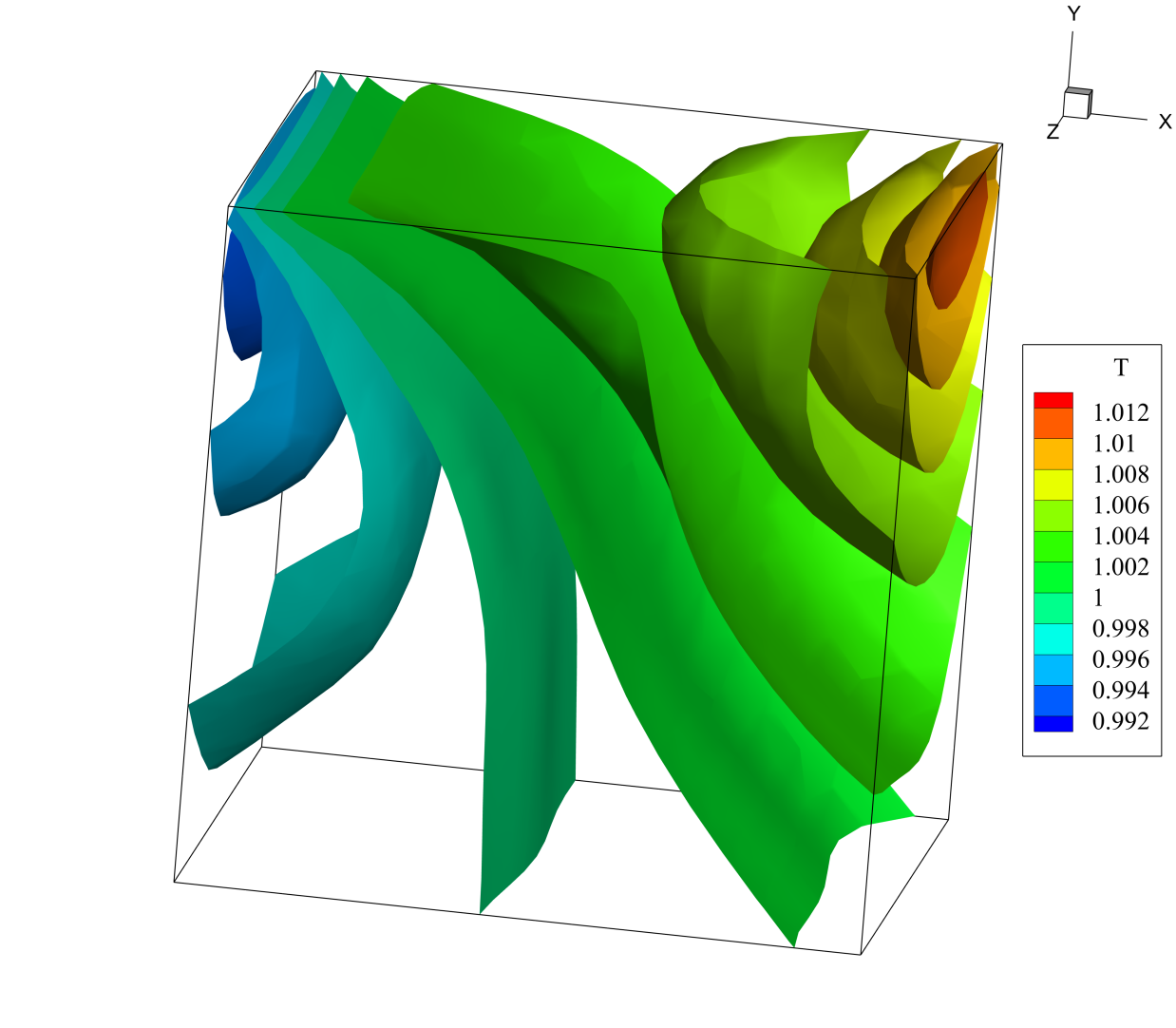}}
	
	\caption{Temperature contours of the 3D lid-driven cavity flow for
		$\mathrm{Kn}=0.1$, $1.0$, and $10$ with different angular
		discretizations. The rows correspond to $\mathrm{Kn}=0.1$, $1.0$,
		and $10$, respectively, while the columns correspond to
		Lat--Lon tensor-product, Lebedev, and Monte Carlo distributions,
		respectively.}
	\label{4.3Fig3}
\end{figure*}

The lid-driven cavity flow is considered to assess the proposed three-dimensional radial--directional quadrature framework, with particular emphasis on its compatibility with different spherical quadrature rules. The computational domain is a cubic cavity with side length \(L\). The top wall moves uniformly in the horizontal direction with a velocity of \(U_w=0.15\sqrt{2RT_{\mathrm{ref}}}\), while the remaining walls are stationary. All walls are maintained at the uniform reference temperature, \(T_w=T_{\mathrm{ref}}\).

Four representative Knudsen numbers, \(Kn=0.01\), \(0.1\), \(1\), and \(10\), are considered to examine the performance of the velocity-space discretization over a broad range of rarefaction conditions. The physical domain is discretized using a uniform \(20\times20\times10\) mesh for all cases. The radial velocity space is discretized using the Tanh-based quadrature with \(\alpha=\beta=1000\), yielding a maximum discrete radial velocity of approximately \(\xi_r\approx4\).
Different spherical quadrature rules are employed according to the degree of rarefaction. For $Kn=0.01$ and $0.1$, the angular discretization uses a $10\times5$ Lat--Lon tensor-product quadrature and an 11th-order Lebedev quadrature with 50 spherical nodes, respectively. For the more strongly rarefied cases, \(Kn=1\) and \(10\), the angular resolution is increased to a \(30\times5\) Lat--Lon quadrature and a 19th-order Lebedev quadrature with 146 spherical nodes, respectively. For comparison, a Monte Carlo (MC) spherical integration with 1000 uniformly distributed points is also employed for all four Knudsen numbers. The corresponding distributions of the spherical quadrature points are shown in Figure.~\ref{4.3Fig1}.

For all four Knudsen numbers, the velocity profiles obtained with both the Lat--Lon and Lebedev quadratures agree well with the reference solutions from the improved implicit DVM of Yang \textit{et al.}~\cite{cavity3D}, as shown in Figure.~\ref{4.3Fig2}. At the lower Knudsen numbers, \(Kn=0.01\) and \(0.1\), the Lebedev quadrature gives slightly closer agreement with the reference solutions. The corresponding isothermal surfaces are also smoother, whereas noticeable corrugations are observed with the Lat--Lon quadrature, as illustrated in shown in the first row of Figure.~\ref{4.3Fig3}. For the more rarefied cases, \(Kn=1\) and \(10\), this behavior is partially reversed: the Lat--Lon quadrature produces slightly smoother isothermal surfaces than the Lebedev quadrature, as shown in the second and third rows of Figure.\ref{4.3Fig3}.

In contrast, the MC spherical integration exhibits substantially larger numerical errors despite using 1000 angular points. At $Kn=0.01$, the computation becomes unstable and eventually diverges, even though the total number of discrete velocity points is approximately 20 times larger than that of the corresponding Lebedev discretization. For $Kn=0.1$, $1$, and $10$, the MC solutions also show appreciable
deviations from the reference solutions and from those obtained with the Lat--Lon and Lebedev quadratures. The dense angular sampling nevertheless produces sufficiently smooth isothermal surfaces,
indicating that visual smoothness alone does not guarantee accurate spherical integration. For the low-order spherical moments relevant to the present flow, the 1000-point MC integration exhibits errors on the order of $\mathcal{O}(N^{-1/2}) \sim 10^{-2}$, whereas the 146-point Lebedev rule integrates these moments to near machine precision. 

Overall, both the Lat--Lon and Lebedev quadratures can be effectively incorporated into the proposed radial--directional framework, yielding accurate predictions over the entire range of Knudsen numbers considered. Notably, at \(\mathrm{Kn}=1\) and \(10\), both approaches require only \(\mathbf{0.2\%}\) of the discrete velocity points used by the reference methods. Their relative performance, however, depends on the degree of rarefaction. The Lebedev quadrature exhibits a slight advantage at low Knudsen numbers, consistent with its highly uniform distribution of discrete directions over the spherical surface. As the flow becomes increasingly rarefied, the Lat--Lon quadrature provides comparable accuracy and slightly smoother isothermal surfaces. The MC results further demonstrate that a large number of angular samples alone is insufficient to ensure accurate and stable velocity-space integration, highlighting the importance of employing deterministic spherical quadratures with appropriate polynomial exactness within the proposed framework.

\subsection{Supersonic Flow Past a Sphere}
\label{sec4.4}

We next examine the applicability of the proposed bell-shaped quadrature framework to three-dimensional rarefied flows at high Mach numbers, with particular emphasis on how different mappings affect the radial
velocity-space discretization and the resulting flow predictions. The supersonic flow past a sphere is selected as a representative test case. In high-Mach-number flows, the particle velocity distribution extends over a relatively broad range, making an adequate representation of the high-velocity tail essential and rendering the required velocity-space discretization a major challenge for deterministic Boltzmann solvers.

\begin{figure*}[!th]
	\centering
	\subfigure[]{
		\includegraphics[width=0.45\textwidth]{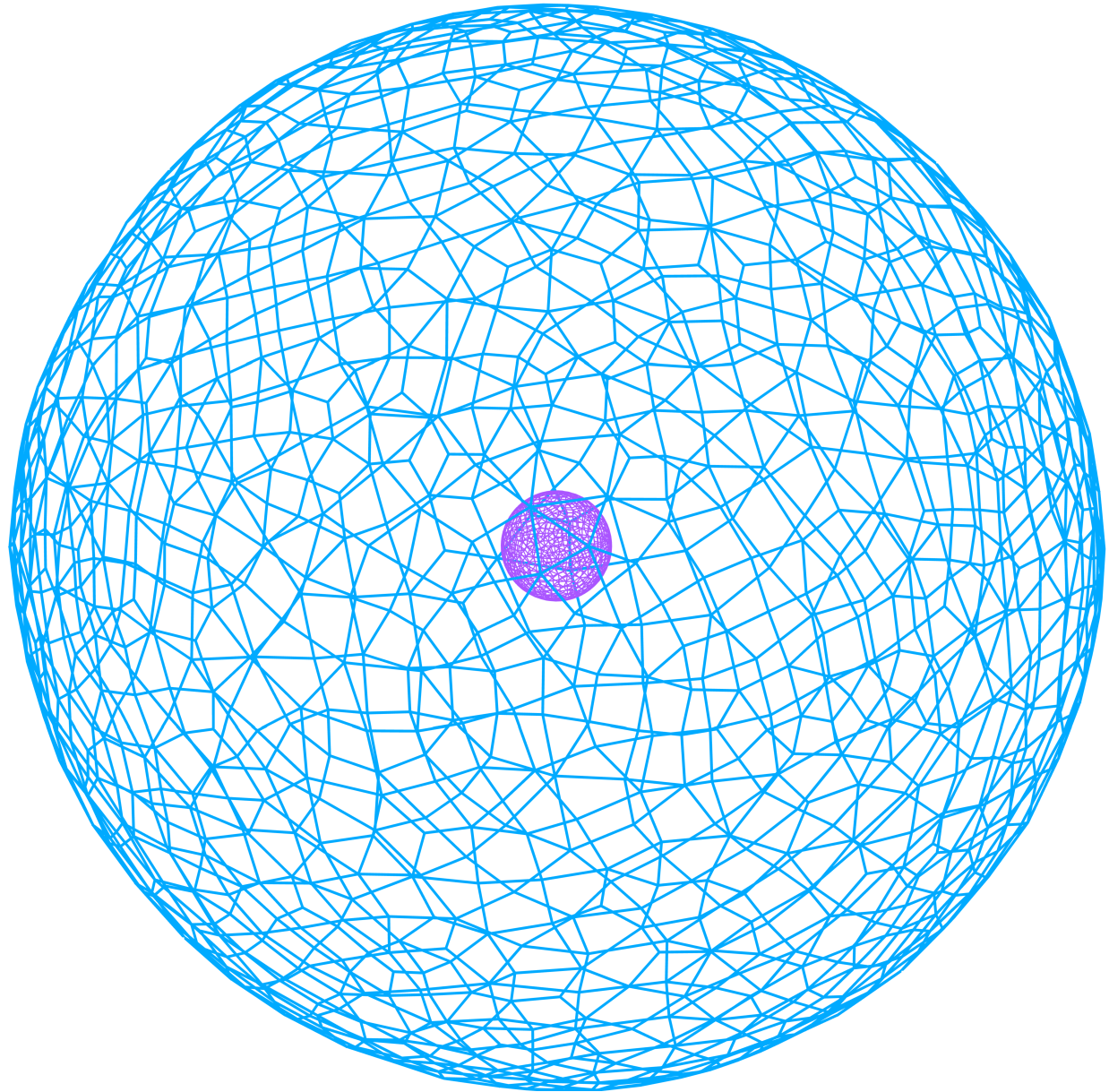}}	
	\hspace{0.01\textwidth}		
	\subfigure[]{
		\includegraphics[width=0.45\textwidth]{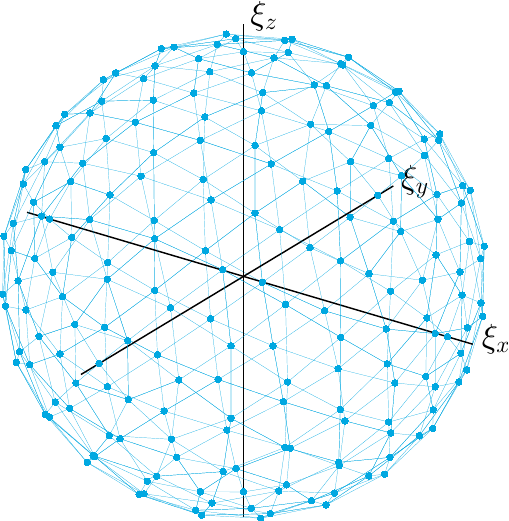}}
	\caption{Discretization of the computational domain for supersonic flow past a sphere: (a) physical domain and (b) directional velocity distribution.}
	\label{4.4Fig_sphere_domain}
\end{figure*}

The sphere has a radius of $r=1$ and is maintained at a uniform temperature of $T_w=1$. The far-field boundary is located at $9r$, with the freestream velocity specified as $U_{\infty}=Ma\sqrt{\gamma R T_{\infty}}$, where $Ma=2$. The freestream density and temperature are set to $\rho_{\infty}=1$ and $T_{\infty}=1$, respectively, and the Knudsen number based on the sphere radius is $Kn=0.1$. The spherical surface is discretized using an unstructured mesh containing approximately 1034 surface cells. The mesh is subsequently stretched in the wall-normal direction over 30 layers, resulting in a total of 31,020 computational cells, as shown in Figure.~\ref{4.4Fig_sphere_domain}(a).

The velocity space is then discretized within the proposed radial--directional framework. The directional space is represented by a 23rd-order Lebedev quadrature with 194 discrete directions, whose distribution over the unit sphere is shown in Figure.~\ref{4.4Fig_sphere_domain}(b). To isolate the effect of the radial mapping, three representative mappings are considered: the exponential (\emph{exp}) mapping, the Gudermannian (\emph{gd}) mapping, and the hyperbolic tangent (\emph{tanh}) mapping. The parameters are chosen as $\alpha=600$ and $\beta=350$, satisfying $\alpha/\beta>1$, so as to provide sufficient coverage of the high-velocity tail under the present high-Mach-number condition. With $N_r=11$ radial points, the resulting discrete radial velocities lie in the range $[-8,8]$.

\begin{figure*}[!th]
	\centering
	\includegraphics[width=0.95\textwidth]{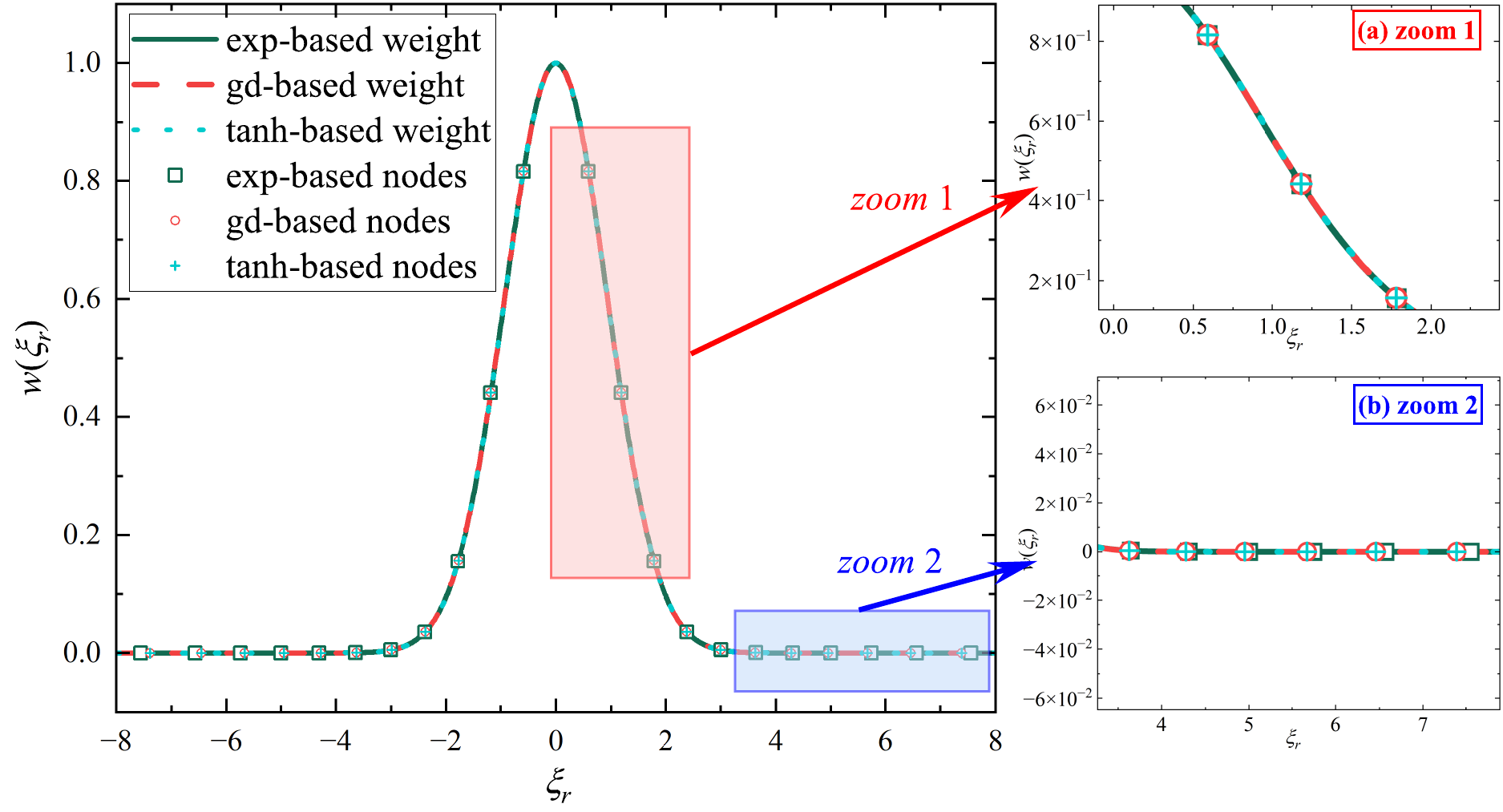}	
	\caption{Radial weight functions and velocity-space discretization for supersonic flow past a sphere.}
	\label{4.4Fig_sphere_radialNodes}
\end{figure*}

Figure~\ref{4.4Fig_sphere_radialNodes} compares the radial weight functions and the corresponding discrete velocity distributions obtained with the three mappings. Despite their different global
forms, the three mappings share the same asymptotic behavior in the vicinity of $\xi=0$, where the resulting radial weights approach a Gaussian form. Accordingly, their discrete points are nearly
indistinguishable in the low-velocity region. The differences become more pronounced toward the high-velocity tail, where the \emph{gd} and \emph{tanh}-based mappings exhibit very similar decay behavior, while both decay more rapidly than the \emph{exp}-based mapping. As a result, the former two mappings produce a relatively narrower distribution of radial discrete velocities. The implications of these differences for the accuracy and efficiency of discrete velocity methods warrant further investigation.

\begin{figure*}[!th]
	\centering
	\subfigure{
		\includegraphics[width=0.32\textwidth]{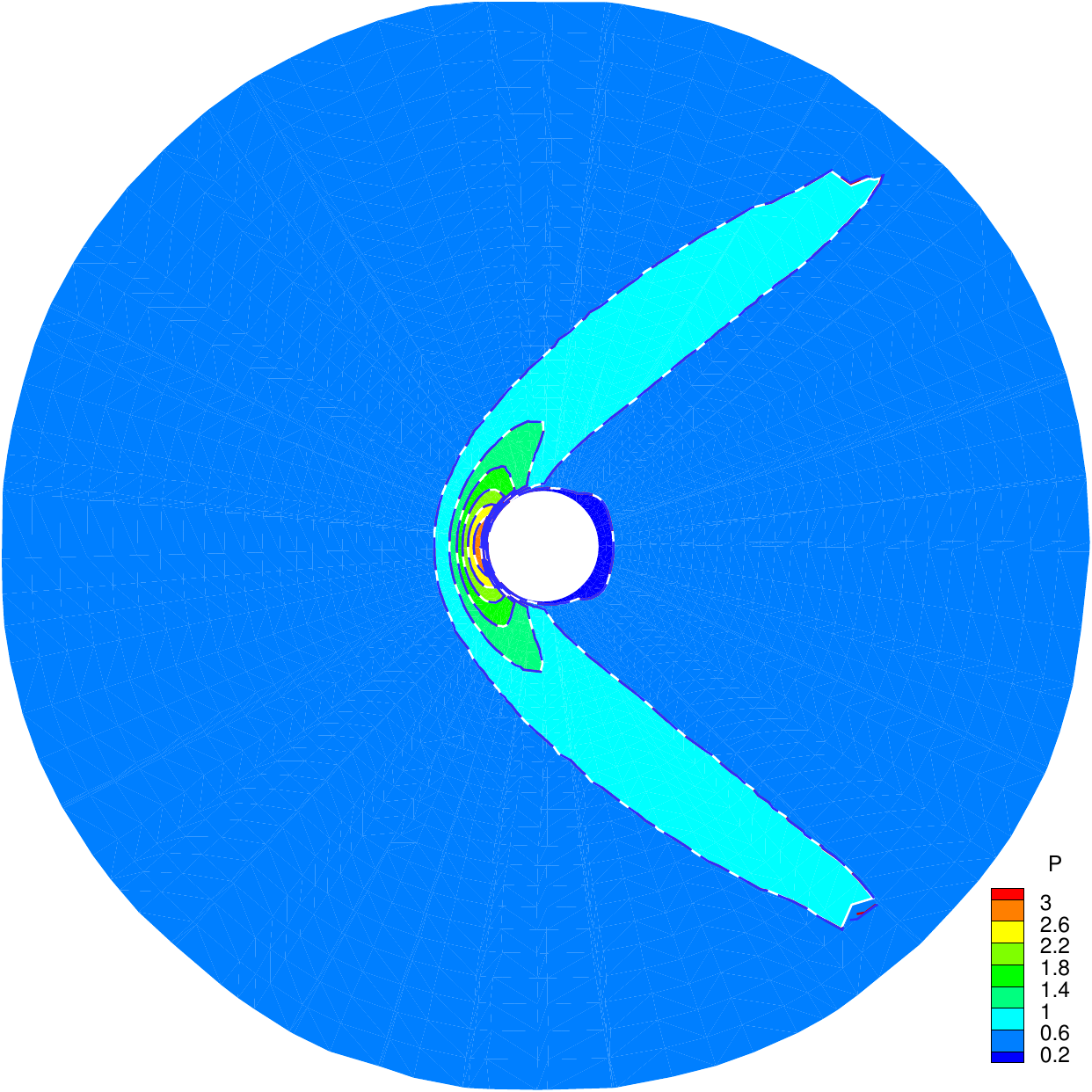}}
	\subfigure{
		\includegraphics[width=0.32\textwidth]{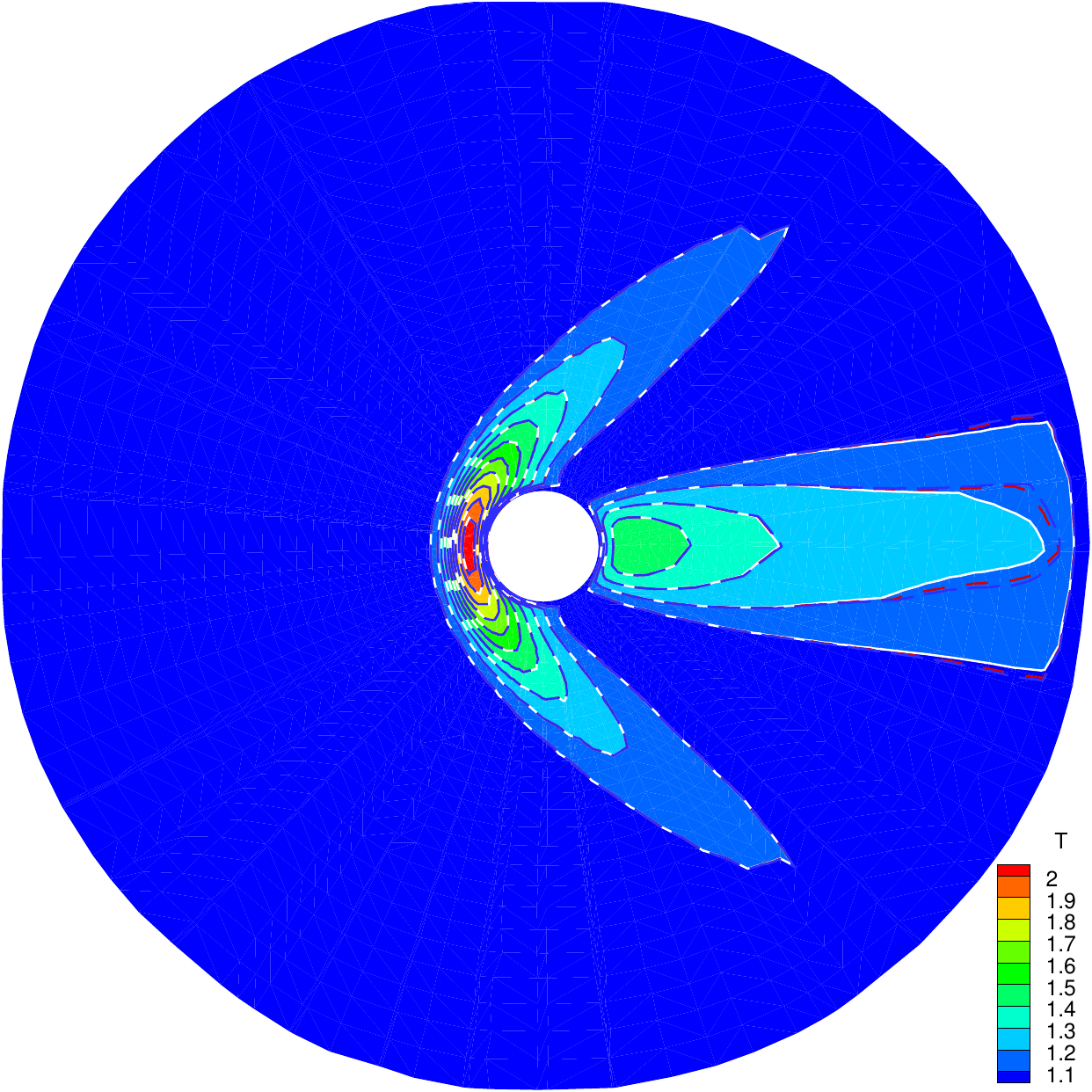}}
	\subfigure{
		\includegraphics[width=0.32\textwidth]{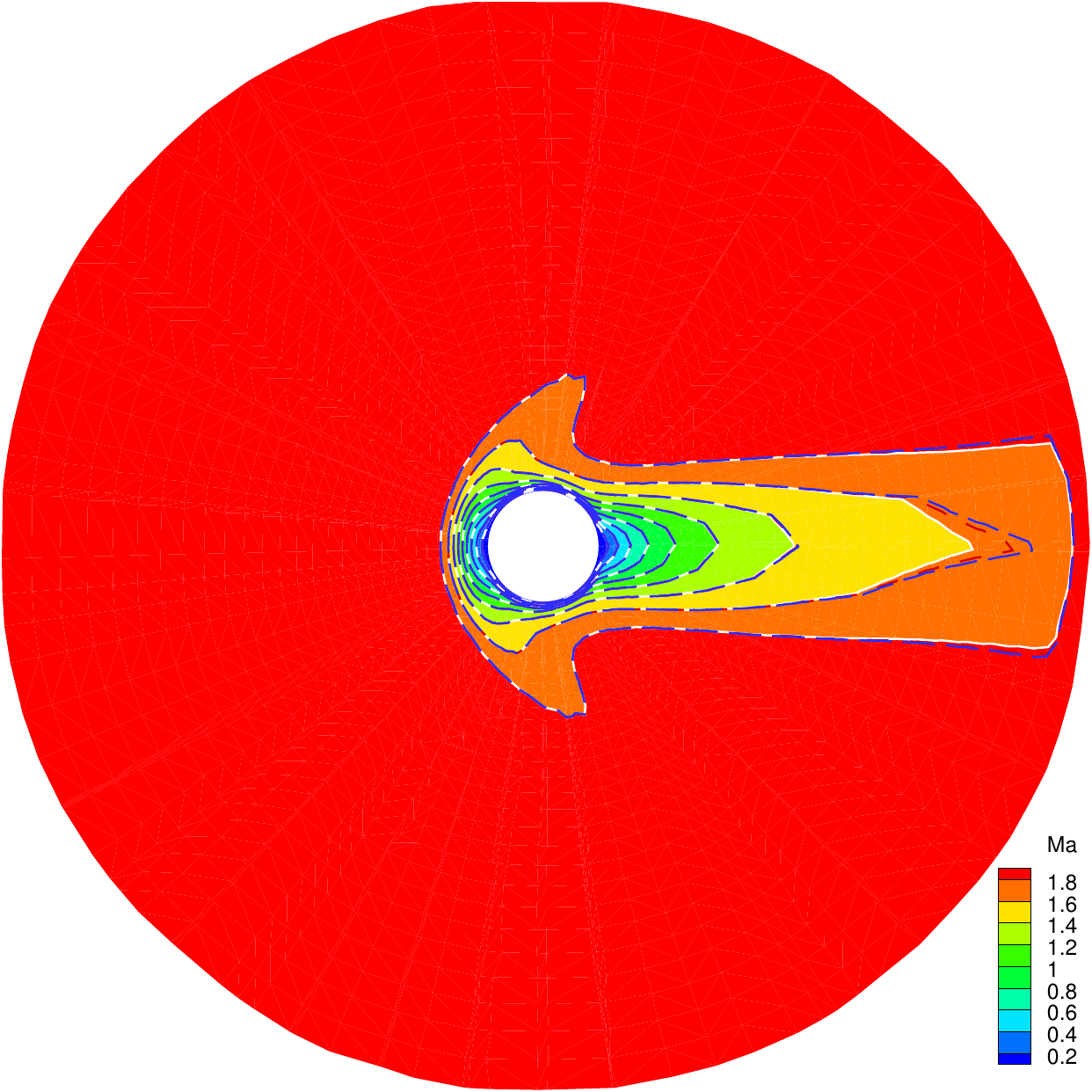}}
	\caption{Comparison of pressure, temperature, and Mach number contours for 3D supersonic flow past a sphere in the $z = 0$ plane at $Ma = 2$ and $Kn = 0.1$. (\emph{exp}-based: coloured background with white solid lines; \emph{gd}-based: red dashed lines; \emph{tanh}-based: blue long dash lines). (For interpretation of the references to colour in this figure legend, the reader is referred to the web version of this article).}
	\label{4.4Fig_velocity_distribution}
\end{figure*}

The influence of these differences on the computed flow field is illustrated in Figure.~\ref{4.4Fig_velocity_distribution}, which presents the pressure, temperature, and Mach-number contours on the $z=0$ plane. 
The flow solutions closely follow the characteristics of the underlying velocity-space discretization: the \emph{gd} and \emph{tanh}-based mappings produce nearly indistinguishable results, whereas the \emph{exp}-based mapping yields noticeable differences, particularly in the downstream region.
These results demonstrate that, although the three mappings exhibit nearly identical behavior near the origin, their treatment of the high-velocity tail can have a measurable impact on the macroscopic solution in high-Mach-number rarefied flows.

\section{Conclusion}
\label{sec5}

This work develops a Gaussian quadrature framework based on radial mappings that induce bell-shaped weight functions and the radial--directional decomposition of velocity-space integrals.
Within this framework, the radial mapping determines the resulting radial weight function and the associated Gauss--Jacobi quadrature, while the directional integration can be selected independently according to the dimensionality and geometric characteristics of the velocity space.
By choosing different radial mappings, several classical and recently developed quadratures can be incorporated into a common formulation. The same construction also provides a systematic means of developing new weight functions and corresponding quadratures.
The independent treatment of the directional integration further enhances the flexibility of the framework, allowing the radial quadrature to be combined with different directional discretizations, including the periodic trapezoidal rule in two dimensions and various spherical integration schemes in three dimensions, such as Lat--Lon, Lebedev, Fibonacci, and Monte Carlo discretizations.
Together, these elements establish a unified theoretical framework for velocity-space discretization in deterministic Boltzmann solvers.

The proposed framework is systematically assessed through a series of steady and unsteady benchmark problems spanning the full Knudsen-number range, including high-Mach-number, highly nonequilibrium flows, and employing both structured and unstructured physical-space meshes. Representative radial quadratures and directional discretizations are systematically compared. The results demonstrate the applicability of the framework to a broad range of flow regimes and computational settings, while allowing the radial and directional resolutions to be adjusted independently according to the requirements of the underlying flow. This flexibility enables efficient velocity-space discretization while preserving accuracy and numerical robustness across different flow regimes.

Future work will explore the properties of radial mappings to develop radial quadrature rules better suited to strongly nonequilibrium and non-Maxwellian velocity distributions. At the same time, high-accuracy spherical integration will be pursued through the optimization of directional point sets and quadrature weights. Advances along these complementary directions may further improve the efficiency and accuracy of velocity-space discretization in kinetic simulations.

\section{Acknowledgements}
Lu Wang acknowledges support from the Zhejiang Key Laboratory of Industrial Carbon Metrology Technology.

\bibliographystyle{model1-num-names}
\bibliography{refs}
\end{document}